\documentclass[10pt,a4paper]{iopart}
\usepackage{epsf}

\def\RRB{restricted repulsive barrier}
\def\EM{Einstein\discretionary{--}{--}{--}Maxwell}
\def\dS{de~Sitter}
\def\adS{anti\discretionary{-}{-}{-}de~Sitter}
\def\KN{Kerr\discretionary{--}{--}{--}Newman}
\def\KdS{Kerr\discretionary{--}{--}{--}de~Sitter}
\def\SdS{Schwarzschild\discretionary{--}{--}{--}de~Sitter}
\def\SadS{Schwarzschild\discretionary{--}{--}{--}%
  anti\discretionary{-}{-}{-}de~Sitter}
\def\RN{Reissner\discretionary{--}{--}{--}Nordstr\"om}
\def\RNdS{Reissner\discretionary{--}{--}{--}%
  Nordstr\"om\discretionary{--}{--}{--}de~Sitter}
\def\KNdS{Kerr\discretionary{--}{--}{--}Newman%
  \discretionary{--}{--}{--}de~Sitter}
\def\KNadS{Kerr\discretionary{--}{--}{--}Newman%
  \discretionary{--}{--}{--}anti\discretionary{-}{-}{-}de~Sitter}
\let\ri=\rm
\def\d{{\rm d}} 
\def\oder#1#2{\frac{\d #1}{\d #2}}
\def\pder#1#2{\frac{\partial #1}{\partial #2}}
\def\pdder#1#2{\frac{\partial^2 #1}{\partial #2^2}}
\def\lpder#1#2{{\partial #1}/{\partial #2}}
\def\lpdder#1#2{{\partial^2 #1}/{\partial #2^2}}
\def\be{\begin{equation}}  \def\ee{\end{equation}}
\def\bea{\begin{eqnarray}} \def\eea{\end{eqnarray}}

\begin{document}
\jl{6}
\title[Equatorial photon motion in the KN spacetimes with $\Lambda\neq0$]%
  {Equatorial photon motion in the Kerr--Newman spacetimes with
  a~non-zero cosmological constant%
  \footnote{Published in Class.\ Quantum Grav.\ \textbf{17} (2000),
    pp.~4541--4576.}}
\author{Z Stuchl\'{\i}k\footnote{E-mail address: Zdenek.Stuchlik@fpf.slu.cz},
  and S Hled\'{\i}k\footnote{E-mail address: Stanislav.Hledik@fpf.slu.cz}}
\address{Institute of Physics, Faculty of Philosophy and Science,
  Silesian University in~Opava, Bezru\v{c}ovo n\'am.~13,
  CZ-746\,01~Opava, Czech Republic}

\begin{abstract}  
  Discussion of the equatorial photon motion in \KN\ black-hole and
  naked-singularity spacetimes with a non-zero cosmological constant is
  presented. Both repulsive and attractive cosmological constants are
  considered. An appropriate `effective potential' governing the photon
  radial motion is defined, circular photon orbits are determined, and
  their stability with respect to radial perturbations is established. The
  spacetimes are divided into separated classes according to the properties
  of the `effective potential'. There is a special class of \KNdS\ 
  black-hole spacetimes with the \RRB. In such spacetimes, photons with
  high positive and all negative values of their impact parameter can
  travel freely between the outer black-hole horizon and the cosmological
  horizon due to an interplay between the rotation of the source and the
  cosmological repulsion.  It is shown that this type of behavior of the
  photon motion is connected to an unusual relation between the values of
  the impact parameters of the photons and their directional angles
  relative to outward radial direction as measured in the locally
  non-rotating frames. Surprisingly, some photons counterrotating in these
  frames have positive impact parameter. Such photons can be both escaping
  or captured in the black-hole spacetimes with the \RRB. For the
  black-hole spacetimes with a standard, divergent repulsive barrier of the
  equatorial photon motion, the counterrotating photons with positive
  impact parameters must all be captured from the region near the
  black-hole outer horizon as in the case of Kerr black holes, while they
  all escape from the region near the cosmological horizon. Further, the
  azimuthal motion is discussed and photon trajectories are given in
  typical situations. It is shown that for some photons with negative
  impact parameter turning points of their azimuthal motion can exist.
\end{abstract}


\maketitle

\section{Introduction}

The \KNdS\ and \KNadS\ solutions of \EM\ equations
represent black holes and naked singularities in spacetimes with a non-zero
cosmological constant $\Lambda$. For a repulsive cosmological constant,
$\Lambda>0$, the geometry is asymptotically \dS, and, generally, contains a
cosmological horizon behind which the geometry must be dynamic. For an
attractive cosmological constant, $\Lambda<0$, the geometry is
asymptotically \adS, and can contain black-hole horizons only.

A wide variety of recent cosmological observations, including, e.g.,
measurements of the present value of the Hubble parameter, measurements of
the anisotropy of the cosmic relic radiation, statistics of the
gravitational lensing of quasars and active galactic nuclei, and the
high-redshift supernovae, suggest that in the framework of the inflationary
cosmology a non-zero repulsive cosmological constant $\Lambda>0$ has to be
considered seriously in order to explain properties of the presently
observed universe \cite{Krauss-Turner,Ostriker-Steinhart,Kochanek,Krauss}.

On the other hand, it was recognized recently that the \adS\ spacetime
plays an important role in the multidimensional string theory
\cite{Horowitz-Myers,Martinec,Sen}. Therefore, solutions of the
\EM\ equations with both positive and negative values of the
cosmological constant deserve attention.

Properties of the \KN\ spacetimes with a non-zero cosmological constant are
appropriately described by their geodesic structure which determines motion
of test particles and photons. The motion of photons in the symmetry plane
(the equatorial plane) of these spacetimes can be considered as a
relatively simple and easily tractable case giving results illustrating the
geometric structure in a highly representative way.  We shall discuss the
properties of the equatorial photon motion for both the black-hole and
naked-singularity spacetimes with both the repulsive and attractive
cosmological constant.

In \KNdS\ black-hole spacetimes with appropriately tuned parameters, an
unusual effect exists due to the interplay between the rotation of the
black hole and the cosmological repulsion
\cite{Stuchlik-Calvani,Stuchlik-Bao-Oestgaard-Hledik}; a \RRB\ of the
equatorial photon motion permits photons with sufficiently high positive
and all negative values of the impact parameter to move freely between the
black hole and cosmological horizons. No such effect exists in the
spherically symmetric \SdS\ and \RNdS\ spacetimes, where the repulsive
barrier always diverges at the black-hole and cosmological horizons. In
\KNdS\ black-hole spacetimes with a standard, divergent repulsive barrier,
the barrier diverges at two radii between the outer black-hole and
cosmological horizons.  We shall show that the effect of the \RRB\ is
related to the fact that the constants of motion and impact parameter of
photons have other asymptotic meaning that we are accustomed to.

The plan of the paper is the following. In Section~\ref{radmot}, the radial
equation of the equatorial photon motion is discussed. Properties of the
radial motion are given in terms of an `effective potential' related to
an appropriately defined impact parameter, and circular photon orbits,
corresponding to local extrema of the effective potential, are determined
for arbitrary values of the parameters of the spacetimes. In
Section~\ref{class}, the \KNdS\ and \KNadS\ spacetimes are classified
according to the properties of the effective potential of the equatorial
photon motion which reflects in an appropriate way the properties of the
spacetime geometry. As the criteria for the classification we use the
number of event horizons, the number of circular photon orbits, and the
number of divergent points of the effective potential. In
Section~\ref{angles}, relation between the impact parameter of the
equatorial photons and their directional angle relative to the outward
radial direction as measured by the family of locally non-rotating
observers is determined. The directions corresponding to the captured
photons, and the counterrotating photons with positive impact parameter,
are established, and related for the Kerr black hole and \KdS\ black holes
with both divergent and restricted repulsive barriers. In
Section~\ref{azimot}, the azimuthal equation of the equatorial photon
motion is considered, and the trajectories of photons are given is some
representative cases. Concluding remarks are presented in
Section~\ref{concl}.

\section{The radial motion}                                  \label{radmot}

In the standard Boyer--Lindquist coordinates with geometric units
($c=G=1$), the \KN\ geometry with a non-zero cosmological constant
$\Lambda$ is described by the line element
\bea
  \d s^2 = &-&\frac{\Delta_r}{I^2\rho^2}(\d t-a\sin^2\theta\,\d\phi)^2 +
    \frac{\Delta_\theta\sin^2\theta}{I^2\rho^2}%
      [a\,\d t-(r^2+a^2)\,\d\phi]^2\nonumber\\
    {}&+&\frac{\rho^2}{\Delta_r}\,\d r^2 +
    \frac{\rho^2}{\Delta_{\theta}}\,\d\theta^2,             \label{metrics}
\eea
where
\bea
  &&\Delta_r =
    \left(1-{\textstyle\frac{1}{3}}\Lambda r^2\right)(r^2+a^2)-2Mr+e^2,\\
  &&\Delta_{\theta} = 1+\textstyle{\frac{1}{3}}\Lambda a^2\cos^2 \theta,\\
  &&I = 1+{\textstyle\frac{1}{3}}\Lambda a^2,\\
  &&\rho^2 = r^2+a^2 \cos^2 \theta.                             \label{rho}
\eea
Here, $M$ is the mass parameter of the spacetime, $a$ is its specific
angular momentum ($a=J/M$), and $e$ is its electric charge.  Note that the
following analysis of the photon equatorial motion holds for dyonic
spacetimes as well, since the magnetic monopole charge $p$ enters the
geometry (\ref{metrics}) in the same way as the electric charge $e$.
Therefore, $e^2$ can be simply replaced by $e^2+p^2$.  It is convenient to
use dimensionless coordinates and parameters. Therefore, we define a new
parameter
\be
  y = {\textstyle\frac{1}{3}}\Lambda M^2,
\ee
and we redefine the following quantities: $s/M \rightarrow s$, $t/M
\rightarrow t$, $r/M \rightarrow r$, $a/M \rightarrow a$, $e/M \rightarrow
e$, i.e., we express all of these quantities in units of $M$. This is
equivalent to putting $M=1$, and leads to
\bea
  &&\Delta_r = (1-yr^2)(r^2+a^2)-2r+e^2,\\
  &&\Delta_\theta = 1+ya^2 \cos^2 \theta,\\
  &&I = 1+ya^2;
\eea
equation~(\ref{rho}) remains the same.

The equations of motion of test particles and photons in the \KN\ 
spacetimes with a non-zero cosmological constant were in the integrated and
separated form given by Carter \cite{Carter}. Using the results of the
discussion of the latitudinal motion \cite{Stuchlik83}, the radial equation
of motion along equatorial null geodesics can be given in the form (see
Eq.~(10) in \cite{Stuchlik83})
\be
  \left(\oder{r}{\lambda}\right)^2 =
  \frac{1}{r^4} R(r;y,a,e),                              \label{radequamot}
\ee
with
\be
  R(r;y,a,e) = I^2 \left\{[(r^2+a^2)E-a\Phi]^2 -
    \Delta_r (aE-\Phi)^2\right\},                             \label{Rryae}
\ee
where $E$ and $\Phi$ are the constants of motion connected with symmetries
of the geometry (\ref{metrics}). They can be expressed as projections of
photon's 4-momentum onto the time Killing vector $\xi_{(t)}=\lpder{}{t}$
and the axial Killing vector $\xi_{(\phi)}=\lpder{}{\phi}$, respectively;
$\lambda$ is the affine parameter along the null geodesics. Recall that the
constants of motion $E$ and $\Phi$ cannot here be interpreted as energy and
the axial component of the angular momentum at infinity, since the geometry
(\ref{metrics}) is not asymptotically flat.  For the equatorial motion of
photons, the last constant of motion, connected with the total angular
momentum of the particle, must be restricted by the condition
$K=I^2(aE-\Phi)^2$, due to the equation of the latitudinal motion
\cite{Stuchlik83}. In this form it enters the equation of the radial motion
(\ref{Rryae}).

The motion of photons is independent of the constant of motion $E$. The
equatorial motion is fully governed by the impact parameter $\ell=\Phi/E$
($E \neq 0$). However, it is convenient to analyze the radial motion of
photons in terms of a redefined impact parameter
\be
  X \equiv \frac{\Phi}{E} - a = \ell - a.
\ee
Then
\be
  R(r;y,a,e,E,X) = I^2 E^2 \left[ (r^2 -aX)^2 - \Delta_r X^2 \right];
\ee
clearly, at the dynamic regions, where $\Delta_r < 0$, there is $R(r) > 0$,
and the radial motion has no turning points there. At the stationary
regions, where $\Delta_r \geq 0$, the turning points of the radial motion,
where $R = 0$, are determined by the `effective potential'
\be
  X_\pm (r;y,a,e) \equiv \frac{r^2}{a\mp\sqrt{\Delta_r}},
\ee
which can be analyzed in a relatively simple way. We will assume $a \geq 0$
in the following.

At the regions, where $a^2 - \Delta_r > 0$ (and $X_+ > 0$), the radial
motion is allowed, if
\be
  X > X_+ (r;y,a,e) \quad \mbox{or} \quad X < X_- (r;y,a,e).
\ee
At the regions, where $a^2 - \Delta_r < 0$ (and $X_+ < 0$), the radial
motion is allowed, if
\be
  X > X_+ (r;y,a,e) \quad \mbox{and} \quad X < X_- (r;y,a,e).
\ee

We have to determine the behavior of the effective potential given by the
functions $X_\pm (r;y,a,e)$. It is necessary to find the regions of reality
of the potential, its local extrema and its divergences. We shall use a
`Chinese boxes' technique; properties of the potentials are given by
families of functions of $r$ and the parameters of the geometry $(y,a,e)$,
the properties of these families of functions are given by other families
of functions of $r$ with the number of parameters lowered by 1, until we
get a function of single $r$. We shall concentrate on the behavior of the
potential in the regions of $r>0$.

First, we consider the reality of the effective potential $X_\pm
(r;y,a,e)$.  Clearly, the potential is well defined in the stationary
regions ($\Delta_r\geq 0$) only.  At the boundaries of the stationary
regions, if they exist, i.e., at the event horizons of the geometry
($\Delta_r = 0$), the common points of $X_+ (r;y,a,e)$ and $X_- (r;y,a,e)$
are located. One more common point is at $r=0$; it is the only point where
$X_\pm=0$. The functions $X_+ (r;y,a,e)$ and $X_-(r;y,a,e)$ have no other
zero point. At the horizons ($r = r_{\ri h}$), there is
\be
  X_\pm (r_{\ri h})=\frac{r^2_{\ri h}}{a}.
\ee
The loci of the event horizons are determined by the condition
\be
  y = y_{\ri h} (r;a,e) \equiv
    \frac{r^2-2r+a^2+e^2}{r^2 \left(r^2+a^2\right)}.          \label{evhor}
\ee
The function $y_{\ri h} (r;a,e)$ diverges at $r=0$, while it approaches
zero from above for $r \rightarrow \infty$. If $a^2>0$ and/or $e^2>0$,
$y_{\ri h} \rightarrow \infty$ for $r \rightarrow 0$. In the special case
$a^2=e^2=0$ (in the \SdS\ geometry) $y_{\ri h} \rightarrow -\infty$ for $r
\rightarrow 0$.

Zeros of the function $y_{\ri h}(r;a,e)$ are determined by the relation
\be
  a^2 = a^2_{\ri z(h)}(r;e) \equiv 2r - r^2 - e^2.
\ee
The function $a^2_{\ri z(h)}(r;e)$ determines loci of the horizons of the
\KN\ black holes with a zero cosmological constant. Zeros of the function
$a^2_{\ri z(h)}(r;e)$ are given by the relation
\be
  e^2 = e^2_{\ri z(z(h))}(r) \equiv r(2-r),
\ee
determining loci of the horizons of the \RN\ black holes.  [All the
characteristic functions will be denoted in this straightforward, although
rather lengthy way. However, this way enables one to obtain an immediate
orientation in relations between the families of the characteristic
functions.] The maximum of the function $a^2_{\ri z(h)}(r;e)$ is at $r=1$,
where $a^2=1-e^2$. This corresponds to the extreme \KN\ black holes. The
motion of photons in the \KN\ spacetimes was extensively discussed in
\cite{Stuchlik81} and \cite{Balek-Bicak-Stuchlik}.
 
The local extrema of the function $y_{\ri h} (r;a,e)$ are determined (due
to the condition $\lpder{y_{\ri h}}{r}=0$) by the relation
\bea
  a^2 = a^2_{{\ri ex(h)}\pm} (r;e)
    &\equiv& \frac{1}{2}\Big\{-2r^2+r-e^2\nonumber\\
  &\pm&\left.\left[r^2(8r+1)-e^2
    \left(4r^2+2r-e^2\right)\right]^{1/2}\right\}.           \label{horext}
\eea
The reality condition of the functions $a^2_{{\ri ex(h)}\pm} (r;e)$ is 
\bea
  e^2 &\geq& e^2_{{\ri r(ex(h))}+},\\
  e^2 &\leq& e^2_{{\ri r(ex(h))}-},
\eea
where
\be
  e^2_{{\ri r(ex(h)}\pm)}(r)\equiv r\left\{2r+1\pm 2[r(r-1)]^{1/2}\right\}.
\ee
Zero points of $a^2_{{\ri ex(h)}\pm} (r;e)$ are determined by the condition
\be
  e^2 = e^2_{{\ri z(ex(h))}}(r) \equiv {\textstyle\frac{1}{2}}r(3-r).
\ee
and the extremal points are given by the relation
\be
  e^2 = e^2_{{\ri ex(ex(h))}\pm}(r)
  \equiv \frac{1}{8}r
  \left\{-16r^2+24r+11\pm |4r-1||4r-5|\right\}.                                  
\ee
Due to the asymptotic behaviour of $y_{\ri h}(r;a,e)$, at least one event
horizon (a cosmological horizon) exists in the spacetimes with $y > 0$. If
fixed values of $a$ and $e$ permit the existence of local extrema of the
function $y_{\ri h}(r;a,e)$, other horizons exist for $y$ located between
the local extrema. Therefore, the Kerr--Newman--de~Sitter spacetimes can
contain two black-hole horizons at $r_-$ (the inner one) and $r_+$ (the
outer one), and a cosmological horizon at $r_{\ri c}$ ($r_- < r_{\ri c} <
r_+$). If the local minimum $y_{\ri min}(a,e)$ enters the region of $y <
0$, Kerr--Newman--anti-de~Sitter black holes with two horizons at $r_-$ and
$r_+$ can exist, if $y_{\ri min} < y < 0$. (Detailed discussion of the
properties of the horizons will be presented in the next section.)

The local extrema of the effective potential determine the loci and impact
parameters of circular photon orbits. They are given by the condition
$\lpder{X_\pm}{r}=0$, which implies the equation
\bea
  y^2a^4r^4 + 2ya^2r^2 (r^2 + 3r - 2e^2) +
  r^2 (r^2 - 6r + 9 + 4e^2 )&&\nonumber \\
  {}-4r (a^2 + 3e^2 ) + 4e^2 (a^2 + e^2 ) &=& 0.
\eea
The extrema are, therefore, determined by the relation
\bea
  y &=& y_{{\ri ex}\pm}(r;a,e)\equiv\frac{1}{a^2 r^2}\nonumber\\
  &\times&\left\{-\left(r^2+3r-2e^2\right) \pm
  2\left[r\left(3r^2+a^2\right)-
  e^2\left(2r^2+a^2\right)\right]^{1/2}\right\}.
\eea
Now, we shall give the relevant properties of the functions $y_{{\ri
    ex}\pm}(r;a,e)$. For $r \rightarrow \infty$, both $y_{{\ri ex}\pm}
\rightarrow - 1/a^2$.  Reality of these functions is determined by the
relations
\be
  a^2 
  \left\{
  \begin{array}{ll}
    \geq a^2_{\ri r(ex)} (r;e) & \mbox{if $r \geq e^2$} \\
    \leq a^2_{\ri r(ex)} (r;e) & \mbox{if $r \leq e^2$}
  \end{array}
  \right. ,                                                 \label{reality}
\ee
where
\be
  a^2_{\ri r(ex)}(r;e) \equiv
  -\frac{r^2\left(3r-2e^2\right)}{\left(r-e^2\right)}.
\ee
The divergence of this function is given by
\be
      e^2 = e^2_{\ri d(r(ex))}(r) \equiv r ,
\ee
and $a^2_{\ri r(ex)} \rightarrow +\infty (-\infty)$ for $r \rightarrow e^2$
from below (above). Zero points of $a^2_{\ri r(ex)}(r;e)$ are determined by
the relation
\be
   e^2 = e^2_{\ri z(r(ex))}(r) \equiv {\textstyle\frac{3}{2}}r .
\ee
One local extremum of $a^2_{\ri r(ex)}$ is at $r=1$ for each $e$, the
others are determined by
\be
  e^2 = e^2_{\ri ex(r(ex))+} \equiv 2r,
\ee
\be
  e^2 = e^2_{\ri ex(r(ex))-} \equiv {\textstyle\frac{3}{4}}r.
\ee
One can immediately see that $a^2_{\ri r(ex)} < 0$ at $r \geq e^2$, and the
upper reality condition (\ref{reality}) is always satisfied at $r \geq
e^2$. Thus, we have to consider only the lower reality condition
(\ref{reality}), restricted to $r < e^2$.  There are no divergent points of
the functions $y_{{\ri ex}\pm}(r;a,e)$.  Their zero points, which determine
photon circular orbits of the \KN\ spacetimes, are given by
\be
  a^2 = a^2_{\ri z(ex)}(r;e) \equiv
    \frac{\left(r^2-3r+2e^2\right)^2}{4\left(r-e^2 \right)}.
\ee
For divergent points of the function $a^2_{\ri z(ex)}$ we find
\bea
  e^2 = e^2_{\ri d(z(ex))}(r) &\equiv& r \nonumber \\
    &=& e^2_{\ri d(r(ex))}(r) ,
\eea
and $a^2_{\ri z(ex)} \rightarrow +\infty (-\infty)$ for $r \rightarrow e^2$
from above (below). Its zero points are given by
\bea
  e^2 = e^2_{\ri z(z(ex))}(r) &\equiv&
      {\textstyle\frac{1}{2}}r(3-r) \nonumber \\
    &=& e^2_{\ri z(ex(h))}(r) .
\eea
One of its extreme points is located at $r=1$ for any value of $e$; the
others are given by
\bea
  e^2 = e^2_{{\ri ex(z(ex))}1}(r) &\equiv&
      {\textstyle\frac{3}{4}}r \nonumber \\
    &=& e^2_{{\ri ex(r(ex))}-}(r) ,
\eea
and
\bea
  e^2 = e^2_{{\ri ex(z(ex))}2}(r) &\equiv&
      {\textstyle\frac{1}{2}}r(3-r) \nonumber \\
    &=& e^2_{\ri z(z(ex))}(r) = e^2_{\ri z(ex(h))}.
\eea
Therefore, the zero points of $a^2_{\ri z(ex)}(r;e)$ are also extreme
points of this function. For $e^2 = \frac{3}{4}$, there is an inflex point
of $a^2_{\ri z(ex)}$ at $r=1$. We shall see that the value of
$e^2=\frac{3}{4}$ plays an important role in the character of the photon
equatorial motion. If $e^2>\frac{3}{4}$, the function $a^2_{\ri z(ex)}$ has
no positive extremum at $r<1$, and it has a minimum at $r=1$. If
$e^2<\frac{3}{4}$, it has a minimum at $r<1$, and a maximum at $r=1$.  In
the special case of $e^2=1$, there is
\be
  a^2_{\ri z(ex)}(r;1) \equiv {\textstyle\frac{1}{4}}(r-1)(r-2)^2 ,
\ee
and the function $a^2_{\ri z(ex)}$ has no divergent points.

Since
\bea
  \pder{y_{{\ri ex}\pm}}{r} &=& (3r-4e^2)\nonumber\\
    &\times&\frac{%
    \left\{\left[r(3r^2+a^2)-e^2(2r^2+a^2)\right]^{1/2}
    \mp (r^2+a^2)\right\}}%
    {a^2 r^3 \left[r(3r^2+a^2)-e^2(2r^2+a^2)\right]^{1/2}},
\eea
we can immediately see that the function $y_{{\ri ex}+}(r;a,e)$ has local
extrema given by the condition
\bea
  a^2 = a^2_{{\ri ex(ex)}\pm}(r;e) &\equiv&
  \frac{1}{2}\Big\{-2r^2+r-e^2\nonumber\\
  &\pm&
  \left[r^2(8r+1)-e^2\left(4r^2+2r-e^2\right)\right]^{1/2}\Big\}\nonumber \\
  &=& a^2_{{\ri ex(h)}\pm}(r;e).
\eea
Therefore, the local extrema of the functions $y_{\ri h} (r,a,e)$ and
$y_{{\ri ex}+}(r,a,e)$ coincide.  Moreover, both $y_{\ri ex+}(r;a,e)$ and
$y_{\ri ex-}(r;a,e)$ have an extreme point determined by the relation
\bea
  e^2 = e^2_{{\ri ex((ex)\pm)}}(r) &\equiv&
      {\textstyle\frac{3}{4}}r \nonumber \\
    &=& e^2_{{\ri ex(z(ex))}1}(r) = e^2_{{\ri (ex(r(ex))}-}.
\eea
Clearly, $y_{\ri ex-}(r=\frac{4}{3}e^2;e)$ is always a minimum.  Because the
extrema of $y_{\ri ex+}(r,a,e)$ coincide with the extrema of $y_{\ri
  h}(r,a,e)$ at some $r_{\ri ex(h)}$, we can conclude that if $y_{\ri
  h}(r;a,e)$ has two extreme points, the function $y_{\ri ex+}(r;a,e)$ has
three extreme points. If $y_{\ri ex+}(r=\frac{4}{3}e^2;e)$ is a maximum, then
$y_{\ri ex+}(r=r_{\ri min(h)};e)$ must be a minimum, and circular photon
orbits can exist under the inner horizon of black holes (with both $y>0$
and $y<0$). If $y_{\ri ex+}(r=\frac{4}{3}e^2;e)$ is a minimum, then $y_{\ri
  ex+}(r=r_{\ri min(h)};e)$ must be a maximum, and two additional circular
photon orbits can exist in the field of naked singularities.

It is easy to determine the character of the extreme point of $y_{\ri
  ex+}(r;a,e)$ at $r = \frac{4}{3}e^2$. We can find that at $r =
\frac{4}{3}e^2$ there is
\bea
  &&\pdder{y_{{\ri ex}\pm}}{r} (r={\textstyle\frac{4}{3}}e^2;e)\nonumber\\
 &&= \frac{3
    \left\{\left[r(3r^2+a^2)-e^2(2r^2+a^2)\right]^{1/2}
    \mp (r^2+a^2)\right\}}%
    {a^2 r^3 \left[r(3r^2+a^2)-e^2(2r^2+a^2)\right]^{1/2}},
\eea
and, substituting for $r=\frac{4}{3}e^2$, the condition
\be
  \pdder{y_{{\ri ex}\pm}}{r}(r={\textstyle\frac{4}{3}}e^2;e) \leq 0 ,
\ee
can be put into the form
\be
  81a^4 + 9e^2(32e^2 - 3)a^2 + 32e^6(8e^2 - 9) \geq 0 . 
\ee
The inflex point ($\lpdder{y_{{\ri ex}\pm}}{r}(r=\frac{4}{3}e^2;e) = 0$)
is given by the relation
\be
  a^2 = a^2_{\ri inf \pm}(e) \equiv
    \frac{1}{18}e^2\left[3 - 32e^2 \pm
    \sqrt{3(320e^2 + 3)}\right].                            \label{inflpts}
\ee
There is $a^2_{\ri inf-}(e) < 0$ for $e^2 < 0$; zeros of $a^2_{\ri inf+}(e)$
are at $e^2 = 0$, and $e^2 = \frac{9}{8}$. The maximum of this function is
\be
  a^2_{\ri max(inf+)} = 0.2666 .
\ee
The maxima of $y_{\ri ex+}(r;a,e)$ at $r=\frac{4}{3}e^2$ ($\lpdder{y_{{\ri
      ex}\pm}}{r}(r=\frac{4}{3}e^2;e) < 0$) are determined by the
conditions
\bea
  &&a^2 > a^2_{\ri inf+}(e) ,                            \label{infplus} \\
  &&a^2 < a^2_{\ri inf-}(e) ,
\eea
while for the minima ($\lpdder{y_{{\ri ex}\pm}}{r}(r=\frac{4}{3}e^2;e) >
0$) we arrive at
\be
  a^2_{\ri inf-} < a^2 < a^2_{\ri inf+} .    
\ee 

Finally, let us determine divergent points of the effective potential. Only
$X_+ (r;y,a,e)$ can diverge; the loci of the divergent points are given by
the relation
\be
  y = y_{\ri d} (r;a,e)\equiv\frac{r^2-2r+e^2}{r^2 (r^2+a^2)}.
\ee
For $r \rightarrow \infty$, this function goes to zero from above; clearly,
there is $y_{\ri h}(r;a,e) > y_{\ri d}(r;a,e)$.  The divergence of $y_{\ri
  d} (r;a,e)$ occurs at $r=0$. Notice that if $e^2>0$, $y_{\ri
  d}(r\rightarrow 0)\rightarrow \infty$, while for $e=0$, $y_{\ri
  d}(r\rightarrow 0) \rightarrow -\infty$. (The case of $e=0$ is discussed
in \cite{Stuchlik-Calvani}.)

Zeros of the function $y_{\ri d}(r;a,e)$ are independent of the parameter
$a$;
\bea
  e^2 = e^2_{\ri z(d)}(r) &\equiv& r(2-r) \nonumber \\
      &=& e^2_{\ri z(z(h))}(r).
\eea
The local extrema of the function $y_{\ri d}(r;a,e)$ are determined by
\be
  a^2 = a^2_{{\ri ex(d)}}(r;e) \equiv
  \frac{r^2 (r^2-3r+2e^2)}{(r-e^2)}.                         \label{localext}
\ee
Divergent points of $a^2_{{\ri ex(d)}}(r;e)$ are given by
\bea
  e^2 = e^2_{\ri d(ex(d))}(r) &\equiv& r \nonumber \\
    &=& e^2_{\ri d(r(ex))}(r) = e^2_{\ri d(z(ex))}(r).  
\eea
In the special case of $e=1$, there is no divergent point since
\be
  a^2_{\ri ex(d)}(r;1) = r^2(r-1).
\ee
Zero points of the function $a^2_{{\ri ex(d)}} (r;e)$ are located where
\bea
  e^2 = e^2_{\ri z(ex(d))}(r) &\equiv&
        {\textstyle\frac{1}{2}}r(3-r) \nonumber \\
        &=& e^2_{\ri z(z(ex))}(r) =
        e^2_{\ri z(ex(h))} (r) =
        e^2_{\ri ex(z(ex))2} (r),
\eea
and its local extrema are determined by the relation
\be
  e^2 = e^2_{{\ri ex(ex(d))}\pm} (r) \equiv
    \frac{1}{8}
    \left\{-4r^2+11r\pm 4r\left|r-\frac{5}{4}\right|\right\}.
\ee

Now, we can discuss properties of the photon equatorial motion using the
formulae presented above. The properties enable us to classify the
spacetimes under consideration.

\section{Classification of the spacetimes} \label{class}

We propose a classification of the \KNdS\ and \KNadS\ spacetimes according
to the properties of the effective potential $X_{\pm}(r;y,a,e)$ governing
photon motion in the equatorial plane. The classification will arise from a
systematic study of the properties of the functions given in the preceding
section. The crucial features of the classification will be the number of
event horizons present in these spacetimes, the number of divergences of
the effective potential, the number of its local extrema, governing loci
and impact parameters of circular photon geodesics, and its asymptotic
behavior.

\begin{figure}[b]
\centering \leavevmode
\epsfxsize=.7\hsize \epsfbox{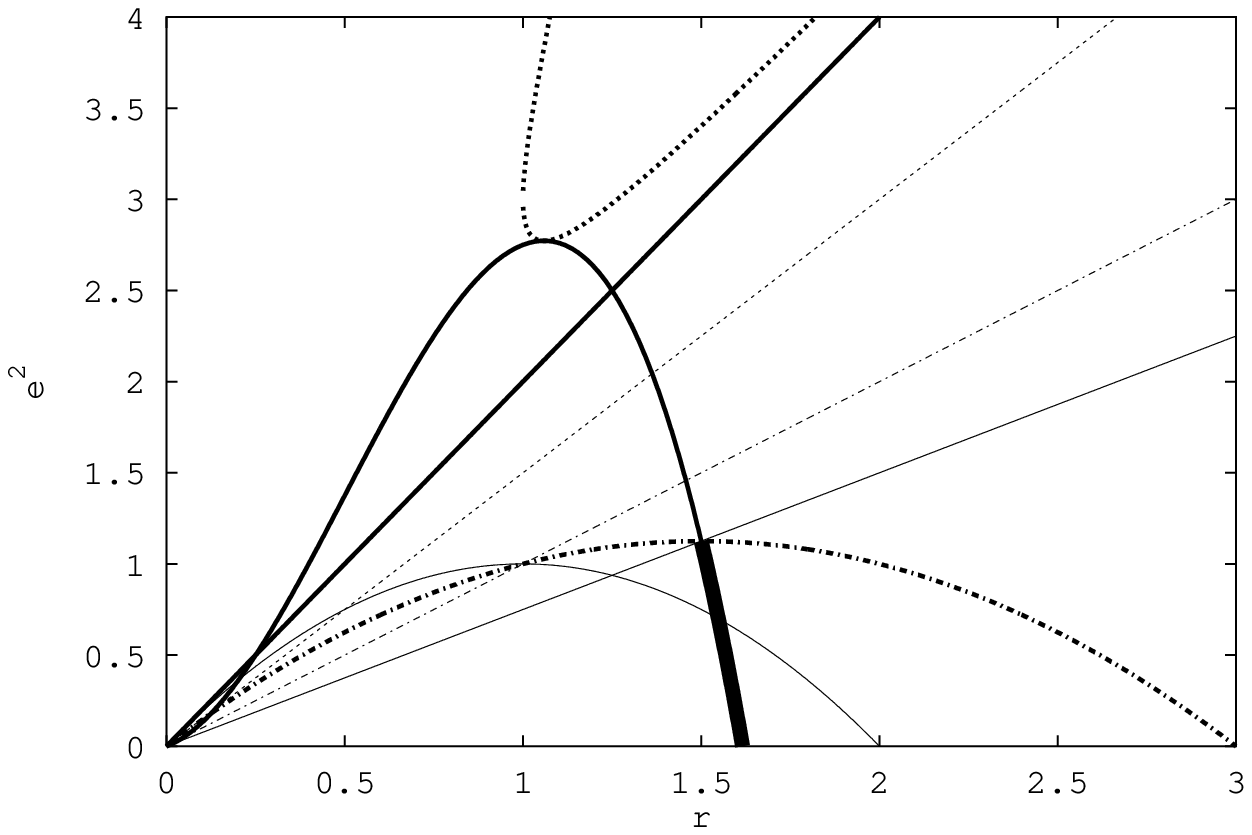}
\caption{The characteristic functions  $e^2(r)$ governing the reality
  conditions (event horizons), local extrema and divergent points of the
  effective potential of the equatorial photon motion in the \KN\ 
  spacetimes with a non-zero cosmological constant. The functions $e^2_{\ri
    ex(ex(h))\pm}(r)$ are represented by the bold solid curves; their linear
  part represents also the function $e^2_{\ri ex(r(ex))+}(r)$. The
  functions $e^2_{\ri r(ex(h))\pm}(r)$ are represented by the bold
  dotted curve. The bold dashed-dotted curve represents the functions $e^2_{\ri
    z(ex(h))}(r) = e^2_{\ri z(z(ex))}(r) = e^2_{\ri ex(z(ex))2}(r) = e^2_{\ri
    z(ex(d))}(r)$. The functions $e^2_{\ri ex(ex(d))\pm}(r)$ are
  represented by the thin solid curves; the linear part of them corresponds
  also to the functions $e^2_{\ri ex(r(ex))-}(r) = e^2_{\ri ex(z(ex))1}(r)
  = e^2_{\ri ex((ex)\pm)}(r)$, the parabolic part corresponds also to the
  functions $e^2_{\ri z(z(h))}(r) = e^2_{\ri z(d)}(r)$. The function
  $e^2_{\ri z(r(ex))}(r)$ is represented by the thin dotted line. The
  functions $e^2_{\ri d(r(ex))}(r) = e^2_{\ri d(z(ex))}(r) = e^2_{\ri
    d(ex(d))}(r)$ are represented by the thin dashed-dotted line. The inflex
  points of the function $y_{\ri h}(r;a,e)$ are given by the branch of the
  function $e^2_{{\ri ex(ex(h))}-} (r)$ located at $r\geq 1.5$ and $e^2\leq
  \frac{9}{8}$ (emphasized by extra bold curve), where $a^2_{{\ri
      ex(h)}+}(r;e)>0$. The inflex points of $y_{\ri d}(r;a;e)$ are given
  by the function $e^2_{{\ri ex(ex(d))}\pm} (r)$ at $r\geq 1.5$ and
  $e^2\geq \frac{9}{8}$, i.e., outside the region of existence of
  black-hole horizons.}
\label{f1}
\end{figure}

All the characteristic functions $e^2(r)$
\begin{itemize}
  \item[$\bullet$] $e^2_{\ri r(ex(h))\pm}(r)$,
  \item[$\bullet$] $e^2_{\ri z(ex(h))}(r) = e^2_{\ri z(z(ex))}(r) =
    e^2_{\ri ex(z(ex))2}(r) = e^2_{\ri z(ex(d))}(r)$,
  \item[$\bullet$] $e^2_{\ri ex(ex(h))\pm}(r)$,
  \item[$\bullet$] $e^2_{\ri z(z(h))}(r) = e^2_{\ri z(d)}(r)$,
  \item[$\bullet$] $e^2_{\ri d(r(ex))}(r) = e^2_{\ri d(z(ex))}(r) =
    e^2_{\ri d(ex(d))}(r)$,
  \item[$\bullet$] $e^2_{\ri z(r(ex))}(r)$,
  \item[$\bullet$] $e^2_{\ri ex(r(ex))-}(r) = e^2_{\ri ex(z(ex))1}(r) =
    e^2_{\ri ex(ex)\pm}(r)$,
  \item[$\bullet$] $e^2_{\ri ex(ex(d))\pm}(r)$,
  \item[$\bullet$] $e^2_{\ri ex(r(ex))+}(r)$,
\end{itemize}
which are relevant in order to determine the properties of the
characteristic functions $a^2(r;e)$
\begin{itemize}
  \item[$\bullet$] $a^2_{\ri ex(h)\pm}(r;e) = a^2_{\ri ex(ex)\pm}(r;e)$,
  \item[$\bullet$] $a^2_{\ri z(ex)}(r;e)$,
  \item[$\bullet$] $a^2_{\ri r(ex)}(r;e)$,
  \item[$\bullet$] $a^2_{\ri z(h)}(r;e)$, 
  \item[$\bullet$] $a^2_{\ri ex(d)}(r;e)$,
\end{itemize}
are illustrated in Fig.\,\ref{f1}. Of course, we must restrict ourselves to
the relevant regions, where the characteristic functions $e^2(r)$, and
$a^2(r;e)$ are non-negative.

\begin{figure}[b]
\centering \leavevmode
\epsfxsize=\hsize \epsfbox{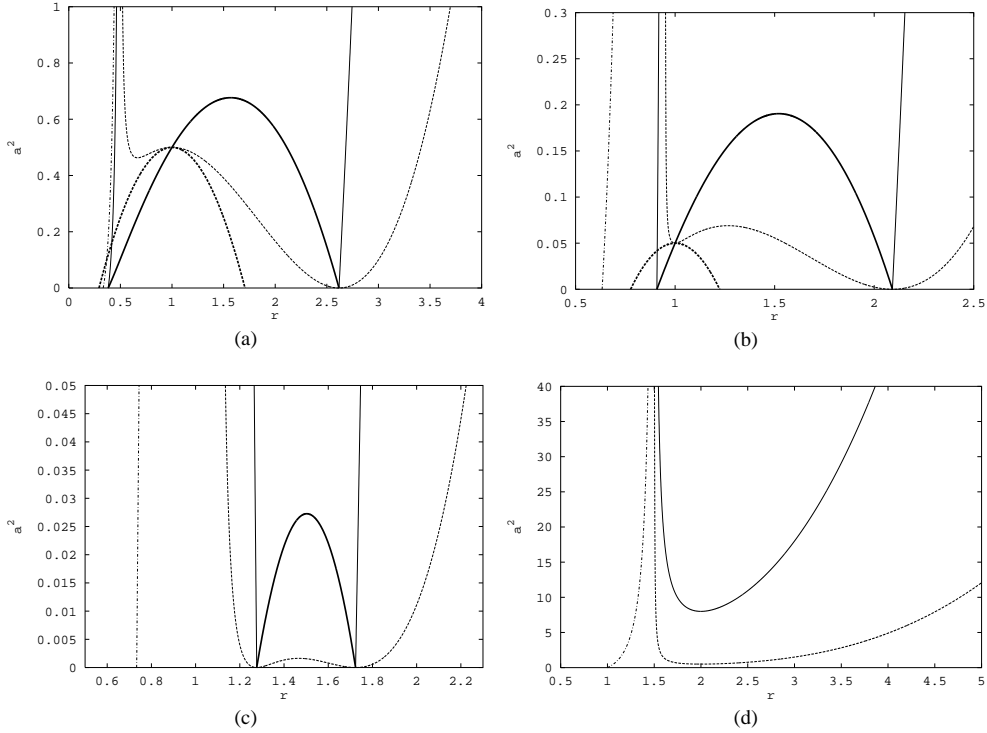}
\caption{The sequence of characteristic functions $a^2(r;e)$ determining
  behavior of functions $y_{\ri h}(r;a;e)$, $y_{\ri ex\pm}(r;a;e)$, $y_{\ri
    d}(r;a;e)$ characterizing properties of the effective potential of the
  equatorial photon motion. The functions are drawn in the physically
  relevant part $a^2(r)\geq 0$ only: $a^2_{\ri ex(h)}$ are drawn as bold
  solid curves, $a^2_{\ri z(h)}$ as bold dashed curves, $a^2_{\ri ex(d)}$
  as thin solid curves, $a^2_{\ri z(ex)}$ as thin dashed curves, and
  $a^2_{\ri r(ex)}$ as thin dashed-dotted curves. The sequence covers all
  the qualitatively different cases (A)--(D) of the behavior of the
  functions $a^2(r;e)$ in the dependence on the parameter $e$, as
  determined in Section~\protect\ref{class}. The following values of
  parameter $e$ are used: (a)~$e^2 = 0.5$, (b)~$e^2 = 0.95$, (c)~$e^2 =
  1.1$, (d)~$e^2 = 1.5$.}
\label{f2}
\end{figure}

\begin{figure}[b]
\centering \leavevmode
\epsfxsize=.675\hsize \epsfbox{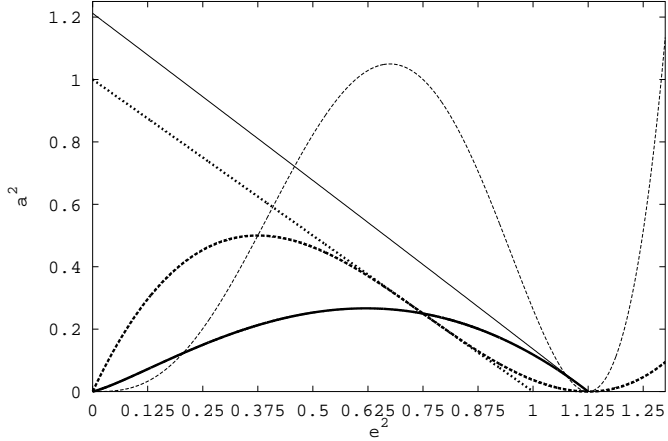}
\caption{The functions $a^2_{\ri inf+}(e^2)$ (bold solid curve), $a^2_{\ri
    ex(z(ex))}(e^2)$ (bold dashed curve), $a^2_{\ri max(z(h))}(e^2)$ (bold
  dotted curve), $a^2_{\ri max(ex(h))+}(e^2)$ (thin solid curve), and
  $a^2_{\ri min(ex(d))}(e^2)$ (thin dashed curve), determining behavior of
  the functions $y_{\ri h}(r;a,e)$, $y_{\ri ex\pm}(r;a,e)$, $y_{\ri
    d}(r;a,e)$. Note that the function $a^2_{\ri max(z(h))}(e^2)$ separates
  the black-hole and naked-singularity \KN\ spacetimes in the parameter
  space ($a^2$-$e^2$), and by itself corresponds to \KN\ extreme black
  holes. Similarly, the function $a^2_{\ri ex(z(ex))}(e^2)$ separates the
  regions of the \KN\ spacetimes admitting different number of circular
  photon geodesics. On the other hand, the function $a^2_{\ri
    max(ex(h))+}(e^2)$ represents the limiting values of parameters $a^2$,
  $e^2$ allowing the existence of the \KNdS\ black holes (with $y$ tuned
  appropriately); above this function, naked-singularity spacetimes exist
  for any $y > 0$.}
\label{f3}
\end{figure}

It follows from the behavior of the characteristic functions $e^2(r)$ that
there are four qualitatively different cases of the behavior of the
characteristic functions $a^2(r;e)$ at the relevant region of $a^2 \geq 0$. We
denote them in the following way:

\medskip

\begin{minipage}{.46\linewidth}
\begin{description}
  \item[(A)]  $e^2 < \frac{3}{4}$ ,
  \item[(B)]  $\frac{3}{4} < e^2 < 1$ ,
\end{description}
\end{minipage}\hfill%
\begin{minipage}{.46\linewidth}
\begin{description}
  \item[(C)]  $1  < e^2 < \frac{9}{8}$ ,
  \item[(D)]  $e^2 > \frac{9}{8}$.
\end{description}
\end{minipage}

\medskip

The behavior of the characteristic functions $a^2(r;e)$ is for the cases
A--D demonstrated in Figs~\ref{f2}a--d. The limiting situations,
corresponding to equalities in conditions A--D, can be inferred in a
straightforward way, and are related to continuous changes of these
characteristic functions.

The characteristic functions enable us to determine the behavior of the
functions $y_{\ri h}(r;a,e)$, $y_{{\ri ex }\pm}(r;a,e)$, $y_{\ri
  d}(r;a,e)$.  However, in order to find the regions of the parameter $a^2$
corresponding to different cases of the behavior of $y_{\ri h}$, $y_{\ri
  ex\pm}$, $y_{\ri d}$, we need the functions $a^2_{\ri inf+}(e)$,
$a^2_{\ri ex(z(ex))}(e)$, $a^2_{\ri max(z(h))}(e)$, $a^2_{\ri
  max(ex(h)+)}(e)$, $a^2_{\ri min(ex(d))}(e)$, which are drawn in
Fig.\,\ref{f3}. The function $a^2_{\ri inf+}(e)$ is given by the relation
(\ref{inflpts}). Further, we can easily find that for the extreme point of
$a^2_{\ri z(ex)}(r;e)$ at $r=\frac{4}{3}e^2$ there is
\be
  a^2_{\ri ex(z(ex))}(e) \equiv {\textstyle\frac{1}{27}}e^2(8e^2 - 9)^2 ,
\ee
and for $a^2_{\ri z(h)}(r;e)$ at $r=1$ there is
\be
  a^2_{\ri max(z(h))}(e) \equiv 1 - e^2 .
\ee
Clearly, $a^2_{\ri max(z(h))}(e)$ governs the extremal \KN\ black holes,
and, together with the function $a^2_{\ri ex(z(ex))}(e)$ yields the
classification of the equatorial photon motion in the \KN\ backgrounds
\cite{Stuchlik81,Balek-Bicak-Stuchlik}. The function $a^2_{\ri
  max(ex(h))+}(e)$ is implicitly given in a parametric form by $a^2_{\ri
  ex(h)+}(r;e)$, and $e^2_{\ri ex(ex(h))-}(r)$ with $r$ being the
parameter.  Similarly, the function $a^2_{\ri min(ex(d))}(e)$ is determined
by $a^2_{\ri ex(d)}(r;e)$, and $e^2_{\ri ex(ex(d))+}(r)$; we find
\be
  a^2_{\ri min(ex(d))}(e) \equiv
    2\left({\textstyle\frac{2}{3}}\right)^5 e^6 (8e^2 - 9)^2. 
\ee
Further, it is useful to establish the common points of $y_{\ri
  ex+}(r;a,e)$ and $y_{\ri d}(r;a,e)$.  They are determined by two
conditions:
\bea
  &&e^2 =
    e^2_{\ri (d\hbox{-}ex)+}(r;a) \equiv
    \frac{a^2r + r^3(3 - r)}{2r^2 + a^2},                \label{commpts1}\\
  &&e^2 = e^2_{\ri (d\hbox{-}ex)-}(r;a) \equiv
    \frac{-4a^4 + a^2r(1 - 4r) +
    r^3(3 - r)}{2r^2 + a^2}.                             \label{commpts2}
\eea
However, the condition (\ref{localext}) for the extrema of $y_{\ri d}(r;a,e)$
can be transferred into the form
\be
  e^2_{\ri ex(d)}(r;a) \equiv \frac{a^2r
  + r^3(3 - r)}{2r^2 + a^2}.                                  \label{e2exd}
\ee
Clearly, the intersections determined by the first condition
(\ref{commpts1}) are just at the extrema of the function $y_{\ri
  d}(r;a,e)$. The intersections determined by the second condition
(\ref{commpts2}) are irrelevant for the character of the photon equatorial
motion.

\begin{figure}[t]
\let\figlabsize=\small
\centering\leavevmode
\begin{minipage}[b]{0.48\hsize}
\centering\leavevmode
\epsfxsize=\hsize \epsfbox{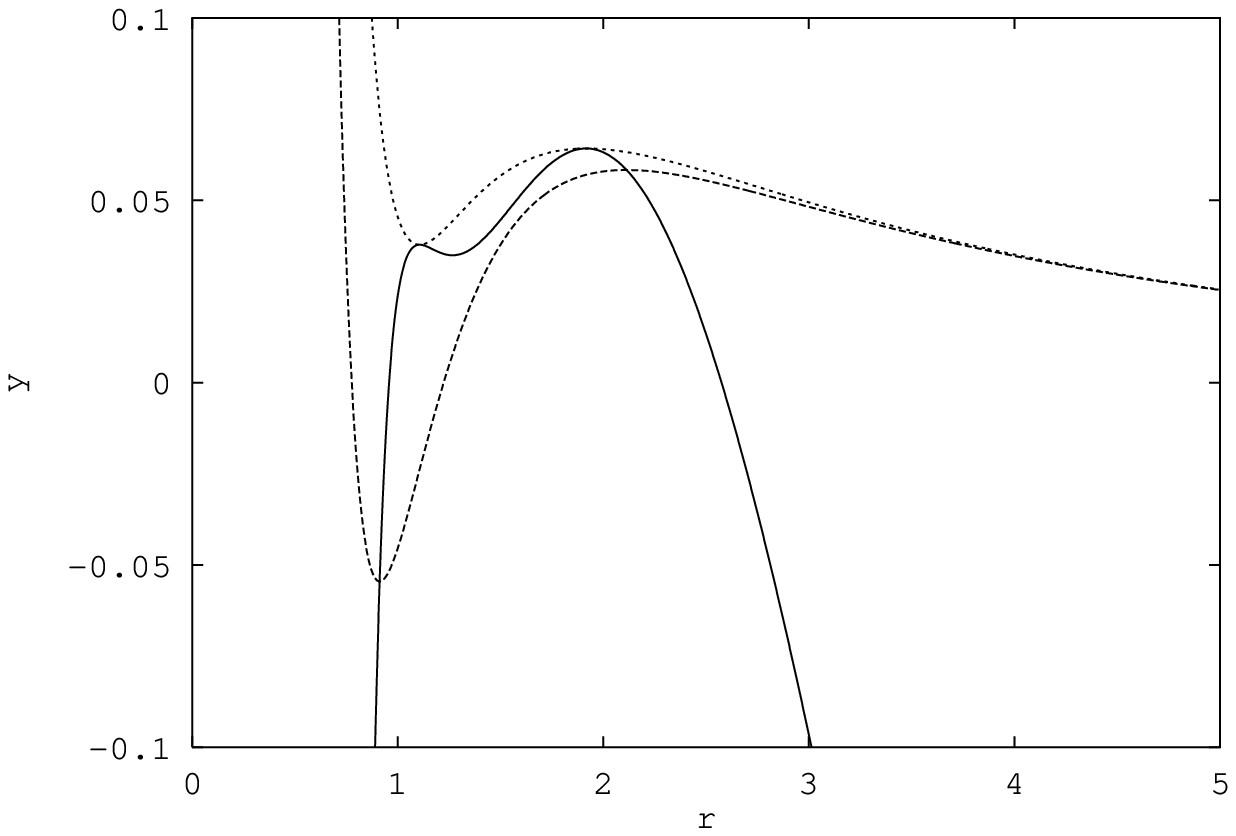}
{\figlabsize (g)}
\end{minipage}\hfill%
\begin{minipage}[b]{0.48\hsize}
\centering\leavevmode
\epsfxsize=\hsize \epsfbox{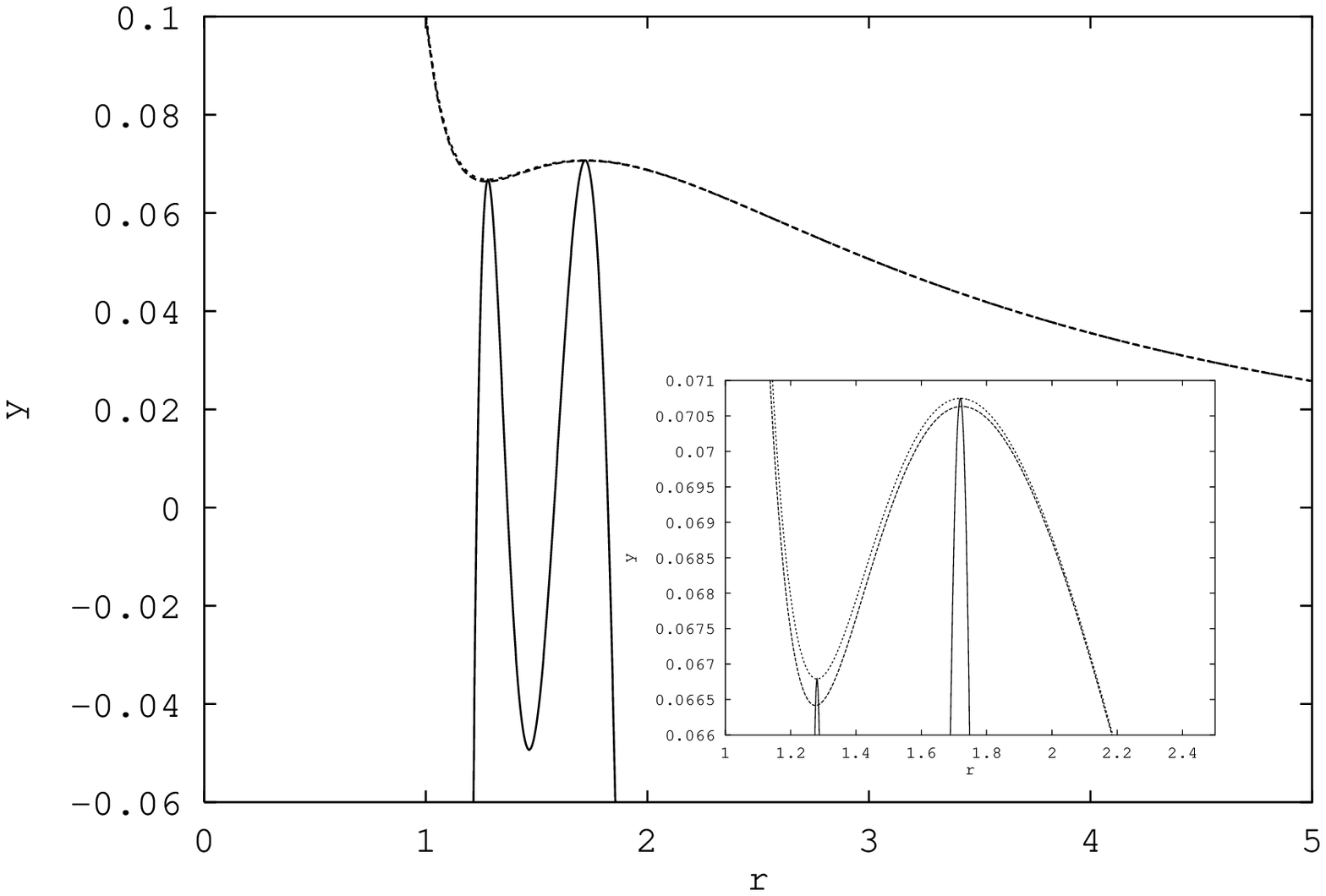}
{\figlabsize (h)}
\end{minipage}
\par\vskip 5mm\par
\begin{minipage}[b]{0.48\hsize}
\centering\leavevmode
\epsfxsize=\hsize \epsfbox{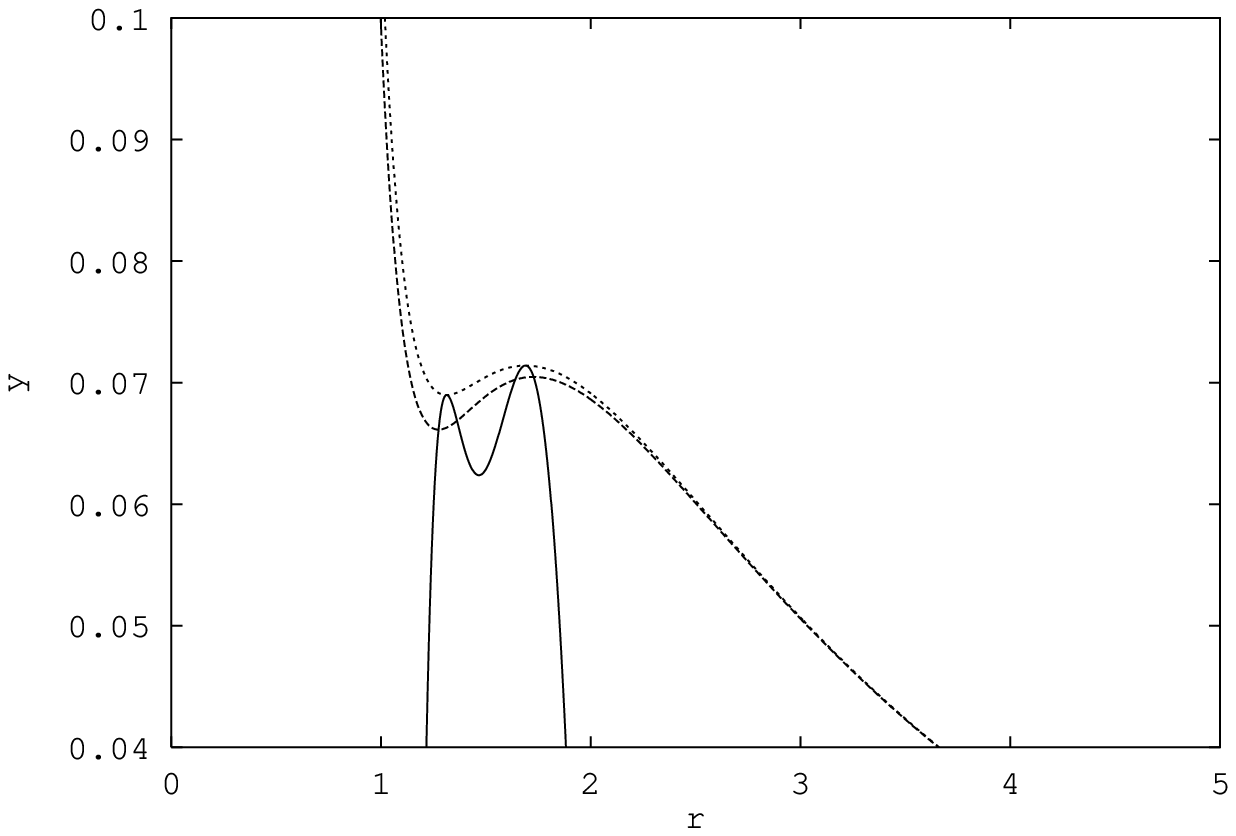}
{\figlabsize (i)}
\end{minipage}\hfill%
\begin{minipage}[b]{0.48\hsize}
\centering\leavevmode
\epsfxsize=\hsize \epsfbox{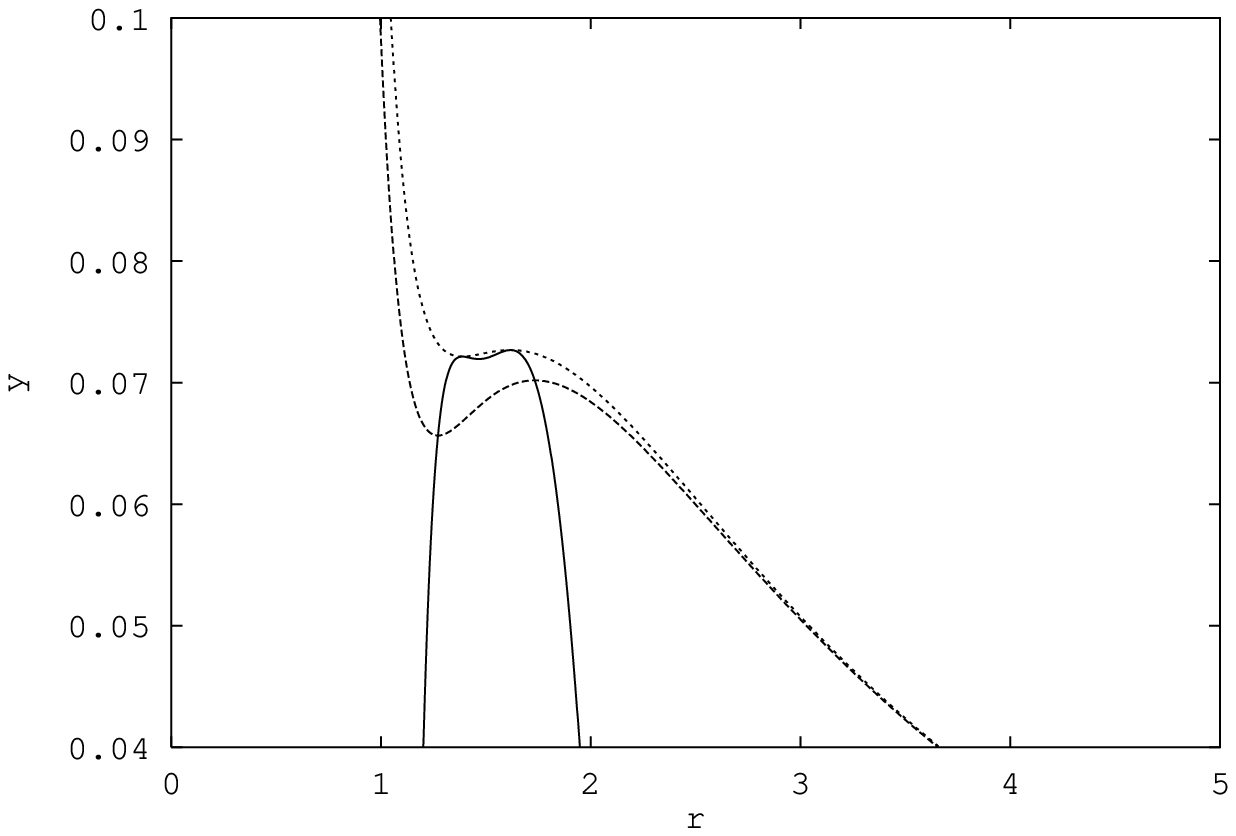}
{\figlabsize (j)}
\end{minipage}
\par\vskip 5mm\par
\begin{minipage}[b]{0.48\hsize}
\centering\leavevmode
\epsfxsize=\hsize \epsfbox{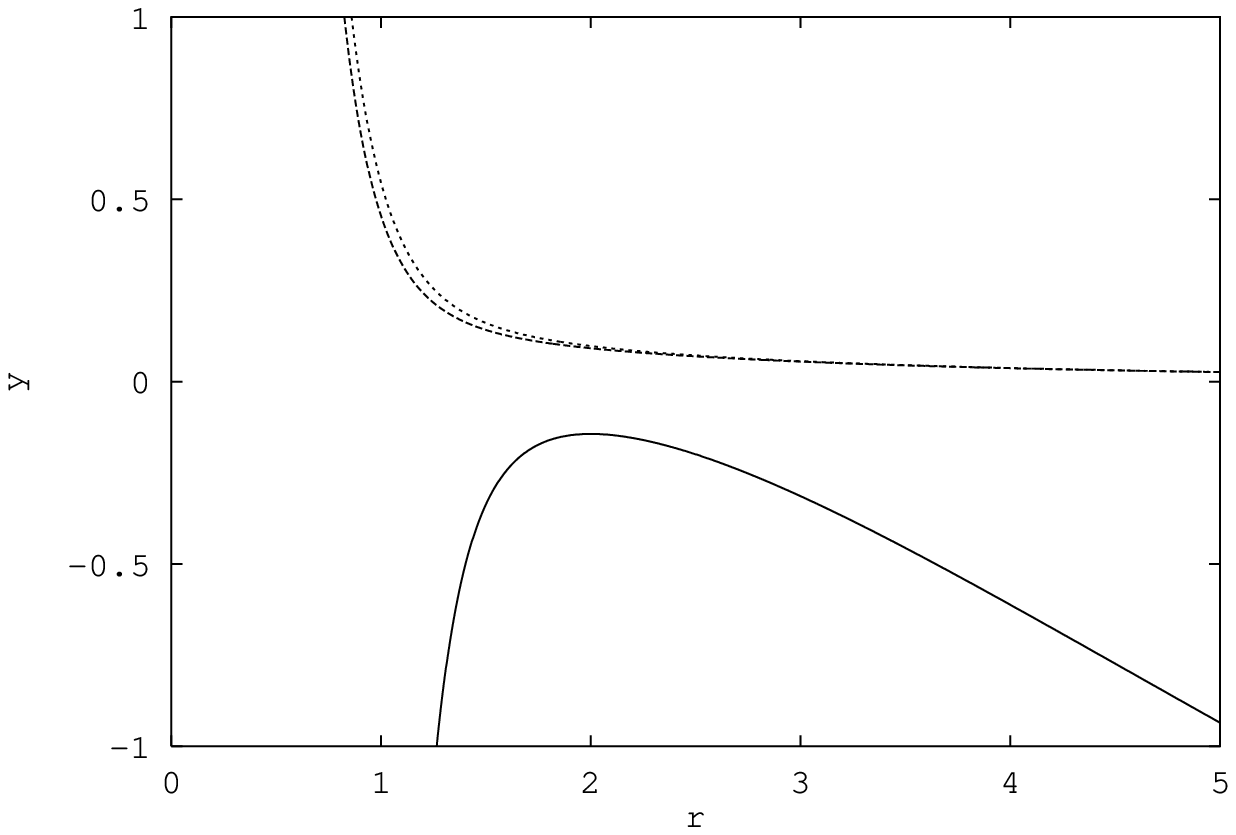}
{\figlabsize (k)}
\end{minipage}\hfill%
\begin{minipage}[b]{0.48\hsize}
\centering\leavevmode
\epsfxsize=\hsize \epsfbox{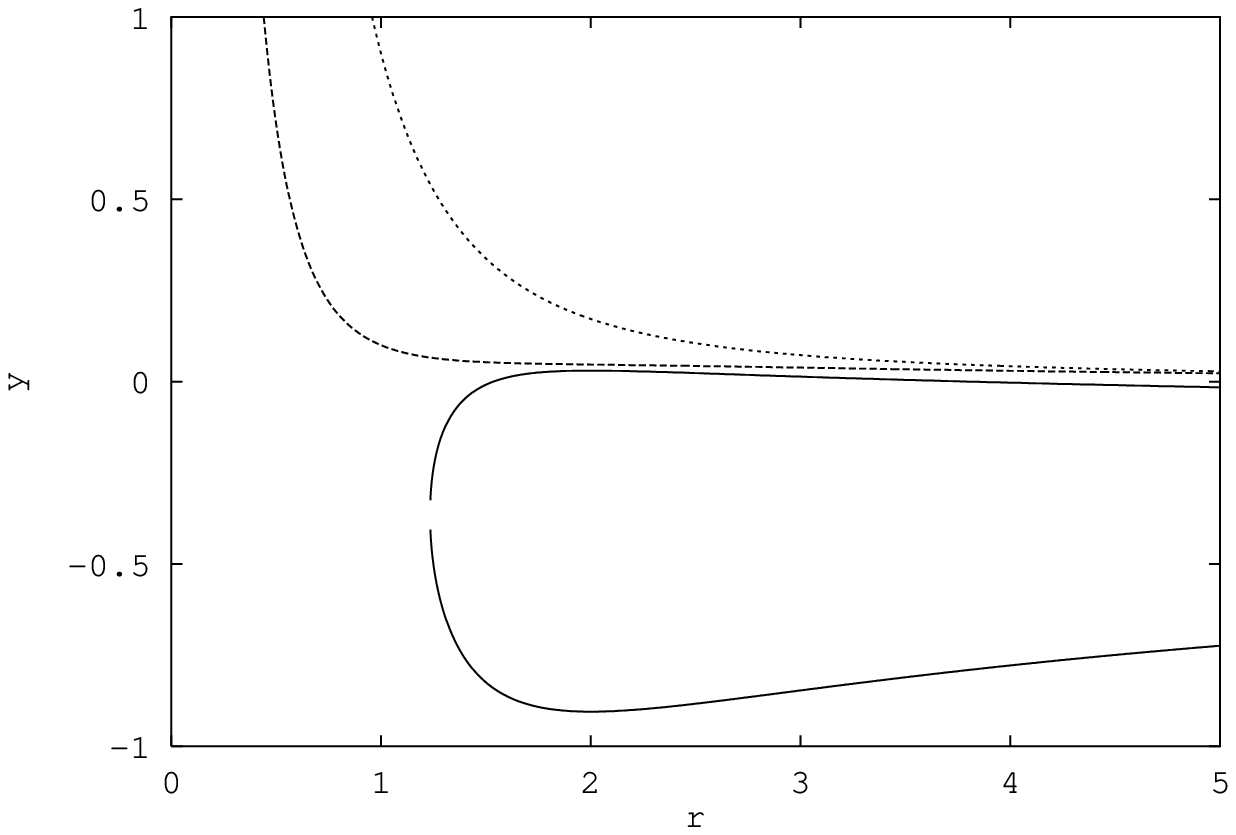}
{\figlabsize (l)}
\end{minipage}
\caption{Different types of the functions $y_{\ri h}(r;a,e)$ (dotted
  curves), $y_{\ri ex\pm}(r;a,e)$ (solid curves), $y_{\ri d}(r;a,e)$
  (dashed curves) determining the behavior of the `effective potential'
  of the photon equatorial motion. The cases illustrated in figures
  (a)--(l) are governed by the behavior of the functions $a^2(r)$ and
  $a^2(e^2)$; see Section \protect\ref{class}, where the situations not
  explicitly illustrated by the sequence of figures (a)--(l) are discussed.
  (a)~$e^2 = 0.5$, $a^2 = 0.16$, (b)~$e^2 = 0.5$, $a^2 = 0.36$, (c)~$e^2 =
  0.5$, $a^2 = 0.49$, (d)~$e^2 = 0.5$, $a^2 = 0.64$, (e)~$e^2 = 0.5$, $a^2
  = 0.81$, (f)~$e^2 = 0.95$, $a^2 = 0.06$, (g)~$e^2 = 0.95$, $a^2 = 0.1$,
  (h)~$e^2 = 1.1$, $a^2 = 0.001$, (i)~$e^2 = 1.1$, $a^2 = 0.01$, (j)~$e^2 =
  1.1$, $a^2 = 0.02$, (k)~$e^2 = 1.5$, $a^2 = 0.1$, (l)~$e^2 = 1.5$, $a^2
  = 4$.}
\label{f4}
\end{figure}

Now, using Figs~\ref{f2}\ and \ref{f3}, the behavior of the functions
$y_{\ri h}(r;a,e)$, $y_{{\ri ex}\pm}(r;a,e)$, $y_{\ri d}(r;a,e)$ can be
given in the following exhaustive scheme. [However, we must stress that in
some of the following cases, there are variants of the behavior of these
functions, determined by different relations between their extremal values
than are those shown in the corresponding figures. These variant cases will
not be drawn explicitly. We will only point out, which variants should be
considered in the explicitly illustrated cases.]
\begin{description}
\item[(A)] {\boldmath{$e^2 < 3/4$}}
\vspace{4pt}
\begin{description}
  \item[(a)] $a^2 < a^2_{\ri inf+}$                        (Fig.\,\ref{f4}a)
  \item[(b)] $a^2_{\ri inf+} < a^2 < a^2_{\ri ex(z(ex))}$  (Fig.\,\ref{f4}b)
  \item[(c)] $a^2_{\ri ex(z(ex))} < a^2 < a^2_{\ri max(z(h))} =
                                             1 - e^2$ (Fig.\,\ref{f4}c)
  \item[(d)] $1 - e^2 = a^2_{\ri max(z(h))} < a^2 <
                          a^2_{\ri max(ex(h))+}$       (Fig.\,\ref{f4}d)
  \item[(e)] $a^2_{\ri max(ex(h))+} < a^2$                 (Fig.\,\ref{f4}e)
\end{description}
\end{description}
In the case Aa one has to compare $y_{\ri min(d)}(a,e)$ with $y_{\ri
  min(ex+)(a,e)}$.  The function $y_{\ri min(d)}(a,e)$ is given implicitly
in a parametric way by $a^2_{\ri ex(d)}(r;e)$ and $y_{\ri d}(r;a,e)$,
with $r$ being the parameter. On the other hand, at $r = \frac{4}{3}e^2$
the minima of $y_{{\ri ex}\pm}(r;a,e)$ at $r=\frac{4}{3}e^2$ (or maximum of
$y_{\ri ex+}$, for $a^2 > a^2_{\ri inf+}$) are determined by the functions
\be
  y_{{\ri ex(ex)}\pm}(a,e) \equiv
    \frac{9}{8a^2e^2}\left[-1 - \frac{8e^2}{9} \pm 
    \left(\frac{32e^2}{9} + \frac{a^2}{3e^2}\right)^{1/2}\right].
\ee   
The minimum $y_{\ri min(ex)+}(a,e)$ and maxima $y_{\ri max(ex)+}(a,e)$ are
separated by the function $a^2_{\ri inf+}(e)$.  In the case Ac the
function $y_{\ri max(ex+)}(a,e)$ have to be related with $y_{\ri
  max(d)}(a,e)$ and $y_{\ri max(h)}(a,e)$, while in the case Ad, all the
functions $y_{\ri max(ex+)}(a,e)$, $y_{\ri max(d)}(a,e)$, $y_{\ri
  min(h)}(a,e)$, and $y_{\ri max(h)}(a,e)$ have to be related. The function
$y_{\ri max(d)}(a,e)$ is parametrically given by $a^2_{\ri ex(d)}(r;e)$ and
$y_{\ri d}(r;a,e)$. The functions $y_{\ri min(h)}(a,e)$ and $y_{\ri
  max(h)}(a,e)$ are parametrically given by $a^2_{\ri ex(h)+}(r;e)$ and
$y_{{\ri h}}(r;a,e)$. All these functions are determined by a
numerical code.
\begin{description}
\item[(B)] {\boldmath{$3/4 < e^2 < 1$}}
\vspace{4pt}
\begin{description}
  \item[(a)] $a^2 < a^2_{\ri max(z(h))} = 1 - e^2$ (equivalent to Aa)
  \item[(b)] $1 - e^2 = a^2_{\ri max(z(h))} < a^2 <
                                a^2_{\ri ex(z(ex))}$ (Fig.\,\ref{f4}f)
  \item[(c)] $a^2_{\ri ex(z(ex))} < a^2 < a^2_{\ri inf+}$ (Fig.\,\ref{f4}g)
  \item[(d)] $a^2_{\ri inf+} < a^2 < a^2_{\ri max(ex(h))+}$ (equivalent to Ad)
  \item[(e)] $a^2_{\ri max(ex(h))+} < a^2$     (equivalent to Ae)
\end{description}
\end{description}
In the case Bb, $y_{\ri min(d)}(a,e)$ have to be related to $y_{\ri
  min(ex+)}(a,e)$, and $y_{\ri min(h)}(a,e)$ to $y_{\ri max(d)}(a,e)$,
while in the case Bc we have to relate $y_{\ri max(d)}(a,e)$ to $y_{\ri
  min(h)}(a,e)$, and $y_{\ri min(ex+)}(a,e)$ to $y_{\ri max(d)}(a,e)$.
\begin{description}
\item[(C)] {\boldmath{$1 < e^2 < 9/8$}}
\vspace{4pt}
\begin{description}
  \item[(a)] $a^2 < a^2_{\ri ex(z(ex))}$                     (Fig.\,\ref{f4}h)
  \item[(b)] $a^2_{\ri ex(z(ex))} < a^2 < a^2_{\ri inf+}$    (Fig.\,\ref{f4}i)
  \item[(c)] $a^2_{\ri inf+} < a^2 < a^2_{\ri max(ex(h))}$   (Fig.\,\ref{f4}j)
  \item[(d)] $a^2_{\ri max(ex(h))+} < a^2$     (equivalent to Ae)
\end{description}
\end{description}
In the case Ca we have to relate $y_{\ri min(h)}(a,e)$ to $y_{\ri
  max(d)}(a,e)$; in the case Cb we should consider relations of $y_{\ri
  min(h)}(a,e)$ with $y_{\ri max(d)}(a,e)$, and of $y_{\ri min(ex+)}(a,e)$
with $y_{\ri min(d)}(a,e)$ and $y_{\ri max(d)}(a,e)$; in the case Cc we
have to relate $y_{\ri min(h)}(a,e)$ to $y_{\ri max(d)}(a,e)$, and $y_{\ri
  max(ex+)}(a,e)$ to $y_{\ri max(h)}(a,e)$.
\begin{description}
\item[(D)] {\boldmath{$9/8 < e^2$}}
\vspace{4pt}
\begin{description}
  \item[(a)] $a^2 < a^2_{\ri min(z(ex))}$                    (Fig.\,\ref{f4}k)
  \item[(b)] $a^2_{\ri min(z(ex))}<a^2<a^2_{\ri min(ex(d))}$ (Fig.\,\ref{f4}l)
  \item[(c)] $a^2_{\ri min(ex(d))} < a^2$                   (equivalent to Ae)
\end{description}
\end{description}
There is no variant in these cases.
Note that in the special case of $e=1$, the cases Ca--d are relevant with
\be
  y_{\ri d} (r;a,e) = \frac{(r - 1)^2}{r^2(r^2 + a^2)} .
\ee

By using the classification criteria presented in the beginning of this
section, we can show that there is 18 types of the behavior of the
effective potential $X_{\pm}(r;y,a,e)$. We shall characterize all the
classes, and the corresponding behavior of the effective potential.
However, we shall determine and illustrate the corresponding range of the
parameters $y, a, e$ only for some selected classes. Of course, using the
same procedure as for the selected classes, the corresponding distribution
in the parameter space can be determined also for the remaining classes. We
shall concentrate on the detailed distribution of the black-hole
spacetimes, and will not consider the naked-singularity spacetimes.

\begin{figure}[b]
\let\figlabsize=\small
\centering\leavevmode
\begin{minipage}[b]{0.48\hsize}
\centering\leavevmode
\epsfxsize=\hsize \epsfbox{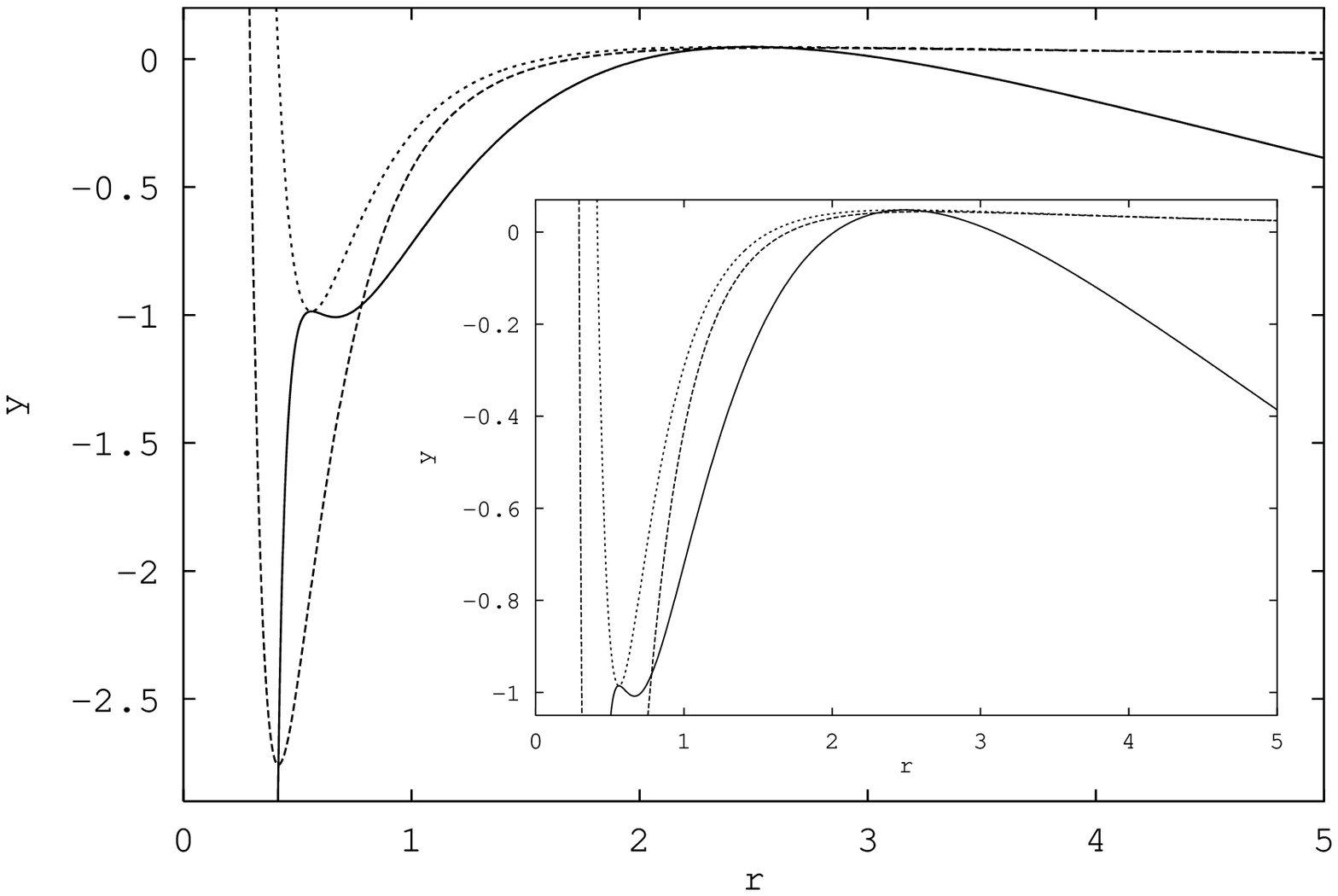}
{\figlabsize (a)}
\end{minipage}\hfill%
\begin{minipage}[b]{0.48\hsize}
\centering\leavevmode
\epsfxsize=\hsize \epsfbox{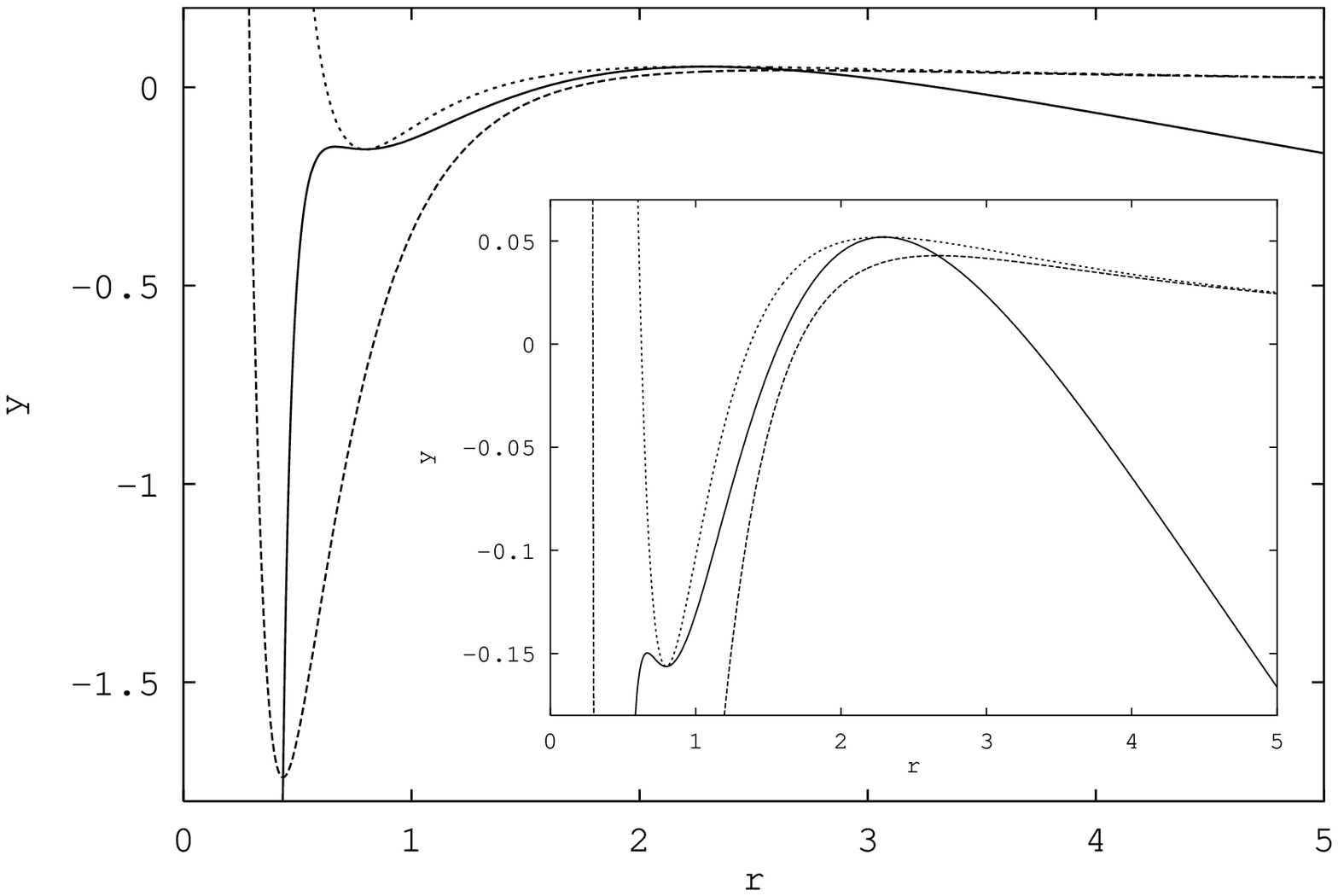}
{\figlabsize (b)}
\end{minipage}
\par\vskip 4mm\par
\begin{minipage}[b]{0.48\hsize}
\centering\leavevmode
\epsfxsize=\hsize \epsfbox{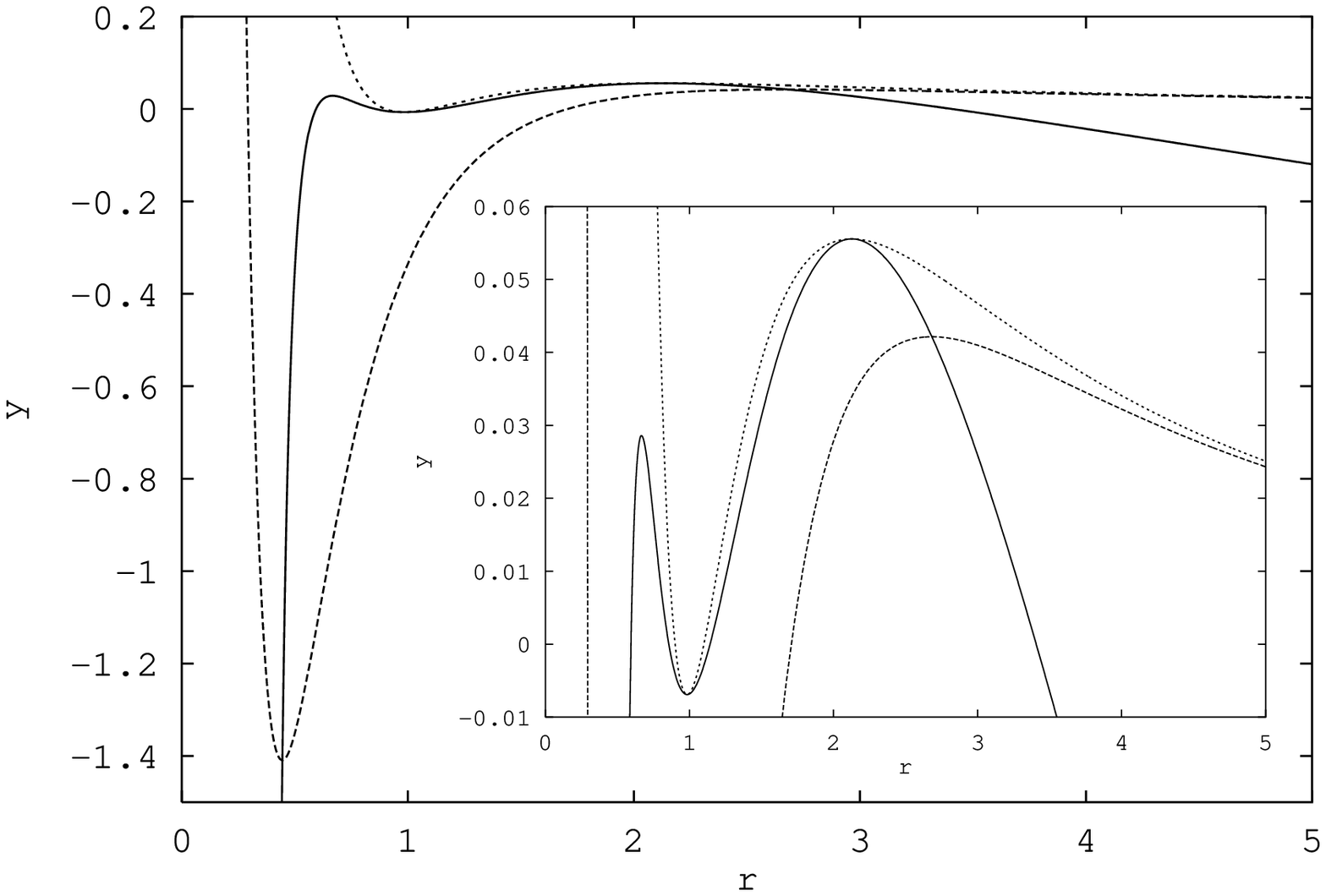}
{\figlabsize (c)}
\end{minipage}\hfill%
\begin{minipage}[b]{0.48\hsize}
\centering\leavevmode
\epsfxsize=\hsize \epsfbox{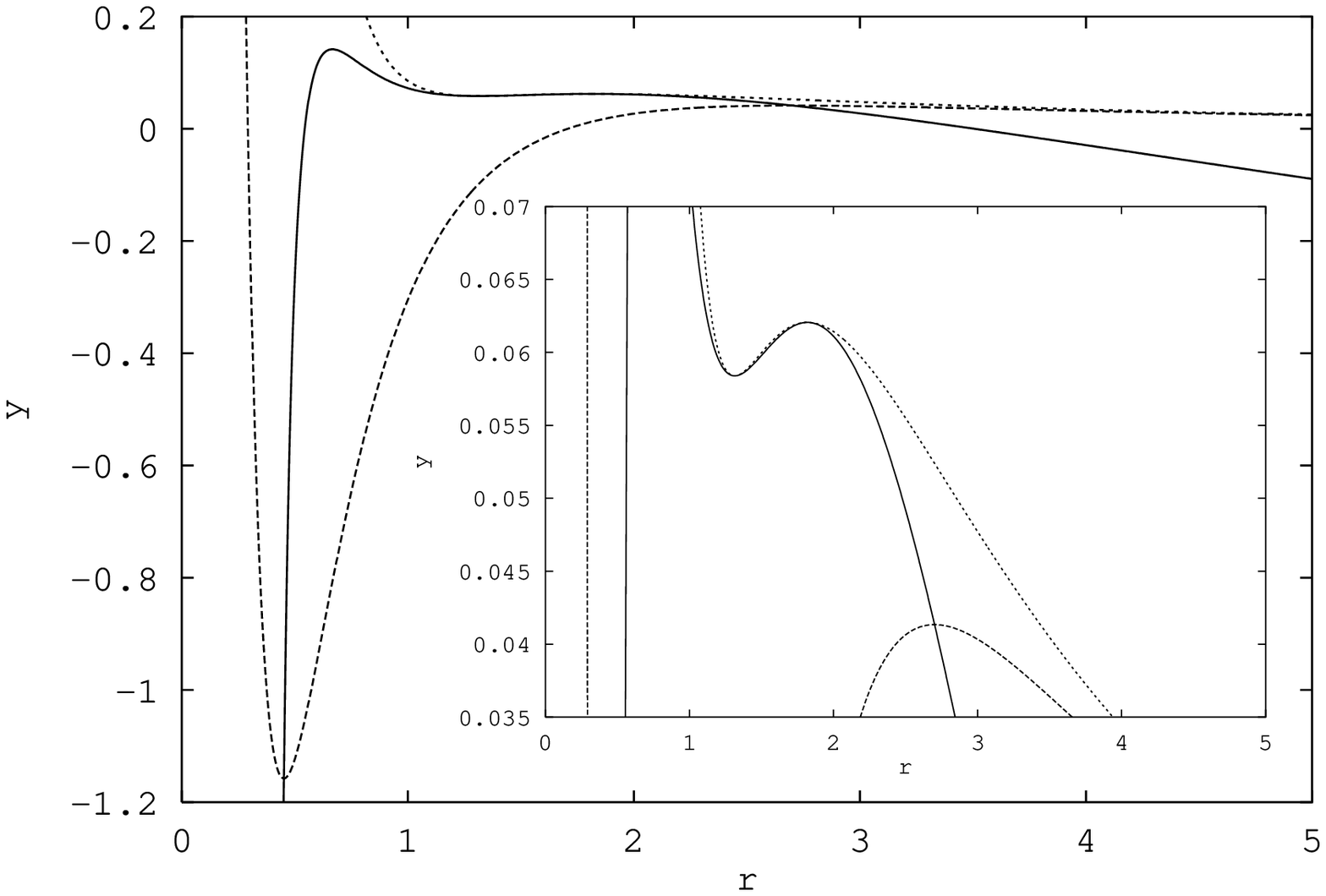}
{\figlabsize (d)}
\end{minipage}
\par\vskip 4mm\par
\begin{minipage}[b]{0.48\hsize}
\centering\leavevmode
\epsfxsize=\hsize \epsfbox{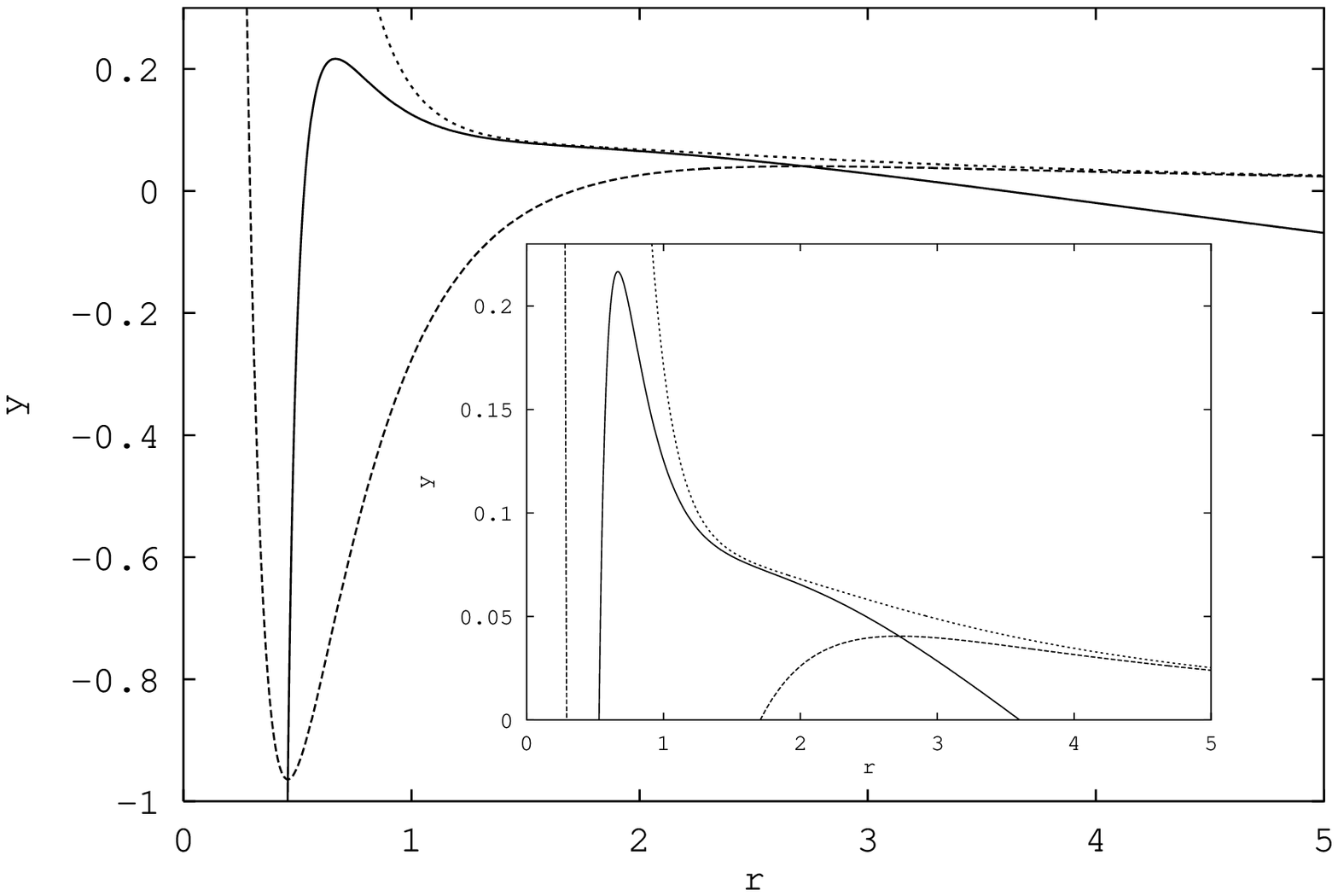}
{\figlabsize (e)}
\end{minipage}\hfill%
\begin{minipage}[b]{0.48\hsize}
\centering\leavevmode
\epsfxsize=\hsize \epsfbox{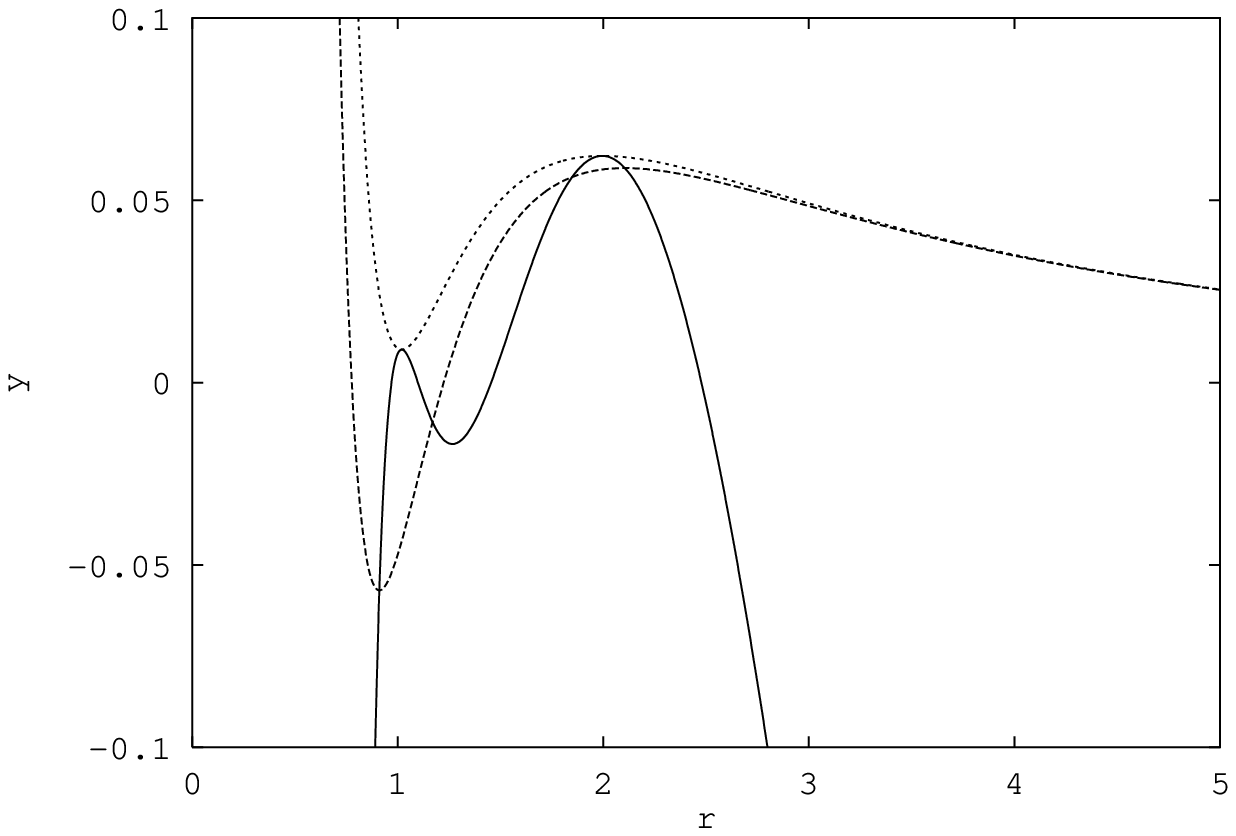}
{\figlabsize (f)}
\end{minipage}
\end{figure}

\begin{figure}[t]
\centering \leavevmode
\epsfxsize=.9\hsize \epsfbox{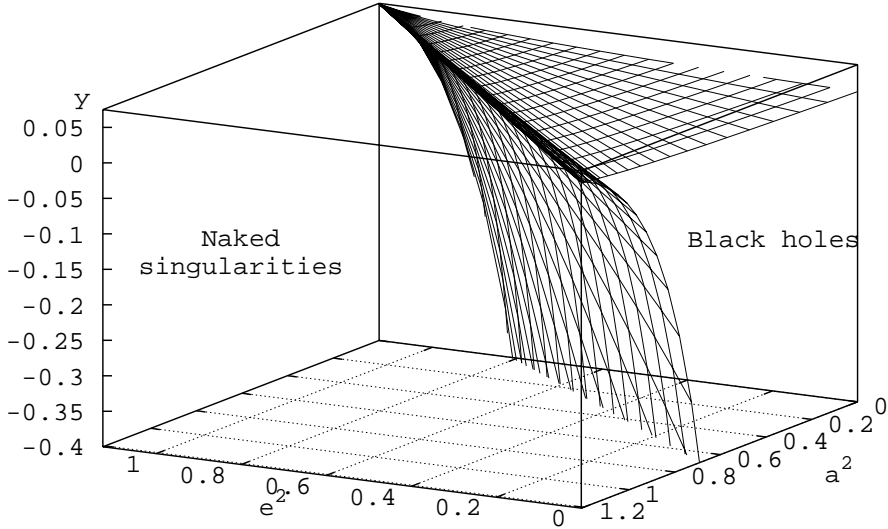}
\caption{Parameter space ($y$-$a^2$-$e^2$) of the \KN\ spacetimes with a
  non-zero cosmological constant separated into parts corresponding to the
  black-hole and naked-singularity spacetimes, respectively. The boundary
  between these states is given by the functions $y_{\ri max(h)}(a,e)$ and
  $y_{\ri min(h)}(a,e)$ determined by a numerical code. The black-hole
  spacetimes can exist for any asymptotically anti-de~Sitter spacetime ($y
  < 0$), while for de~Sitter spacetimes ($y > 0$), they are restricted by
  the upper limit $y_{\ri max} = 2/27$. In the limiting case, $y = y_{\ri
    max}$, we obtain a \RNdS\ extreme black hole with $e^2 = 9/8$ (and $a^2
  = 0$).}
\label{f5}
\end{figure}

The starting point of our classification is separation of \KNdS\ ($y > 0$)
and \KNadS\ ($y < 0$) spacetimes. This basic separation reflects different
asymptotic character of these spacetimes, which is represented by different
asymptotic behavior of $X_{\pm}(r;y,a,e)$. For $y > 0$, a cosmological
horizon exist behind which the spacetime is dynamic. The effective
potential is well-defined up to the cosmological horizon. For $y < 0$,
there is no cosmological horizon, and for $r \rightarrow \infty$ there is
\be
  X_{\pm} \approx  \pm \frac{1}{\sqrt{-y}}.
\ee
Further, we have to separate the black-hole and naked-singularity
spacetimes, i.e., we use the criterion of the number of event horizons
representing reality limits of the effective potential.  We give the
discussion in full detail -- the other cases will be considered in much
briefer form. The event horizons are determined by the function $y_{\ri
  h}(r;a,e)$. Due to the behavior of this function at $r \rightarrow 0$ and
$r \rightarrow \infty$, at least one event horizon (cosmological) exist in
spacetimes with $a^2 > 0$, $e^2 > 0$. The black-hole horizons can exist, if
$y_{\ri h}(r;a,e)$ has local extrema. The relevant extrema of $a^2_{{\ri
    ex(h)}+} (r;e)$ are given by $e^2_{{\ri ex(ex(h))}-}(r)$ at the branch
lying under the curve $e^2_{{\ri z(ex(h))}}$. Therefore, the relevant
extrema of $a^2_{{\ri ex(h)}+} (r;e)$ exist for $e^2 < \frac{9}{8}$. In the
limiting case of $e=0$, the function $a^2_{{\ri ex(h)}+} (r)$ has its
maximum at $r_{\ri crit} = 1.61603$, with a corresponding critical value of
the rotation parameter corresponding to the marginal black-hole spacetime
\be
  a^2_{\ri crit} = \left( \frac{3+2\sqrt{3}}{8} \right)
  \sqrt{\left(7+4\sqrt{3}\right)} - \frac{\sqrt{3}}{16}
  \left(5\sqrt{3}+8\right) = 1.21202,
\ee
and the critical value of the cosmological parameter
\be
  y_{\ri crit} = \frac{16\left[\sqrt{\left(7+4\sqrt{3}\right)}-3\right]}%
    {3\left(7+4\sqrt{3}\right)
    \left[\sqrt{7+4\sqrt{3}} + 1\right]} = 0.0592.
\ee
If $0 < e^2 < \frac{9}{8}$, the critical value $a^2_{\ri max(ex(h))}(e)$,
governing an inflex point of $y_{\ri h}(r;a,e)$, is determined by
$e^2_{{\ri ex(ex(h))}-}(r)$. For $a^2 < a^2_{\ri max(ex(h))}(e)$, the
function $y_{\ri h} (r;a,e)$ has two local extrema $y_{\ri min(h)}(a,e)$
and $y_{\ri max(h)}(a,e)$, determined by (\ref{horext}) and (\ref{evhor}),
with a given value of the parameter $e$. For $a^2 = a^2_{\ri
  max(ex(h))}(e)$, these extrema coincide at $y_{\ri crit}(e)$ which is the
limiting value for black-hole spacetimes with a fixed parameter $e$. The
black-hole spacetimes exist for $y_{\ri h(min)}(a,e) < y < y_{\ri
  h(max)}(a,e)$. If $y = y_{\ri min(h)}(a,e)$, the two black-hole horizons
coincide and the geometry determines an extreme black hole; for $y < y_{\ri
  min(h)}(a,e)$ it determines a naked singularity. Certain kind of
`instability' occurs at $y_{\ri max(h)}(a,e)$. If $y = y_{\ri
  max(h)}(a,e)$, the outer black-hole and cosmological horizons coincide,
keeping the role of the cosmological horizon in a spacetime with an extreme
black hole. For $y > y_{\ri max(h)}(a,e)$, the geometry describes a naked
singularity, and the cosmological horizon is determined be the branch of
$y_{\ri h}(r;a,e)$ determining the inner black-hole horizon for $y < y_{\ri
  max(h)}(a,e)$. The \KNadS\ black holes correspond to the range of
parameters
\be
  y_{\ri min(h)}(a,e) < y < 0.
\ee
Distribution of black-hole and naked-singularity spacetimes in the
parameter space is given by the functions $y_{\ri min(h)}(a,e)$, $y_{\ri
  max(h)}(a,e)$, and can be determined by a numerical code. The results are
given in Fig.\,\ref{f5}. We can see that black-hole spacetimes can exist
for all values of the attractive cosmological constant ($y < 0$), contrary
to the case of repulsive cosmological constant ($y > 0$), when black-hole
spacetimes must have $y \leq \frac{2}{27}$. The extremal value of $y = \frac{2}{27}$
corresponds to the extreme \RNdS\ geometry with the extremal value of $e^2
= \frac{9}{8}$ (and $a^2 = 0$) \cite{Stuchlik-Calvani}.

Note that in the special case of the Schwarzschild spacetimes ($a^2 = e^2 =
0$) with $y \neq 0$, there is $y_{\ri h} \rightarrow -\infty$ for $r
\rightarrow 0$. Then the black-hole and cosmological horizon exist for $0 <
y < y_{\ri crit} = \frac{1}{27}$.  They are determined by the relations
\bea
  &&r_{\ri h} = \frac{2}{\sqrt{3y}}\,\cos \frac{\pi + \xi}{3}, \\
  &&r_{\ri c} = \frac{2}{\sqrt{3y}}\,\cos \frac{\pi - \xi}{3},
\eea
where
\be
  \xi = \cos^{-1} \left(3 \sqrt{3y}\right).                      \label{xi}
\ee
If $y > \frac{1}{27}$, the spacetime is dynamic at all $r > 0$, and
represents certain kind of naked singularity. On the
other hand, in any \SadS\ geometry with $y<0$ there is a black-hole horizon
located at $r_{\ri h}$ determined by the relation
\bea
  &&r_{\ri h} = \left(  -\frac{1}{y} \right)^{1/3}\nonumber\\
  &&\times\left\{ \left[ 1+\left( 1-\frac{1}{27y}  \right)^{1/2}  \right]^{1/3} +
    \left[ 1-\left( 1-\frac{1}{27y}  \right)^{1/2}  \right]^{1/3} \right\}.
\eea
Clearly, $r_{\ri h} \rightarrow 2$ for $y \rightarrow 0$, while $r_{\ri h}
\rightarrow 0$ for $y \rightarrow -\infty$.

The other criterion of the classification is given by the number of
divergent points of the effective potential $X_\pm(r;y,a,e)$.
There exist \KNdS\ black-hole spacetimes with an unusual property of the
effective potential, namely with a \RRB\ allowing
photons with high positive-valued and any negative-valued impact parameter
$X$ to move freely between the outer black-hole and cosmological horizons.
These spacetimes were extensively studied in
\cite{Stuchlik-Bao-Oestgaard-Hledik}. Their character is a non-standard
one, because from the photon motion in other black-hole spacetimes we are
accustomed to the existence of a divergent barrier repelling photons with
high values of impact parameter.  Really, if $y=0$, the effective
potential diverges at the horizons and at infinity in the spherically
symmetric Schwarzschild and \RN\ black-hole spacetimes.
When the rotation is `switched on', i.e., in the Kerr and \KN\ black-hole
spacetimes, the effective potential is finite at the horizons, but it
diverges at infinity and at some loci between the horizon and infinity. In
the case of spherically symmetric \SdS\ and \RNdS\ geometries, the
effective potential diverges at the horizons again, as can be inferred
directly from the formula
\be
  X_{\pm} (r;y,e) = \mp \frac{r^2}{\sqrt{\Delta_r}}.
\ee
Therefore, in all these cases, a repulsive barrier does exist for photons with
a high magnitude of the impact parameter.

\begin{figure}[t]
\centering \leavevmode
\epsfxsize=.9\hsize \epsfbox{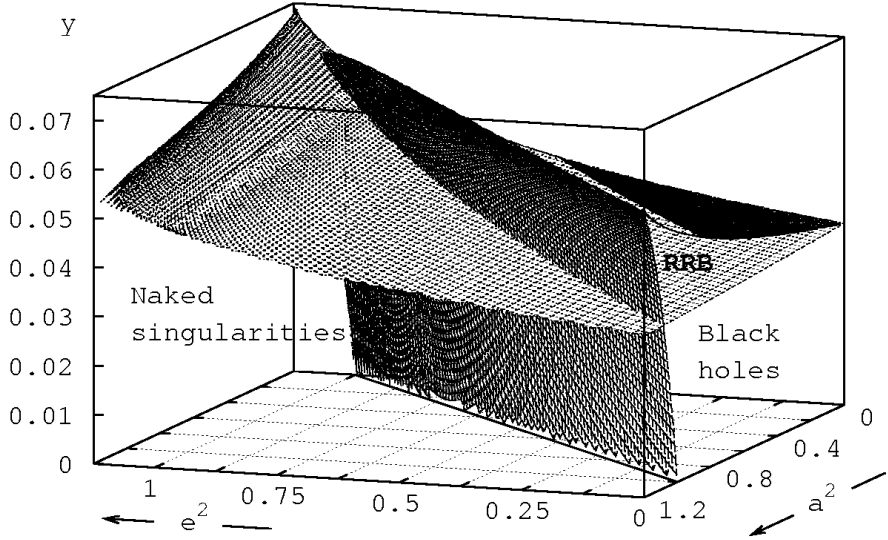}
\caption{Parameter space of the \KNdS\ black-hole spacetimes separated into
  classes Ia+IIa (with divergent repulsive barrier) and Ib+IIb (with \RRB,
  see below). The boundary surface, determined by a numerical code, is
  given by the function $y_{\ri max(d)}(a,e)$ illustrated by the light
  surface.%
}
\label{f6}
\end{figure}

The black-hole spacetimes with a \RRB\ must have the
cos\-mol\-o\-gi\-cal\discretionary{-}{-}{-}constant parameter in the
interval
\be
  y_{\ri max(d)}(a,e) < y < y_{\ri max(h)}(a,e) .
\ee
Using a numerical code, the region of the parameter space corresponding to
these spacetimes can be determined. The result is shown in
Fig.\,\ref{f6}.

\begin{figure}[t]
\centering \leavevmode
\epsfxsize=.9\hsize \epsfbox{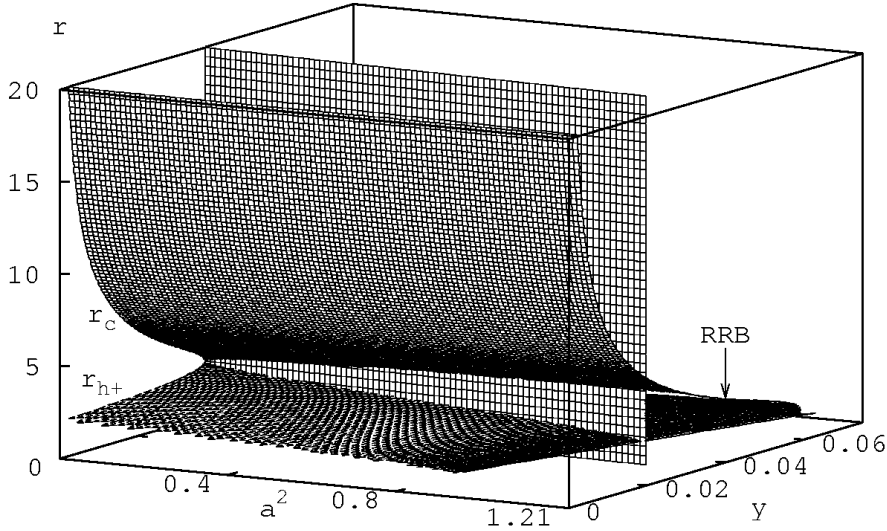}
\caption{Illustration of dependence of the outer black-hole
  and cosmological horizon radii of the \KNdS\ spacetimes with a \RRB\ on
  cosmological parameter $y$ and rotation parameter $a^2$. The light plane
  separates the \KdS\ spacetimes with divergent and \RRB. The cosmological
  parameter $y$ is so high for these spacetimes that the horizon radii are
  comparable.}
\label{f7}
\end{figure}

Extension of the region of the parameter space corresponding to the \KNdS\ 
black-hole spacetimes with the \RRB\ of the photon motion depends strongly
on the parameter $e$. The region is suppressed with increasing values of
$e$, and it disappears for $e^2 > \frac{9}{8}$.  We can easily find
\cite{Stuchlik-Calvani} that, for $a=0$, the critical value $y_{\ri crit}
(e)$, corresponding to the boundary between the black-hole and
naked-singularity spacetimes, shifts from the value $y_{\ri crit}(e=0) =
\frac{1}{27}$ for the extreme \SdS\ geometry to the value $y_{\ri crit}
(e^2=\frac{9}{8}) = \frac{2}{27}$ for the extreme \RNdS\ geometry.  We can
intuitively expect such kind of behavior. Since the rotation parameter $a$
is responsible for the existence of the \RRB, we understand that the
corresponding region of the parameter space will be largest for the
smallest restrictions coming from the other parameter $e$.  The minimal
values of the parameter $y$, allowing the black-hole spacetimes with the
\RRB\ are given by the common points of $y_{\ri max(d)} (a,e)$ and $y_{\ri
  min(h)} (a,e)$. If $e=0$, it reaches its minimum value
\be
  y_{\ri rrb(min)} (e=0) = 0.033185,
\ee
at the rotation parameter
\be
  a^2_{\ri rrb(max)} (e=0) = 1.08316.
\ee
With $e$ increasing, the value of $y_{\ri rrb(min)} (e)$ also increases,
while $a^2_{\ri rrb(min)} (e)$ decreases.  Further, we should note that
inspecting the geometries allowing the \RRB, we
find that the radii of their outer black-hole and cosmological horizons
must be comparable (see Fig.\,\ref{f7}).

The values of $y_{\ri rrb(min)} (e)$ are very high, and correspond to black
holes with an enormously high mass parameter. Considering the recent
estimate~\cite{Krauss,Zeldovich-Novikov} on the relict cosmological constant
\be
  \Lambda_0\sim  8\pi\times 0.65\,\rho_{\ri crit}\sim
    1.1\times10^{-56}\,{\ri cm}^{-2},
\ee
we obtain the limit on the mass of black holes with \RRB\ to be
\be
  M \geq M_{\ri rrb(0)} \sim 2\times 10^{22} M_\odot.
\ee

Of course, radically different estimates on the mass $M_{\ri rrb}$ could be
obtained for primordial black holes in the very early stages of expansion
of the Universe, when phase transitions connected to symmetry breaking of
physical interactions due to Higgs mechanism could take place. For example,
the electroweak symmetry breaking at $T \sim 100$\,GeV could correspond to
an effective cosmological constant \cite{Kolb-Turner}
\be 
  \Lambda_{\ri ew}\sim 0.028\,{\ri cm}^{-2},
\ee
and the related limiting mass is
\be
  M_{\ri rrb(ew)} \sim 2.5\times 10^{28}\,{\ri g}\sim
    1.3\times 10^{-5} M_\odot.
\ee
The phenomenon of the \RRB\ is related to the fact
that, similarly to the constants of motion $E$ and $\Phi$, also the impact
parameters $X$ or $\ell$ have other asymptotical meaning than we are
accustomed to because of the asymptotically \dS\ structure of the
spacetimes. Nevertheless, the physical meaning of the impact parameters $X$
and $\ell$ can be given by their relation to directional angles as measured
by physically well defined stationary observers located between the
black-hole and cosmological horizons.  It will be shown in the next
section, how directional angles of captured and escaping photons measured
by locally non-rotating observers, are related to the impact parameters
having positive values for photons counterrotating relative to these
observers.

The last criterion for the classification of \KN\ spacetimes with $\Lambda
\neq 0$ is given by the local extrema of the effective potential, i.e., it
is given by the number of the circular geodesics.

The behavior of the functions $y_{{\ri ex}\pm}(r;a,e)$ implies that there
are 0, 2, or 4 circular photon orbits present in the \KN\ spacetimes with
$y \neq 0$, except the situations corresponding to the existence of inflex
point of these functions.  Because the local extrema of the functions
$y_{\ri h} (r,a,e)$ and $y_{{\ri ex}+}(r,a,e)$ coincide, we can conclude
that in spacetimes with both $y > 0$ and $y < 0$ two circular photon orbits
always exist outside the outer black-hole horizon. Two additional circular
photon orbits can exist under the inner black-hole horizon.  On the other
hand, in the field of naked singularities, there can exist no, two, or four
circular photon orbits. Stability of the photon circular orbits against
radial perturbation can be directly inferred from the effective potential.

Now, we give the summary of the classification of the \KN\ spacetimes with
$y \neq 0$ according to the properties of the `effective potential'.  We
make the basic separation according to the asymptotic character of the
spacetime (and the potential).  The numbers of the event horizons and
circular photon orbits are considered as main criteria of the
classification. The divergent points of the effective potential are used as
an additional criterion.

\medskip

\begin{figure}[b]
\let\figlabsize=\small
\centering\leavevmode
\begin{minipage}[b]{0.48\hsize}
\centering\leavevmode
\epsfxsize=\hsize \epsfbox{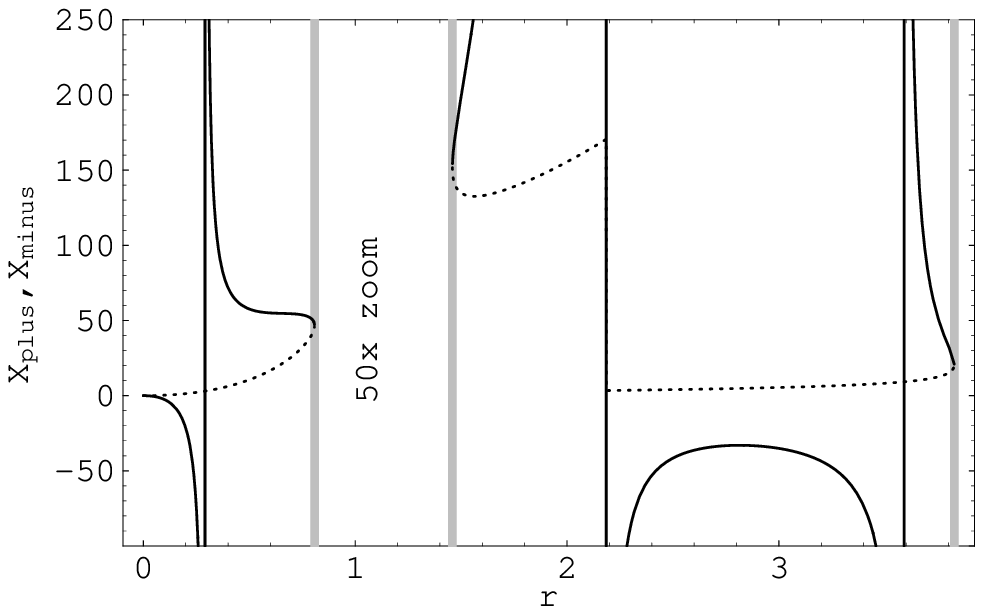}
{\figlabsize (a) Ia: $y = 0.036$, $a^2 = 0.49$, $e^2 = 0.5$}
\end{minipage}\hfill%
\begin{minipage}[b]{0.48\hsize}
\centering\leavevmode
\epsfxsize=\hsize \epsfbox{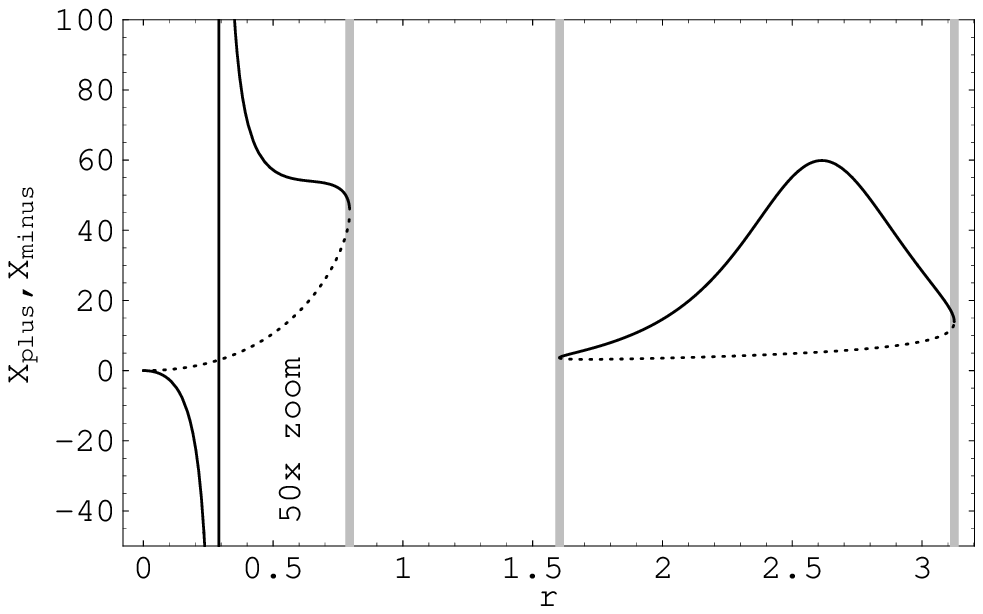}
{\figlabsize (b) Ib: $y = 0.045$, $a^2 = 0.49$, $e^2 = 0.5$}
\end{minipage}
\par\vskip 5mm\par
\begin{minipage}[b]{0.48\hsize}
\centering\leavevmode
\epsfxsize=\hsize \epsfbox{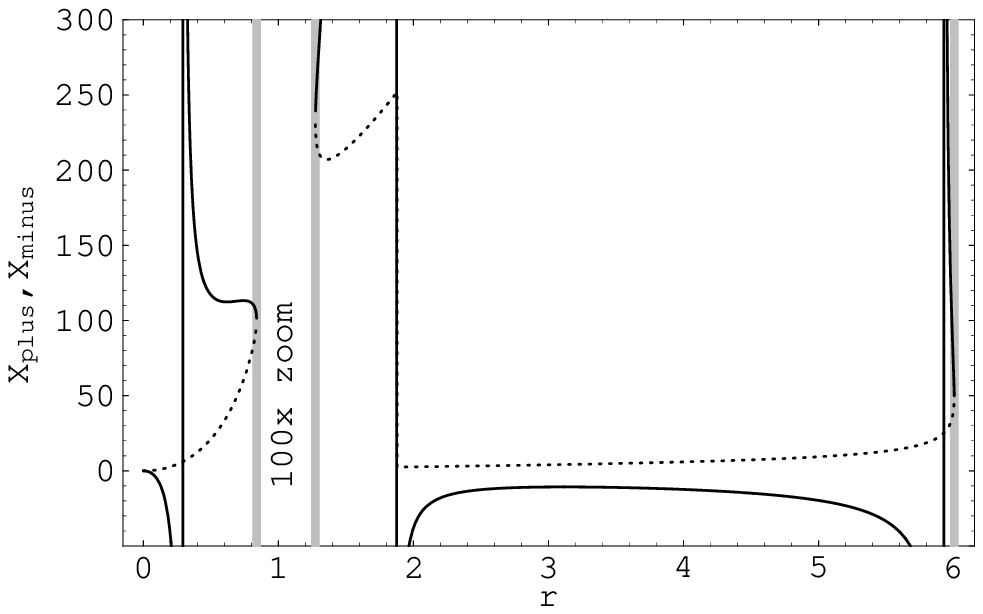}
{\figlabsize (c) IIa: $y = 0.019$, $a^2 = 0.49$, $e^2 = 0.5$}
\end{minipage}\hfill%
\begin{minipage}[b]{0.48\hsize}
\centering\leavevmode
\epsfxsize=\hsize \epsfbox{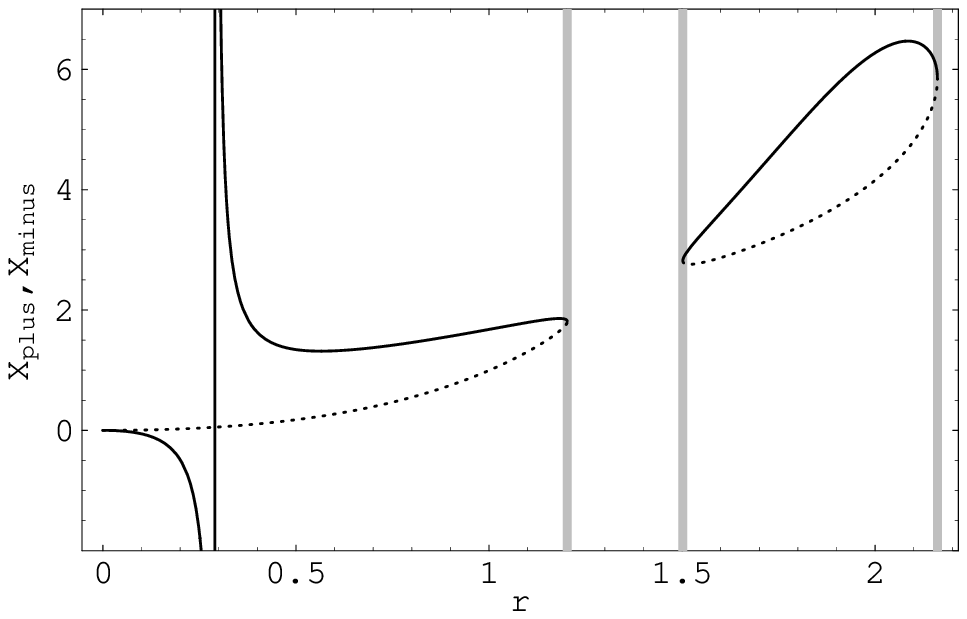}
{\figlabsize (d) IIb: $y = 0.06$, $a^2 = 0.64$, $e^2 = 0.5$}
\end{minipage}
\caption{Typical behavior of effective potential $X_\pm$ for \KNdS\ black-hole
  spacetimes. Stability of the photon circular geodesics can easily be
  inferred from the character of the effective potential. The vertical bars
  of the same thickness as the curves are the vertical asymptotes at the
  points of divergence, the horizons are depicted as thick gray bars. In
  order to clearly display the structure of the curves (especially the
  existence/nonexistence of extremes), certain portions of them are
  vertically zoomed.}
\label{f8}
\end{figure}

\begin{figure}[p]
\let\figlabsize=\small
\begin{minipage}[b]{0.48\hsize}
\centering\leavevmode
\epsfxsize=.95\hsize \epsfbox{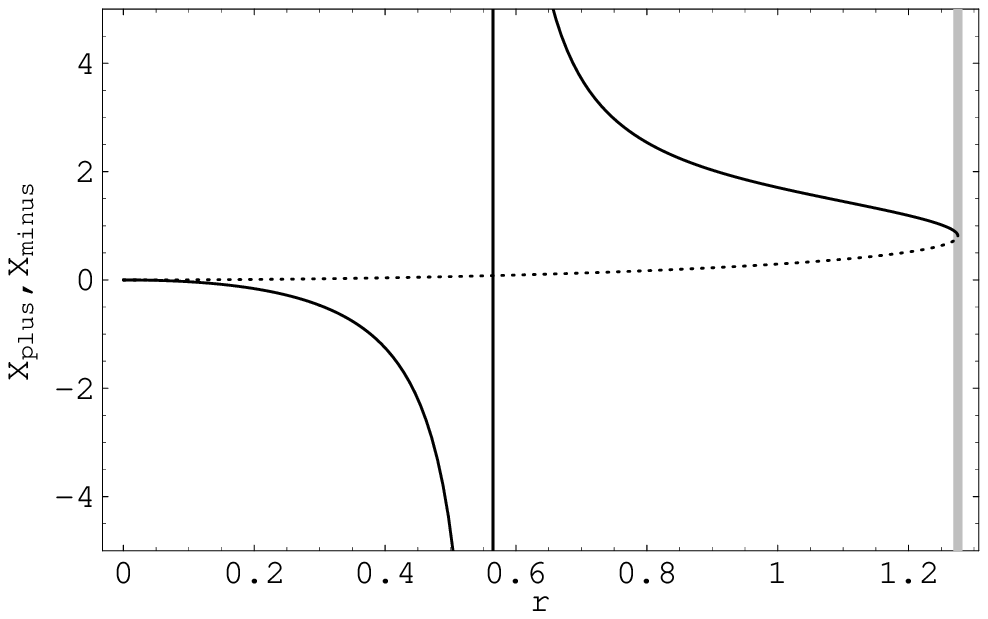}
{\figlabsize (a) III: $y = 0.5$, $a^2 = 4.0$, $e^2 = 1.5$}
\end{minipage}\hfill%
\begin{minipage}[b]{0.48\hsize}
\centering\leavevmode
\epsfxsize=.95\hsize \epsfbox{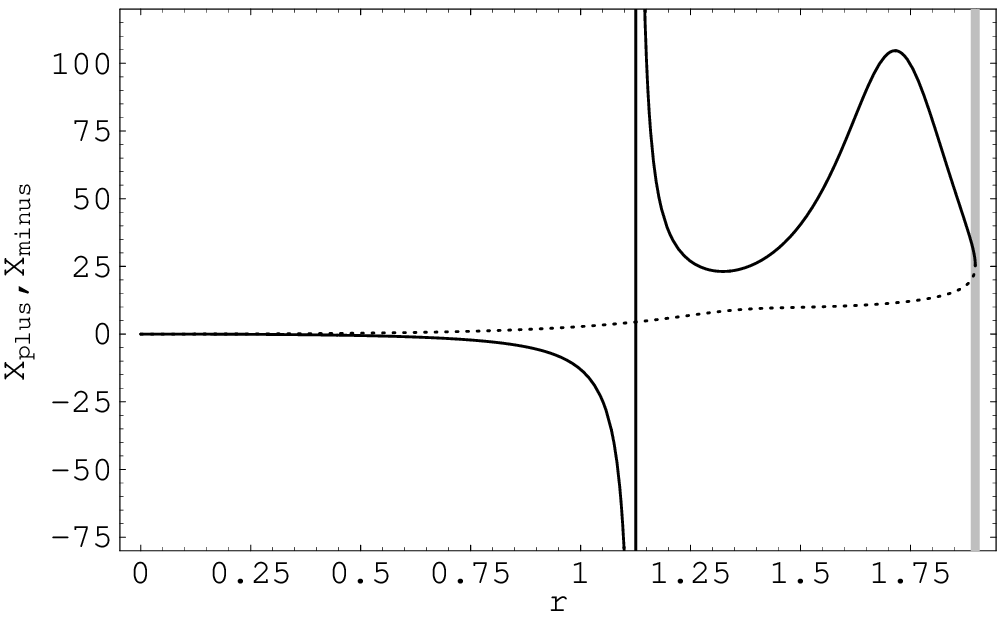}
{\figlabsize (b) IVa: $y = 0.071$, $a^2 = 0.02$, $e^2 = 1.1$}
\end{minipage}
\par\vskip 2.2mm\par
\begin{minipage}[b]{0.48\hsize}
\centering\leavevmode
\epsfxsize=.95\hsize \epsfbox{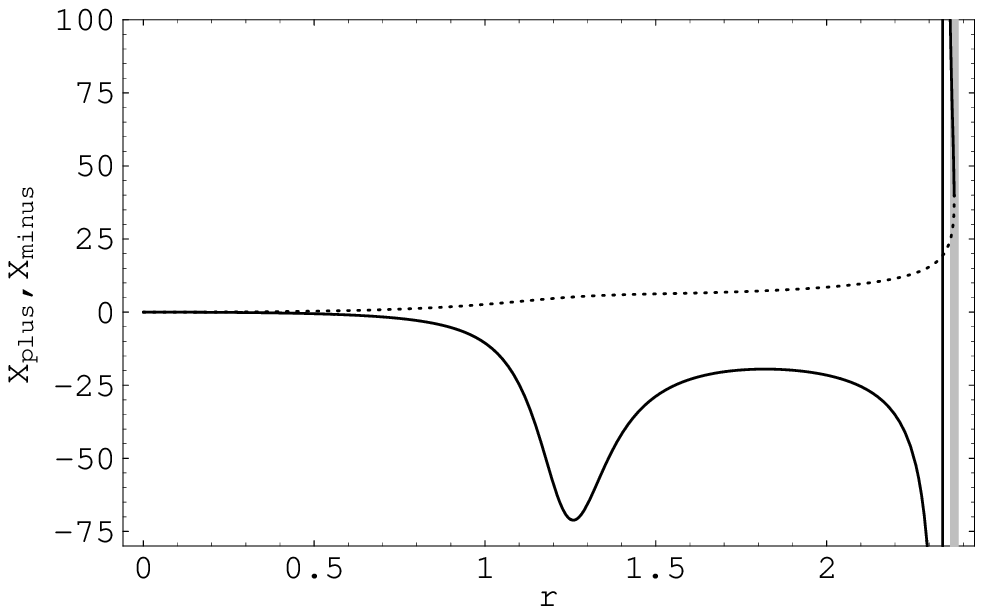}
{\figlabsize (c) IVb: $y = 0.063$, $a^2 = 0.02$, $e^2 = 1.1$}
\end{minipage}\hfill%
\begin{minipage}[b]{0.48\hsize}
\centering\leavevmode
\epsfxsize=.95\hsize \epsfbox{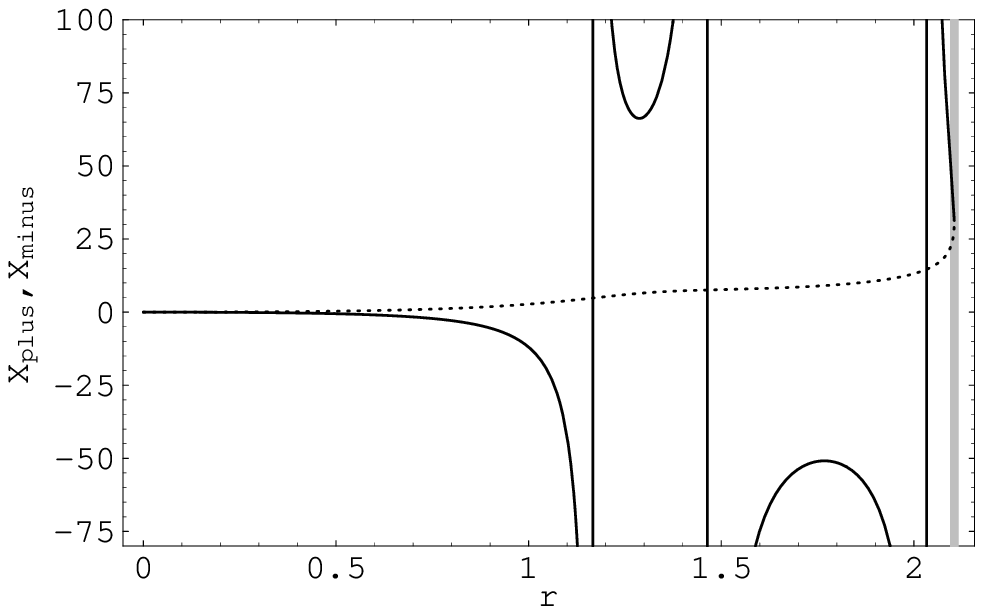}
{\figlabsize (d) IVc: $y = 0.068$, $a^2 = 0.02$, $e^2 = 1.1$}
\end{minipage}
\par\vskip 2.2mm\par
\begin{minipage}[b]{0.48\hsize}
\centering\leavevmode
\epsfxsize=.95\hsize \epsfbox{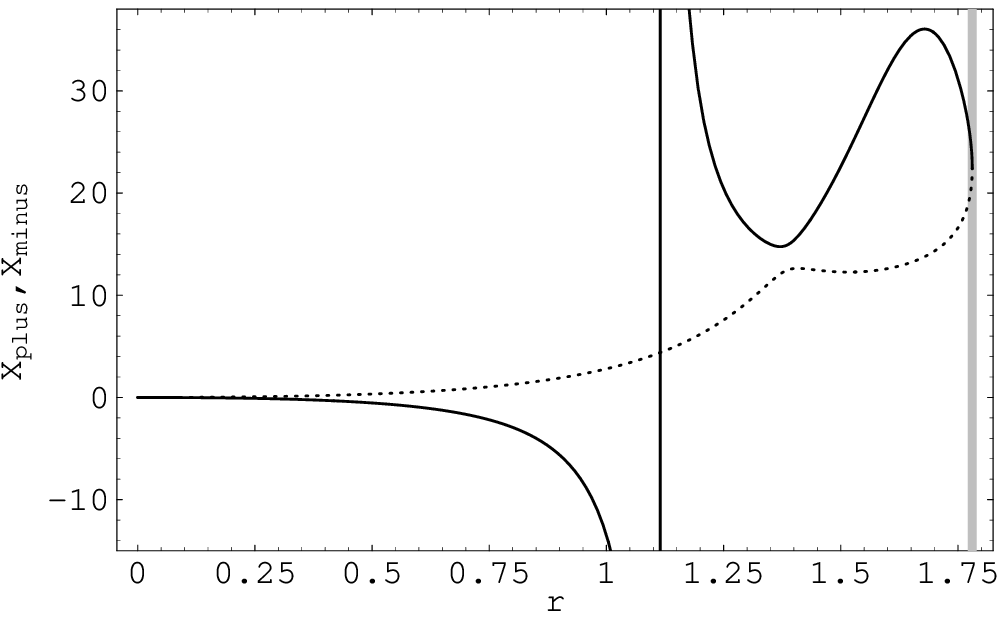}
{\figlabsize (e) Va: $y = 0.0721$, $a^2 = 0.02$, $e^2 = 1.1$}
\end{minipage}\hfill%
\begin{minipage}[b]{0.48\hsize}
\centering\leavevmode
\epsfxsize=.95\hsize \epsfbox{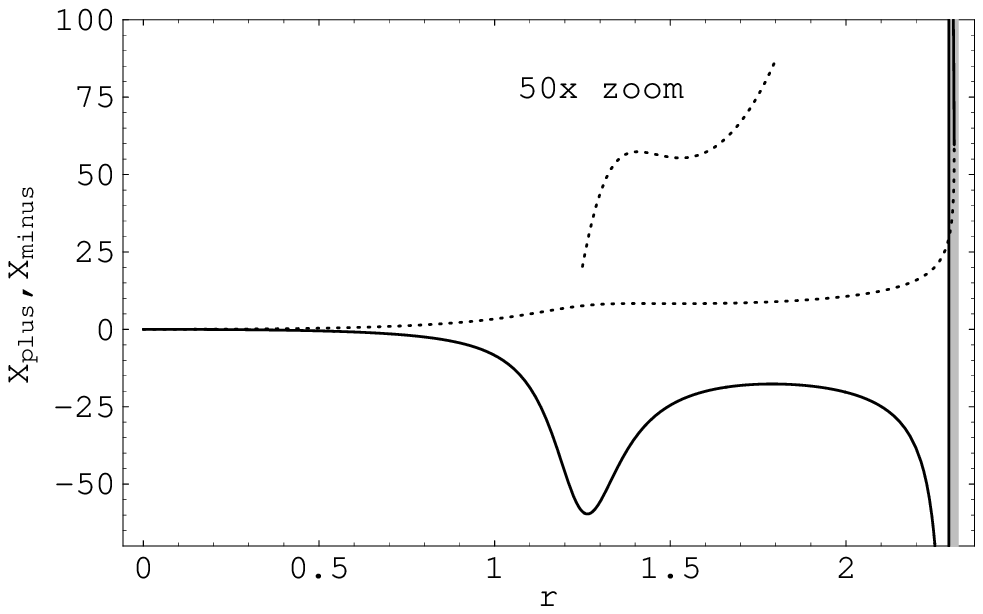}
{\figlabsize (f) Vb: $y = 0.064$, $a^2 = 0.008$, $e^2 = 1.1$}
\end{minipage}
\par\vskip4.2mm\par
\begin{minipage}[c]{0.48\hsize}
\centering\leavevmode
\epsfxsize=.95\hsize \epsfbox{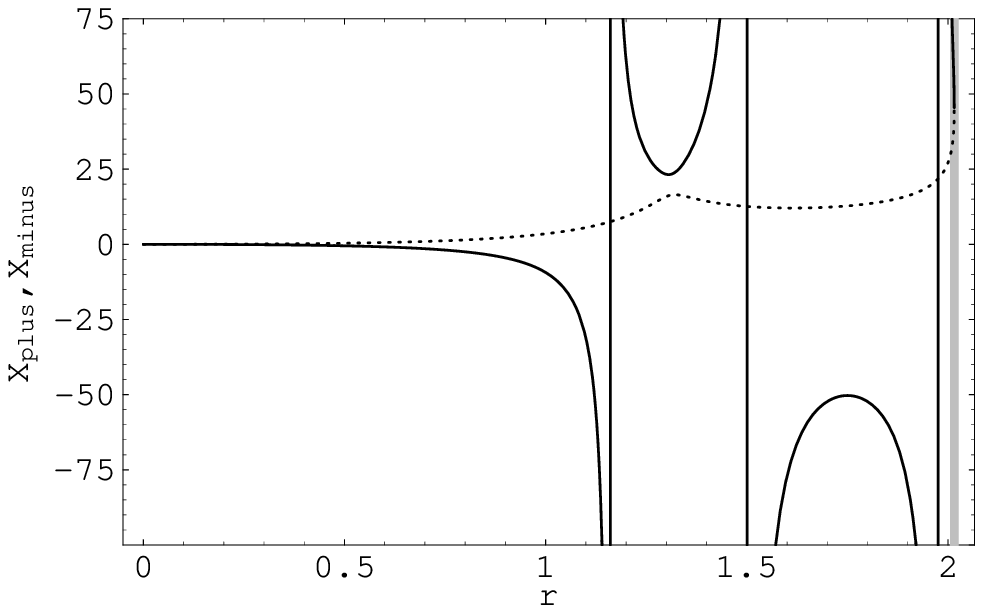}
{\figlabsize (g) Vc: $y = 0.0689$, $a^2 = 0.008$, $e^2 = 1.1$}
\end{minipage}\hfill%
\begin{minipage}[c]{0.48\hsize}
\mbox{}
\end{minipage}
\caption{\label{f9}Typical behavior of effective potential $X_\pm$ for \KNdS\
  naked-singularity spacetimes. Stability of the photon circular geodesics
  can easily be inferred from the character of the effective potential. The
  vertical bars of the same thickness as the curves are the vertical
  asymptotes at the points of divergence, the cosmological horizon is
  depicted as thick gray bar. In order to clearly display the structure of
  the curves (especially the existence/nonexistence of extremes), certain
  portions of them are vertically zoomed.} 
\end{figure}

\begin{figure}[b]
\let\figlabsize=\small
\centering\leavevmode
\begin{minipage}[b]{0.48\hsize}
\centering\leavevmode
\epsfxsize=\hsize \epsfbox{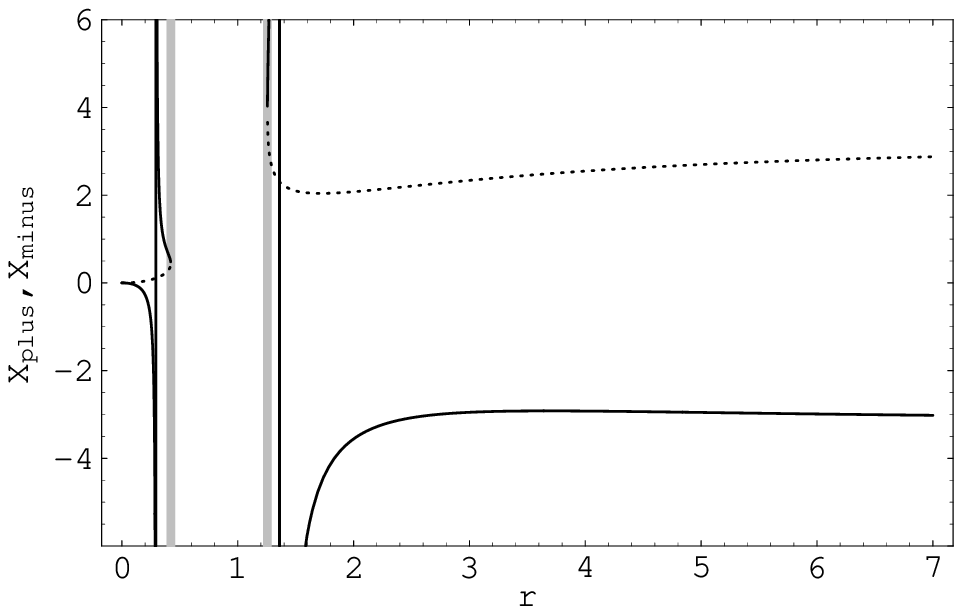}
{\figlabsize (a) VI: $y = -0.1$, $a^2 = 0.16$, $e^2 = 0.5$}
\end{minipage}\hfill%
\begin{minipage}[b]{0.48\hsize}
\centering\leavevmode
\epsfxsize=\hsize \epsfbox{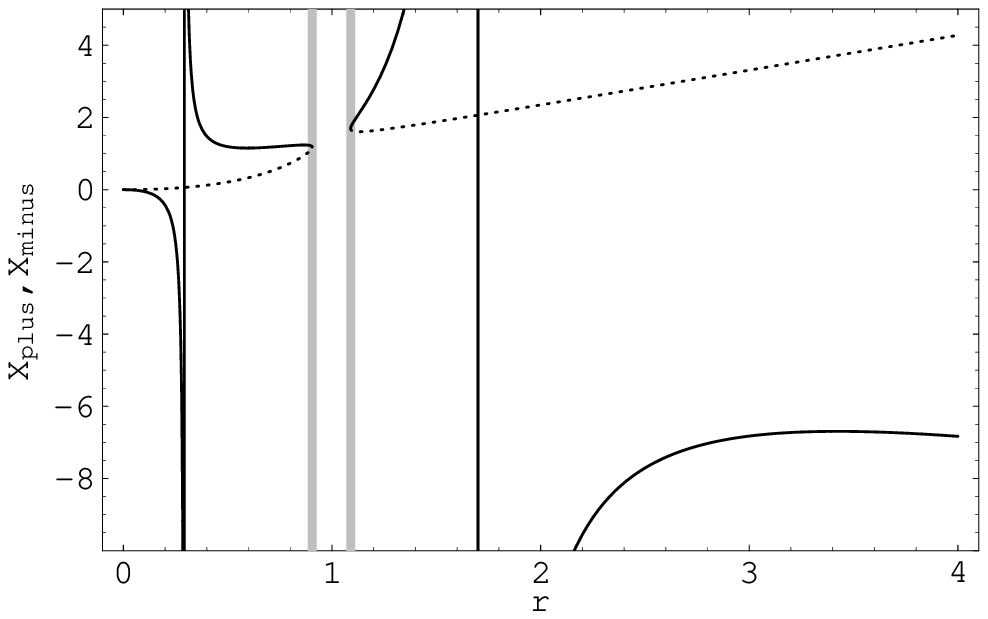}
{\figlabsize (b) VII: $y = -0.001$, $a^2 = 0.49$, $e^2 = 0.5$}
\end{minipage}
\caption{Typical behavior of effective potential $X_\pm$ for \KNadS\
  black-holes. Stability of the photon circular geodesics can easily be
  inferred from the character of the effective potential. The vertical bars
  of the same thickness as the curves are the vertical asymptotes at the
  points of divergence, the black-hole horizons are depicted as thick gray
  bar.}
\label{f10}
\end{figure}

\begin{figure}[t]
\let\figlabsize=\small
\centering\leavevmode
\begin{minipage}[b]{0.48\hsize}
\centering\leavevmode
\epsfxsize=\hsize \epsfbox{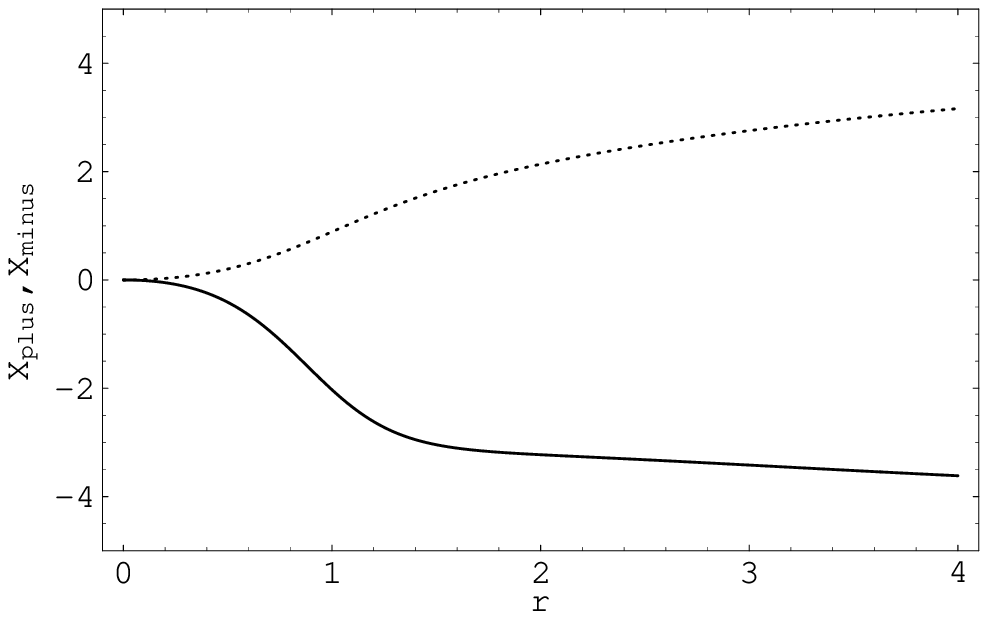}
{\figlabsize (a) VIII: $y = -0.05$, $a^2 = 0.1$, $e^2 = 1.5$}
\end{minipage}\hfill%
\begin{minipage}[b]{0.48\hsize}
\centering\leavevmode
\epsfxsize=\hsize \epsfbox{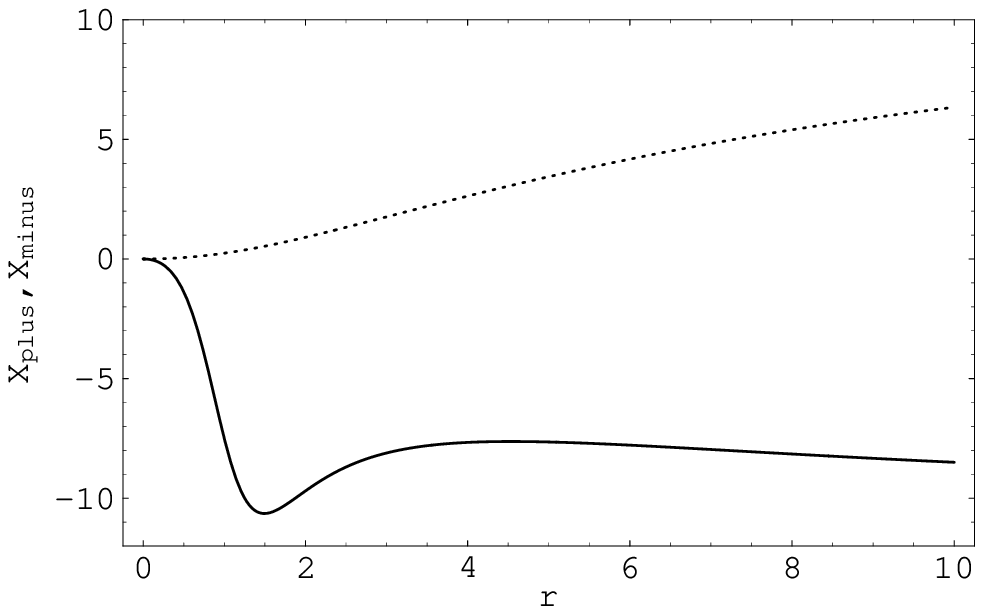}
{\figlabsize (b) IXa: $y = -0.01$, $a^2 = 4.0$, $e^2 = 1.5$}
\end{minipage}
\par\vskip 4mm\par
\begin{minipage}[b]{0.48\hsize}
\centering\leavevmode
\epsfxsize=\hsize \epsfbox{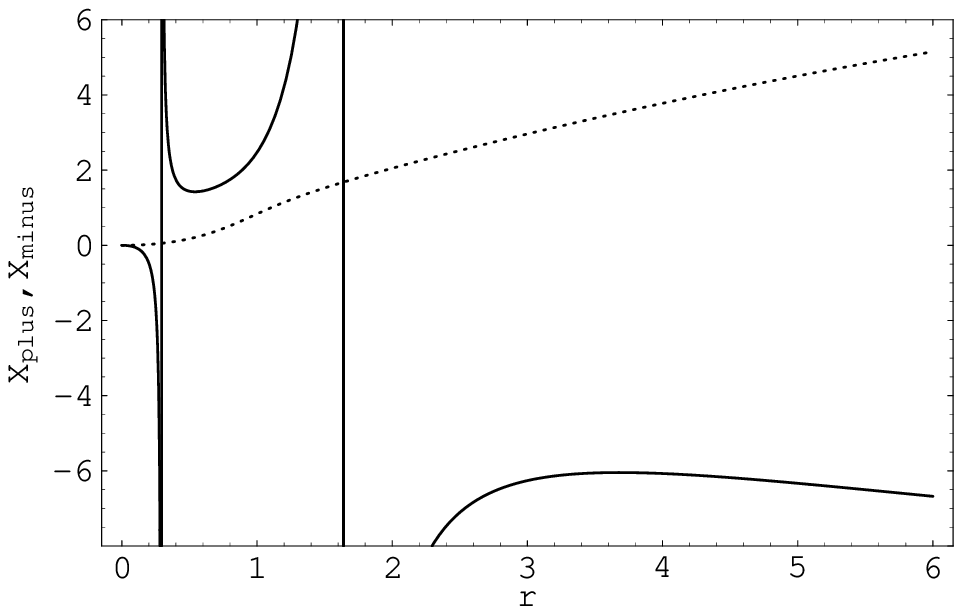}
{\figlabsize (c) IXb: $y = -0.01$, $a^2 = 0.64$, $e^2 = 0.5$}
\end{minipage}\hfill%
\begin{minipage}[b]{0.48\hsize}
\centering\leavevmode
\epsfxsize=\hsize \epsfbox{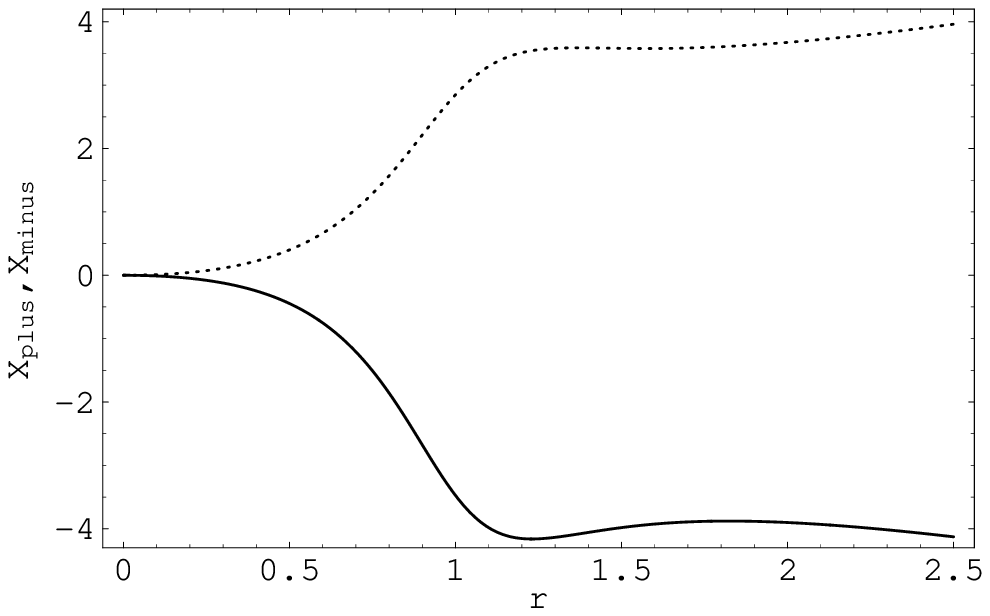}
{\figlabsize (d) Xa: $y = -0.001$, $a^2 = 0.001$, $e^2 = 1.1$}
\end{minipage}
\par\vskip 4mm\par
\begin{minipage}[b]{0.48\hsize}
\centering\leavevmode
\epsfxsize=\hsize \epsfbox{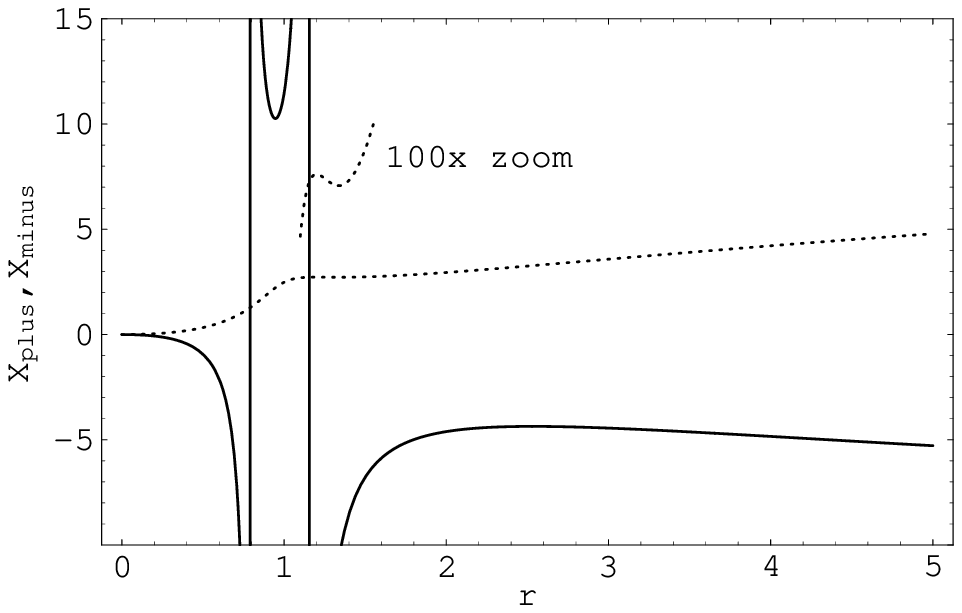}
{\figlabsize (e) Xb: $y = -0.014$, $a^2 = 0.06$, $e^2 = 0.95$}
\end{minipage} 
\caption{Typical behavior of effective potential $X_\pm$ for \KNadS\
  naked-singularity spacetimes. Stability of the photon circular geodesics
  can easily be inferred from the character of the effective potential. The
  vertical bars of the same thickness as the curves are the vertical
  asymptotes at the points of divergence. In order to clearly display the
  structure of the curves (especially the existence/nonexistence of
  extremes), certain portions of them are vertically zoomed.
  \protect\rule[-6mm]{0pt}{6mm}}
\label{f11}
\end{figure}

\noindent{\bf \KNdS\ spacetimes ({\boldmath{$y > 0$}})}
\begin{description}
\item[{\bf Ia:}] Black holes with two photon circular orbits and a divergent
  repulsive barrier between the outer black-hole and cosmological horizons
  (Fig.\,\ref{f8}a). Both circular orbits are unstable relative to radial
  perturbations.
  
\item[{\bf Ib:}] Black holes with two photon circular orbits and a \RRB\ 
  between the outer black-hole and cosmological horizons (Fig.\,\ref{f8}b).
  Both circular orbits are unstable.
  
\item[{\bf IIa:}] Black holes with four photon circular orbits and a
  divergent repulsive barrier (Fig.\,\ref{f8}c). The innermost circular
  orbit is stable, the others are unstable.
  
\item[{\bf IIb:}] Black holes with four photon circular orbits and a \RRB\ 
  (Fig.\,\ref{f8}d). The innermost circular orbit is stable, the others are
  unstable.
  
\item[{\bf III:}] Naked singularities with no circular photon orbit
  (Fig.\,\ref{f9}a).
  
\item[{\bf IVa:}] Naked singularities with two circular photon orbits
  located above the divergent point (Fig.\,\ref{f9}b). The inner circular
  orbit is stable, the outer one is unstable.
  
\item[{\bf IVb:}] Naked singularities with two circular photon orbits
  located under the divergent point (Fig.\,\ref{f9}c). The inner circular
  orbit is stable, the outer one is unstable.
  
\item[{\bf IVc:}] Naked singularities with two circular photon orbits and
  three divergent points (Fig.\,\ref{f9}d). The inner circular orbit
  is stable, the outer one is unstable.
  
\item[{\bf Va:}] Naked singularities with four circular photon orbits
  located above the divergent point (Fig.\,\ref{f9}e). Two circular orbits
  are stable, the others are unstable.
  
\item[{\bf Vb:}] Naked singularities with four circular photon orbits
  located under the divergent point (Fig.\,\ref{f9}f). Two circular orbits
  are stable, the others are unstable.
  
\item[{\bf Vc:}] Naked singularities with four circular photon orbits and
  three divergent points (Fig.\,\ref{f9}g). Two circular orbits are stable,
  the others are unstable.
\end{description}

\smallskip

\noindent{\bf \KNadS\ spacetimes ({\boldmath{$y < 0$}})}
\begin{description}
\item[{\bf VI:}] Black holes with two photon circular orbits
  (Fig.\,\ref{f10}a).  Both circular orbits are unstable.
  
\item[{\bf VII:}] Black holes with four photon circular orbits
  (Fig.\,\ref{f10}b).  The innermost orbit is stable, the others are
  unstable.
  
\item[{\bf VIII:}] Naked singularities with zero circular photon orbits
  (Fig.\,\ref{f11}a).
  
\item[{\bf IXa:}] Naked singularities with two circular photon orbits
  (Fig.\,\ref{f11}b).  The inner circular orbit is stable, the outer one is
  unstable.
  
\item[{\bf IXb:}] Naked singularities with two circular photon orbits and
  two divergent points (Fig.\,\ref{f11}c). The inner circular orbit is
  stable, the outer one is unstable.
  
\item[{\bf Xa:}] Naked singularities with four circular photon orbits
  (Fig.\,\ref{f11}d). Two inner orbits are stable, two outer orbits are
  unstable.
  
\item[{\bf Xb:}] Naked singularities with four circular photon orbits and
  two divergent points (Fig.\,\ref{f11}e). Two inner orbits are stable, two
  outer orbits are unstable.
\end{description}

\begin{figure}[b]
\let\figlabsize=\small
\begin{minipage}{.48\linewidth}
\centering \leavevmode
\epsfxsize=\hsize \epsfbox{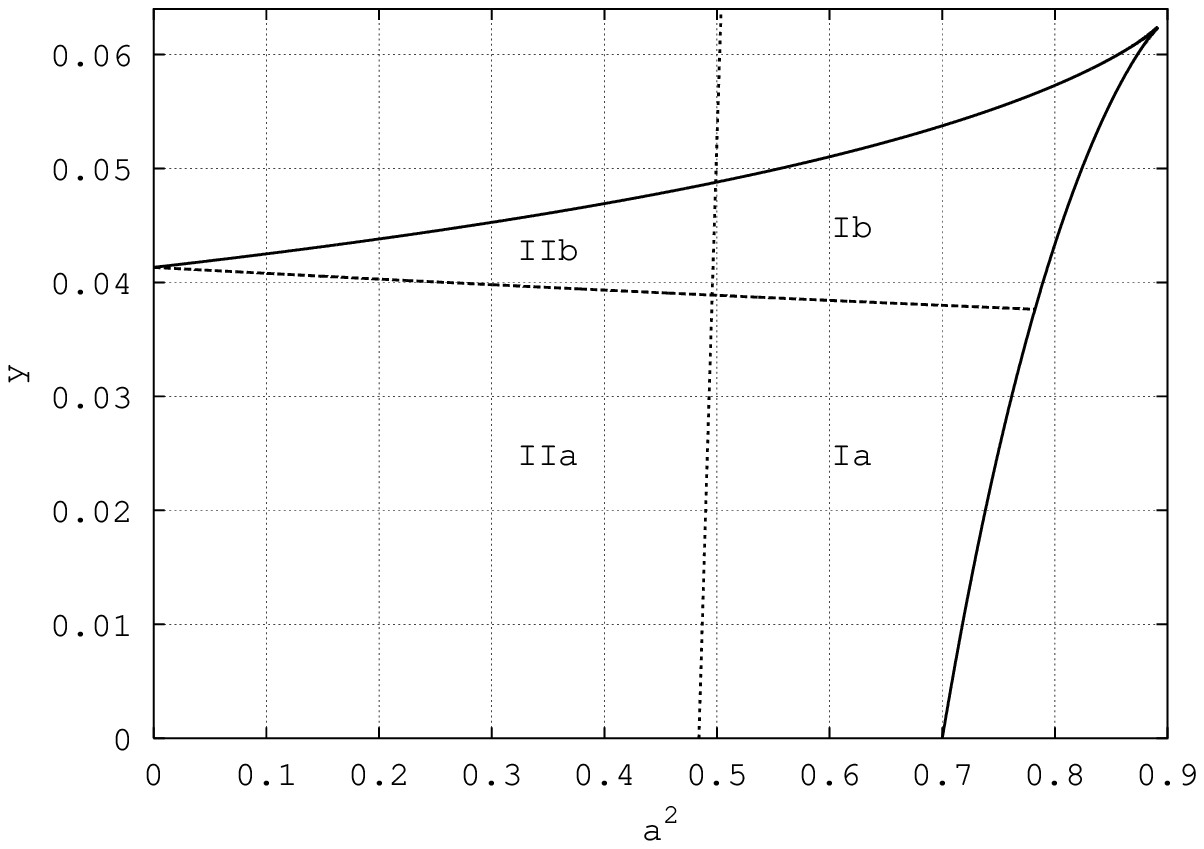}
\par{\figlabsize (a)}\par
\end{minipage}\hfill%
\begin{minipage}{.48\linewidth}
\centering \leavevmode
\epsfxsize=\hsize \epsfbox{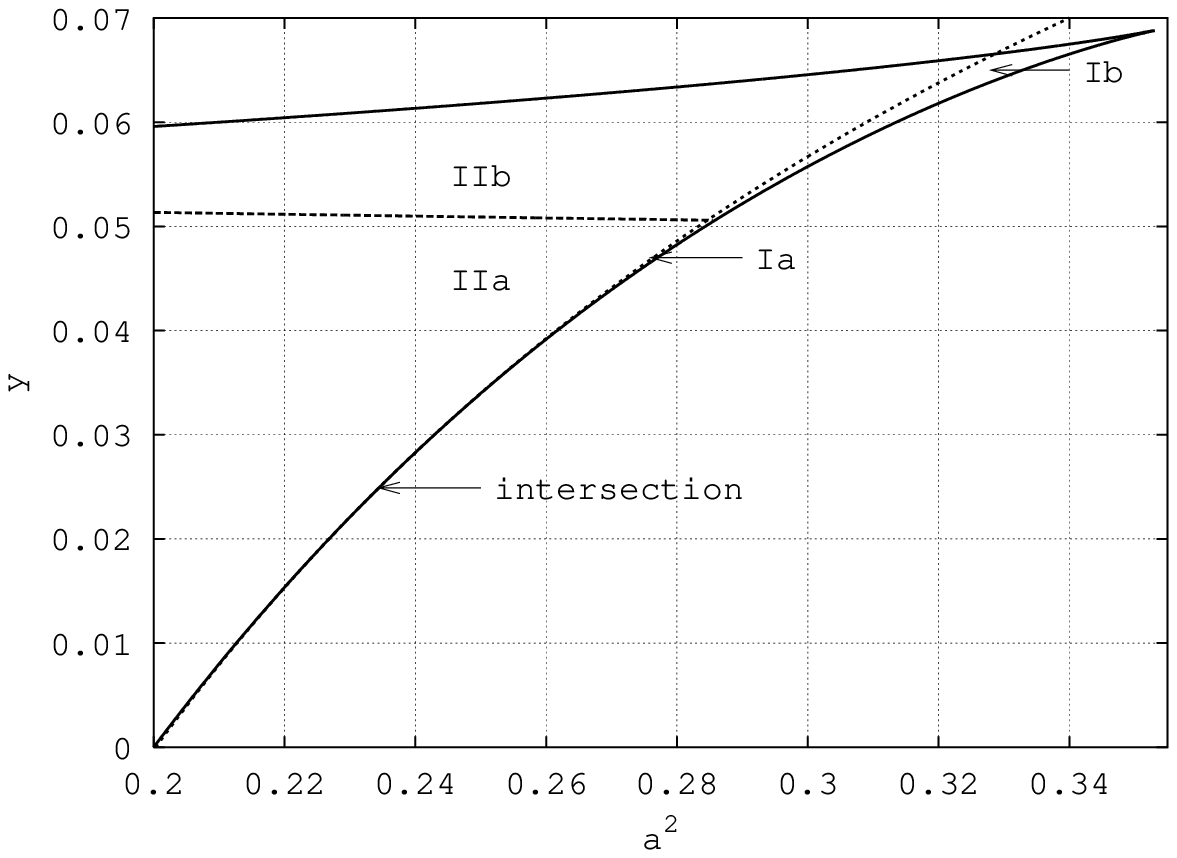}
\par{\figlabsize (b)}\par
\end{minipage}\hfill%
\begin{minipage}{.48\linewidth}
\centering \leavevmode
\epsfxsize=\hsize \epsfbox{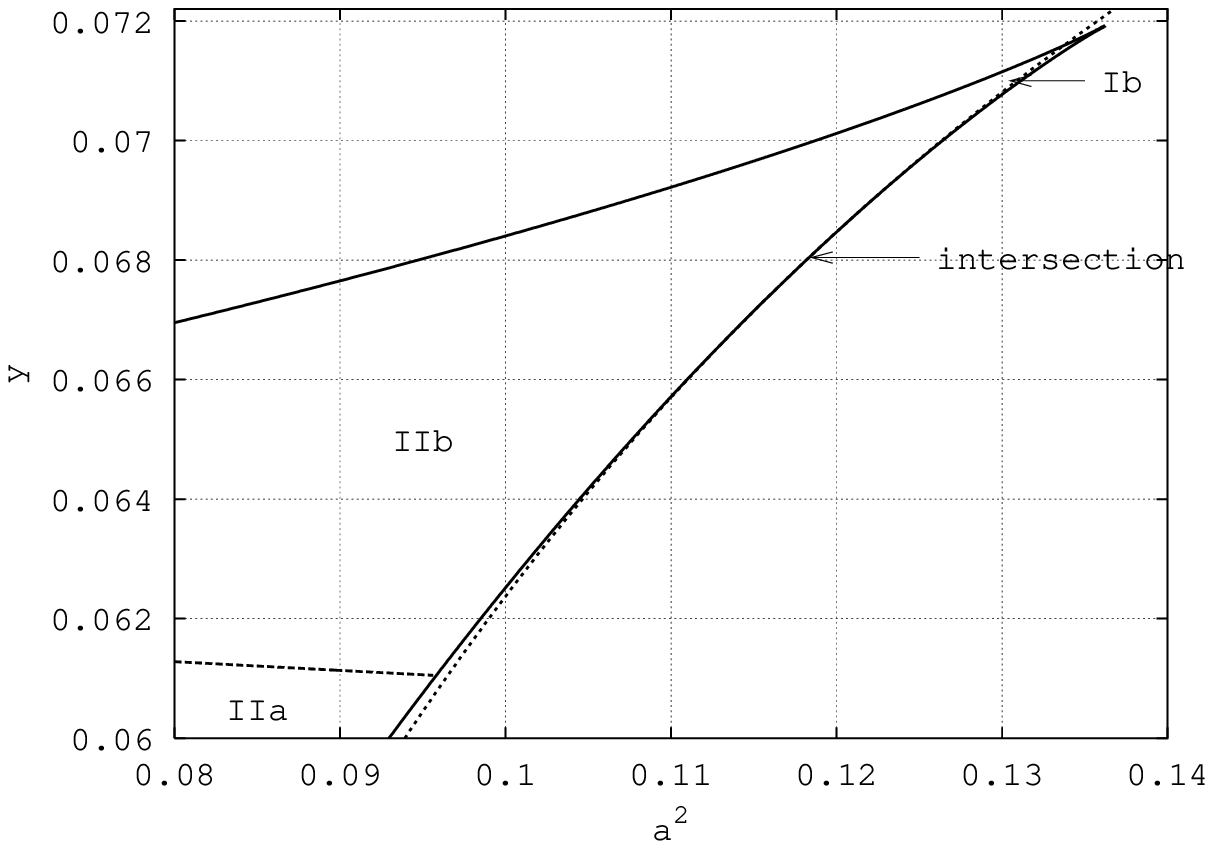}
\par{\figlabsize (c)}
\end{minipage}\hfill%
\begin{minipage}{.48\linewidth}
\mbox{}
\end{minipage}
\caption{Parameter space of the \KNdS\ black-hole space\-times separated into
  classes Ia,b and IIa,b. The distribution in the parameter space
  $y$-$a^2$-$e^2$ is represented by three qualitatively different $e^2 =
  {\ri const}$ slices of the space, since the 3D plot with three different
  surfaces looks rather messy. The typical sections are given for (a)~$e^2
  = 0.3$, (b)~$e^2 = 0.8$, (c)~$e^2 = 1.0$. In the cases (b) and (c), the
  intersection points of $y_{\ri max(ex+)}$ (dotted curve) and $y_{\ri
  min(h)}$ (full curve crossing the $a^2$ axis) are
  explicitly shown. Dashed curve separates spacetimes with divergent and \RRB.}
\label{addf1}
\end{figure}

\begin{figure}[t]
\centering \leavevmode
\epsfxsize=.7\hsize \epsfbox{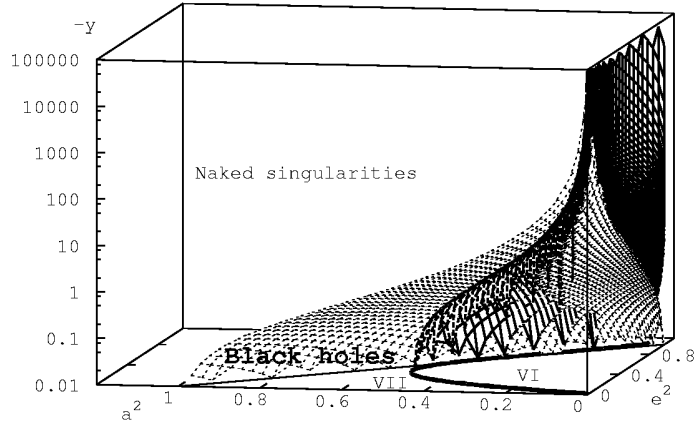}
\caption{Parameter space of the \KNadS\ black-hole spacetimes separated into
  regions corresponding to the class VI and VII in a 3D diagram. We use the
  logarithmic scale on the $y$-axis because black-hole states are allowed
  for all negative values of $y$. The light surface ($y_{\ri min(h)}$)
  separates black-hole and naked-singularity spacetimes, while the dark
  surface ($y_{\ri max(ex+)}$) separates spacetimes allowing different
  number of circular photon orbits. Note that the function $y_{\ri
    max(ex+)}$ diverges for $a^2 \rightarrow 0$. (For $e^2 \rightarrow 0$,
  there is $y_{\ri max(ex+)}(a,e) \rightarrow +\infty$.) The surface
  $y_{\ri max(ex+)}(a,e)$ intersects the $y = 0$ plane at the curve
  $a^2_{\ri ex(z(ex))}(e^2)$ separating the \KN\ spacetimes with different
  number of circular photon geodesics (bold curve). The part of the surface
  $y_{\ri max(ex+)}(a,e)$ at small values of $e^2$ is not visualized. It
  enables to show the intersection of the surfaces $y_{\ri max(ex+)}(a,e)$
  and $y_{\ri min(h)}(a,e)$.}
\label{addf2}
\end{figure}

Now we determine regions of the parameter space corresponding to the
black-hole spacetime classes defined above. Region of the parameter space
corresponding to black holes is given in Fig.\,\ref{addf1} (classes Ia,b and
IIa,b, for spacetimes with a repulsive cosmological constant) and in
Fig.\,\ref{addf2} (classes VI and VII, for spacetimes with an attractive
cosmological constant). For the \KNdS\ black holes with four photon
circular orbits (classes IIa,b), the parameter space is determined by the
condition
\be
  0 < y_{\ri min(h)} < y < y_{\ri max(ex+)} < y_{\ri max(h)},
\ee
where $y_{\ri max(ex+)}$ corresponds to $y_{\ri ex(ex+)}$, i.e., to the
maxima of the function $y_{\ri ex+}(r;a,e)$ taken at $r = \frac{4}{3}e^2$.
They are determined by the `$+$' branch of the function (\ref{e2exd}).
(Note that the conditions determining black-hole spacetimes with a
restricted and divergent repulsive barrier are given by the relation
(\ref{xi}), and the distribution of black-hole spacetimes in the parameter
space is given completely.)  For the \KNadS\ black holes with four photon
circular orbits (class VII), the parameter space is determined by the
condition
\be
  y_{\ri min(h)} < y < y_{\ri max(ex+)} < 0,
\ee
together with the condition (\ref{infplus}) which guarantees that the
extremum of $y_{\ri ex+}(r;a,e)$ at $r= \frac{4}{3}e^2$ is a maximum. Black
holes of class VI (with two photon circular orbits) are determined by the
condition
\be
  y_{\ri max(\ri ex+)} <y <0,
\ee
if relation (\ref{infplus}) is valid; if $a^2 < a^2_{\ri inf+}(e)$, the
spacetimes of class VI are determined by the relation
\be
  y_{\ri min(h)} < y < 0.
\ee
For the classes of the naked-singularity spacetimes, the parameter space
can be divided into the corresponding separated parts in an analogous
manner.

\section{Directional angles of photons in black-hole spacetimes with a
  repulsive cosmological constant} \label{angles}

In order to understand the character of the spacetimes with a \RRB, we
investigate the behavior of directional angles of equatorial photons as
measured by a family of stationary observers in these spacetimes.  We
determine properties of photon escape/capture cones, and relations between
the directional angle of a photon and its impact parameter. It is useful to
compare the results with the situation held in the spacetimes with a
divergent repulsive barrier, and, especially, with the case of pure Kerr
black hole.  Because the effects are caused by the rotation parameter of
the spacetime, we put $e=0$ for simplicity.

The most convenient family of local stationary observers in the rotating
back\-ground is the family of locally non-rotating observers, introduced by
Bardeen \cite{Bardeen}. In the \KdS\ spacetimes, the tetrad of differential
forms corresponding to this family of observers is given by
\bea
  &&\omega^{(t)} \equiv
    \left(
      \frac{\Delta_r \Delta_{\theta} \rho^2}{I^2 A}
    \right)^{1/2}\,\d t, \\
  &&\omega^{(r)} \equiv
    \left(
      \frac{\rho^2}{\Delta_r}
    \right)^{1/2}\,\d r, \\
  &&\omega^{(\theta)} \equiv
    \left(
      \frac{\rho^2}{\Delta_{\theta}}
    \right)^{1/2}\ \d\theta, \\
  &&\omega^{(\phi)} \equiv
    \frac{A^{1/2} \sin \theta}{I \rho}
    (\d\phi - \Omega\,\d t),
\eea
where
\be
  A = (r^2 + a^2)^2 - a^2 \Delta_r,
\ee
and the angular velocity of such observers
\be
  \Omega = \Omega(r,\theta;y,a) = \oder{\phi}{t} =
    \frac{a \left[-\Delta_r + (r^2+a^2)
    \Delta_{\theta}\right]}{A}.                             \label{LNRFobs}
\ee
We can convince ourselves easily that both the functions $A$ and $\Omega$
are positive at the stationary regions of the spacetime at $r>0$. For
completeness, we present also the tetrad of vectors dual to the
differential forms:
\bea
  &&e_{(t)} =
    \left(
      \frac{I^2 A}{\Delta_r \Delta_{\theta}\rho^2}
    \right)^{1/2}
    \left(\pder{}{t} + \Omega\pder{}{\phi}\right), \\
  &&e_{(r)} =
    \left(
      \frac{\Delta_r}{\rho^2}
    \right)^{1/2}
    \pder{}{r}, \\
  &&e_{(\theta)} =
    \left(
      \frac{\Delta_{\theta}}{\rho^2}
    \right)^{1/2}
    \pder{}{\theta}, \\
  &&e_{(\phi)} =
    \frac{I \rho}{A^{1/2} \sin \theta}
    \pder{}{\phi}.
\eea
Locally measured components of photon's 4-momentum $p^{\mu} \equiv \d
x^{\mu}/\d\lambda$, are given by projections onto the tetrads:
\be
  p^{(\alpha)} = p^\mu \omega^{(\alpha)}_\mu; \quad
  p_{(\beta)} = p_\mu e^\mu_{(\beta)}.
\ee
The locally measured components are then related in the simple
special-relativistic way,
\be
  p^{(t)} = -p_{(t)}; \quad p^{(\phi)} = p_{(\phi)}.
\ee
Now we shall restrict our attention to the equatorial motion of photons.
The directional angle $\psi$ of these photons, related to the outward
radial direction is generally determined by the relations
\bea
  &&\sin\psi = \frac{p^{(\phi)}}{p^{(t)}},\\
  &&\cos\psi = \frac{p^{(r)}}{p^{(t)}},
\eea
In terms of the impact parameter $X$, the equatorial components of photon's
4-momentum are given by
\bea
  &&p^r = \oder{r}{\lambda} =
    \pm \frac{I}{r^2} \sqrt{(r^2 - aX)^2 - \Delta_r X^2}, \label{pr} \\
  &&p^\phi = \oder{\phi}{\lambda} =
    \frac{I^2}{r^2} \left[X+ \frac{a\left(r^2-aX \right)}{\Delta_r} \right],\\
  &&p^t = \oder{t}{\lambda} =
    \frac{I^2}{r^2}\left[aX + \frac{(r^2+a^2)(r^2-aX)}{\Delta_r}\right],
\eea
where the $+(-)$ sign in Eq.~(\ref{pr}) corresponds to the outward (inward)
photon's motion. Then we arrive at
\bea
  &&p^{(r)} =
    \pm \frac{I}{r\Delta_r^{1/2}} \sqrt{(r^2 - aX)^2 - \Delta_r X^2},\\
  &&p^{(\phi)} = \frac{Ir}{A^{1/2}} (X+a),\\
  &&p^{(t)} =
    \frac{I}{r} \left(\frac{A}{\Delta_r} \right)^{1/2}[1- \Omega (X+a)],
\eea
and
\bea
  &&\sin\psi =
    \frac{r^2 \Delta_r^{1/2} (X+a)}{A[1- \Omega (X+a)]}, \label{phiX} \\
  &&\cos\psi =
    \pm\frac{\sqrt{(r^2 - aX)^2 - \Delta_r X^2}}{A^{1/2}[1- \Omega (X+a)]},
\eea
or in terms of impact parameter $\ell$, we find the relations
\bea
  &&\sin\psi =
    \frac{r^2 \Delta^{1/2}_r\ell}{A(1- \Omega \ell)}, \label{phil}\\
  &&\cos\psi =
    \pm\frac{\sqrt{1-2\Omega\ell + A^{-1}(a^2-\Delta_r)\ell^2}}{1-\Omega\ell}.
\eea
The angular velocity of the locally non-rotating frames
$\Omega(r,\theta=\pi/2;y,a)$ is given by Eq.~(\ref{LNRFobs}).  Now we are
able, using the properties of the radial motion, to determine equatorial
sections of photon escape (capture, respectively) cones. It is useful to
invert the relations (\ref{phiX}) and (\ref{phil}), and write
\be
  X(\psi;r,y,a) = \frac{A(1-a \Omega)\sin\psi -
    r^2 a \Delta^{1/2}_r}{A \Omega \sin \psi + r^2 \Delta^{1/2}_r},
\ee
and
\be
  \ell(\psi;r,y,a) = \frac{A\sin\psi}{A\Omega\sin\psi + r^2\Delta_r^{1/2}}.
\ee
We can immediately see that, as expected, for the radially directed photons
with $\psi = 0$, or $\psi = \pi$, the impact parameter $\ell = 0$, and $X =
-a$.

\begin{figure}[t]
\let\figlabsize=\small
\centering\leavevmode
\begin{minipage}[b]{0.48\hsize}
\centering\leavevmode
\epsfxsize=\hsize \epsfbox{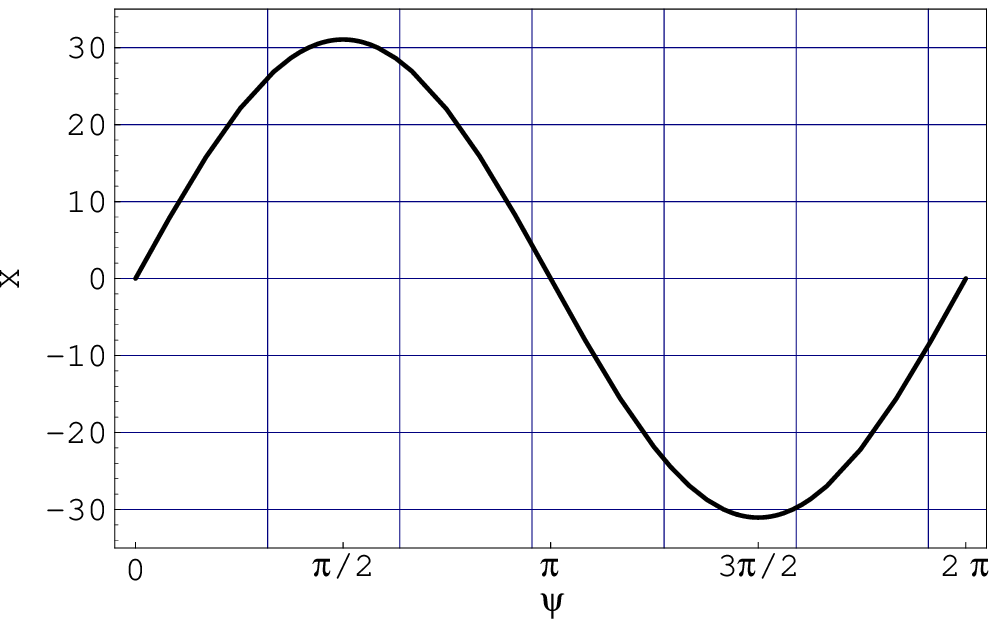}
{\figlabsize (a)}
\end{minipage}\hfill%
\begin{minipage}[b]{0.48\hsize}
\centering\leavevmode
\epsfxsize=\hsize \epsfbox{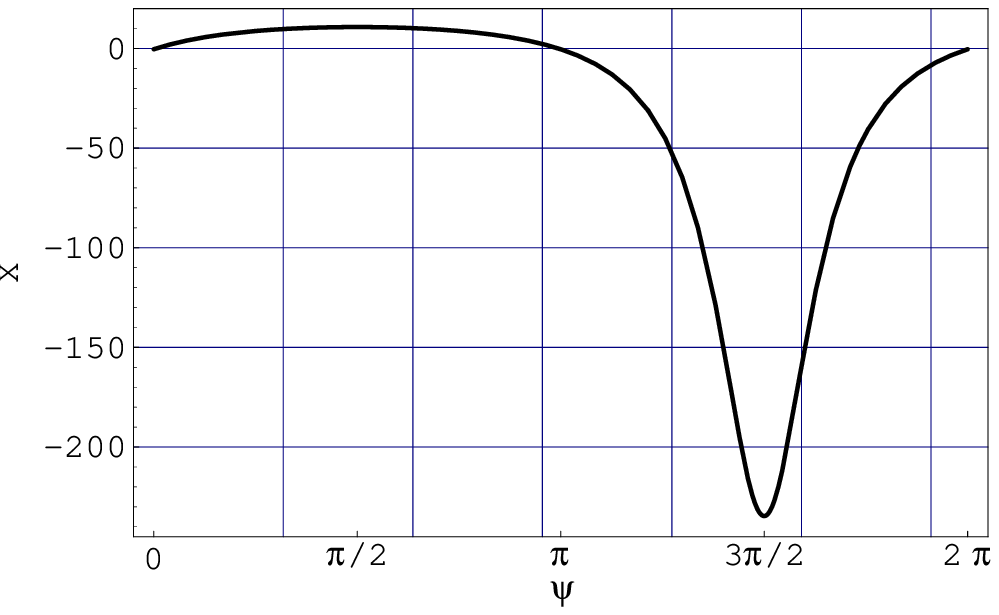}
{\figlabsize (b)}
\end{minipage}
\par\vskip 4mm\par
\begin{minipage}[b]{0.48\hsize}
\centering\leavevmode
\epsfxsize=\hsize \epsfbox{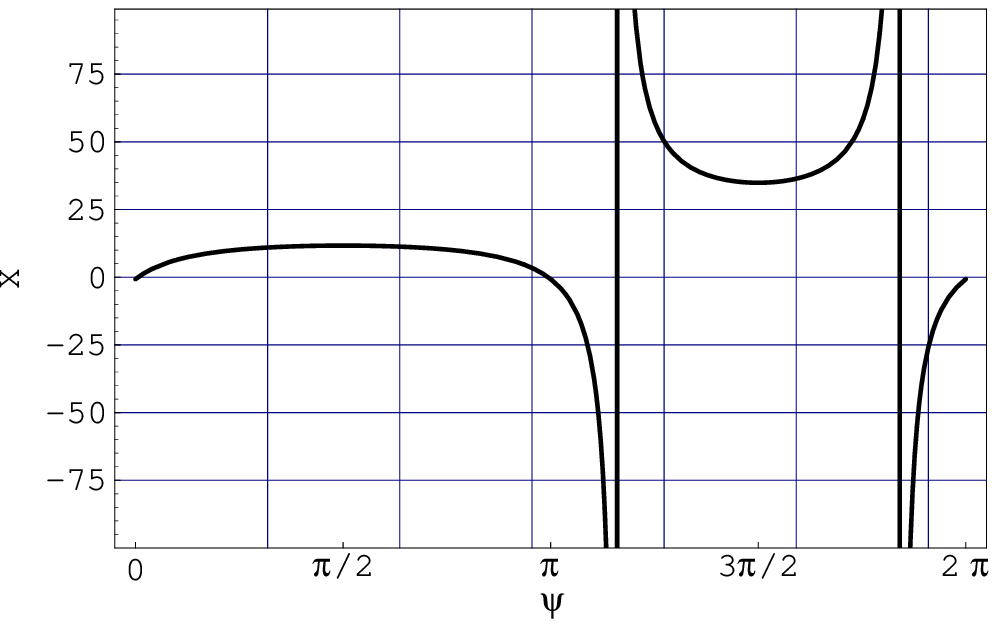}
{\figlabsize (c)}
\end{minipage} 
\caption{The dependence of photon's impact parameter $X(\psi;r,y,a)$ on the
  directional angle $\psi$ measured relative to the outward direction in
  the locally non-rotating frames. The dependence is illustrated in three
  typical cases which are possible in the \KdS\ black-hole spacetimes.
  (a)~$y = 0.036$, $a^2 = e^2 =
  0$, $r = 3$ -- sinus-like symmetric shape; (b)~$y = 0.036$, $a^2 = 0.16$,
  $e^2 = 0$, $r = 3$ -- the shape is distorted, but the function remains
  continuous; (c)~$y = 0.036$, $a^2 = 0.49$, $e^2 = 0$, $r = 3.5$ -- if the
  \RRB\ occurs, discontinuities appear in the function $X(\psi;r,y,a)$.
  \protect\rule[-3mm]{0pt}{3mm}}
\label{f14}
\end{figure}

The function $X(\psi;r,y,a)$ (or $\ell(\psi;r,y,a)$) enables us to
determine photon escape (or capture) cones in a straightforward way by
using the effective potential $X_{\pm}(r;y,a)$ of the radial motion. The
escape cones are given by the directional angles corresponding to the
marginally escaping photons having the impact parameters $X_{\ri c}$
corresponding to the unstable circular photon orbits (see Fig.\,\ref{f14}).

\begin{figure}[b]
\footnotesize
\rule{\textwidth}{.8pt}

\medskip 

\centering
\begin{minipage}[b]{.19\hsize}
\centering $r=1.44=r_{{\ri h}+,1}+\delta$\par\smallskip\par
\leavevmode\epsfxsize=\hsize \epsfbox{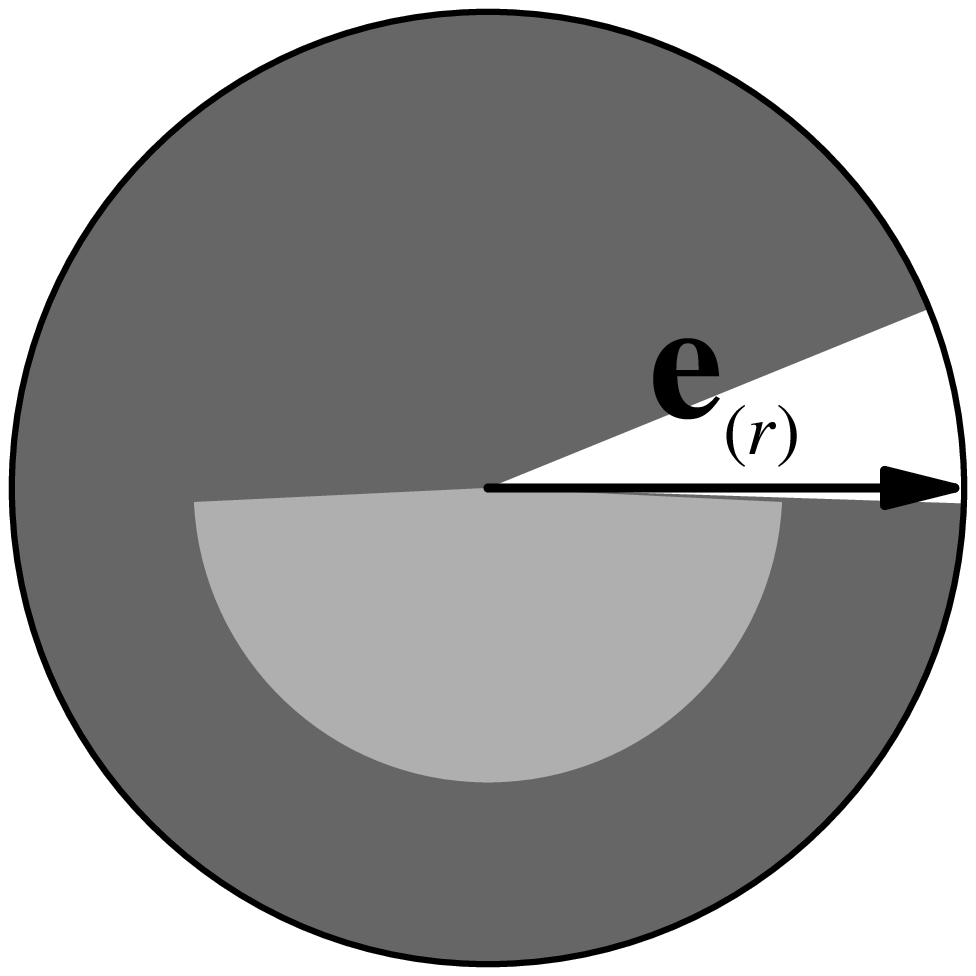}
\end{minipage}\hfill%
\begin{minipage}[b]{.19\hsize}
\centering $r=1.5$\par\smallskip\par
\leavevmode\epsfxsize=\hsize \epsfbox{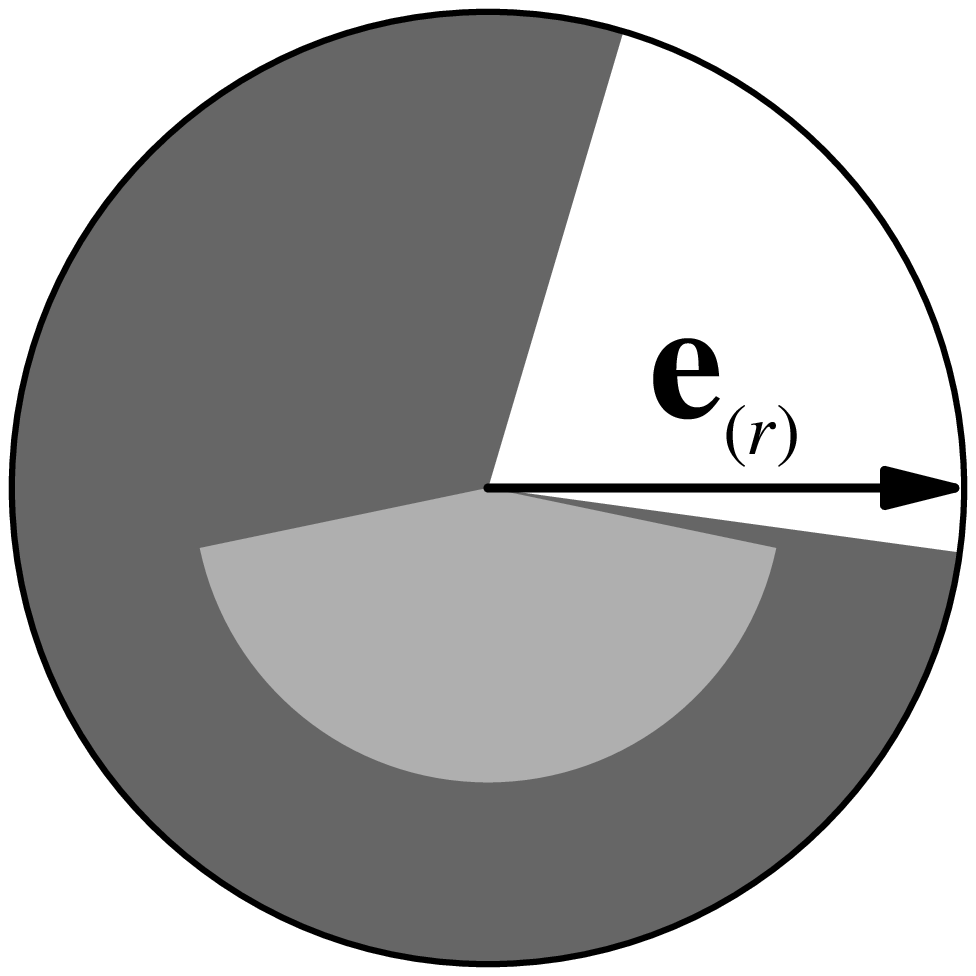}
\end{minipage}\hfill%
\begin{minipage}[b]{.19\hsize}
\centering $r=r_{{\ri min}-,1}$\par\smallskip\par
\leavevmode\epsfxsize=\hsize \epsfbox{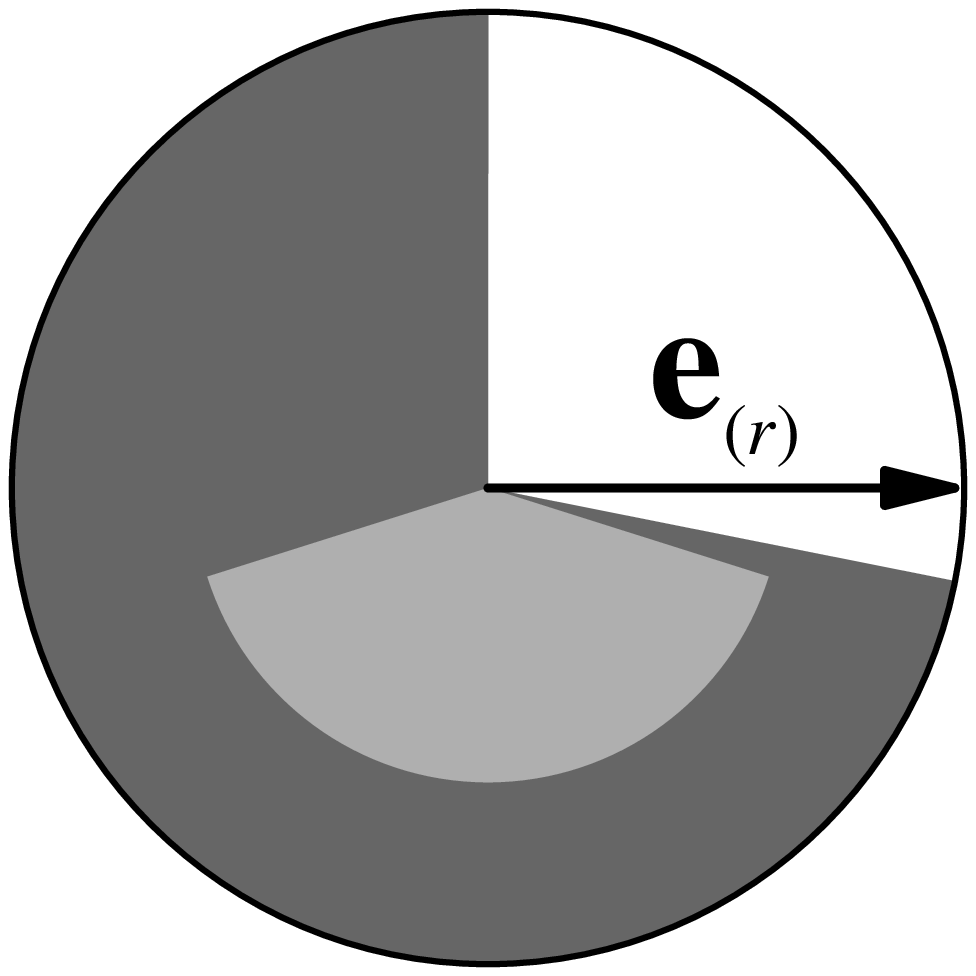}
\end{minipage}\hfill%
\begin{minipage}[b]{.19\hsize}
\centering $r=1.73=r_{{\ri h}+,2}+\delta$\par\smallskip\par
\leavevmode\epsfxsize=\hsize \epsfbox{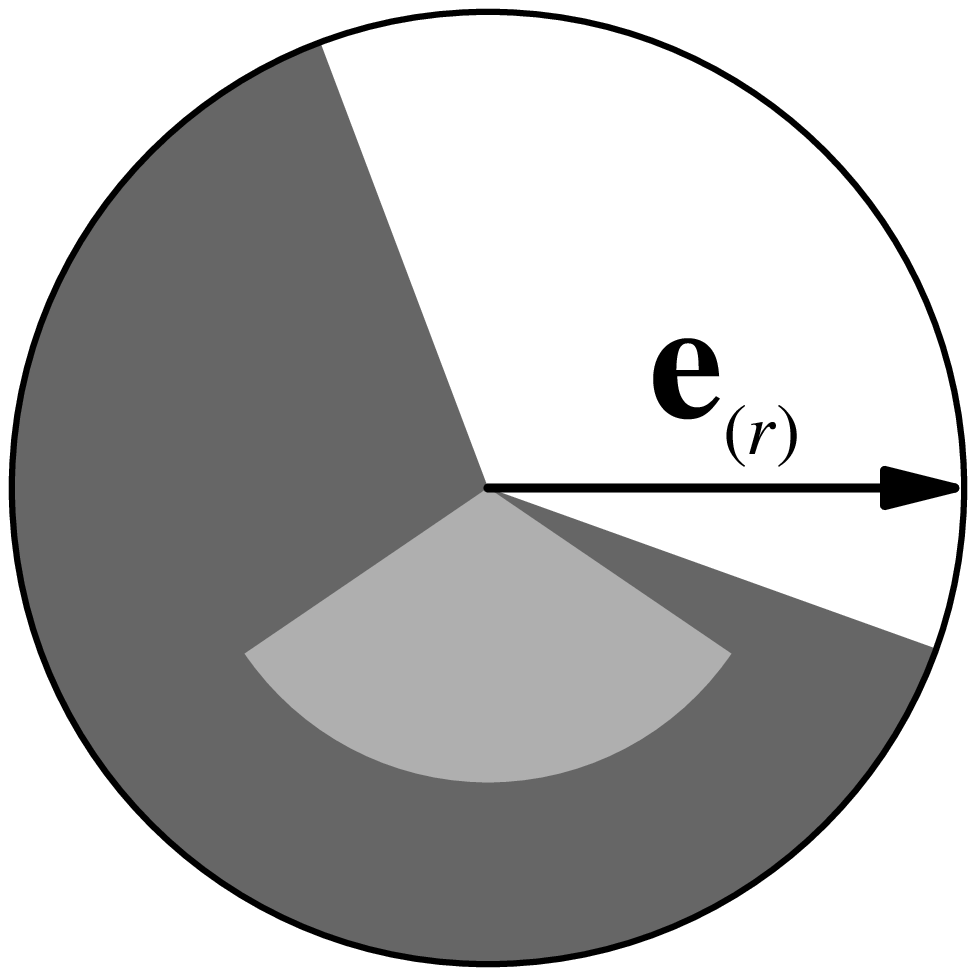}
\end{minipage}\hfill%
\begin{minipage}[b]{.19\hsize}
\centering $r=1.75$\par\smallskip\par
\leavevmode\epsfxsize=\hsize \epsfbox{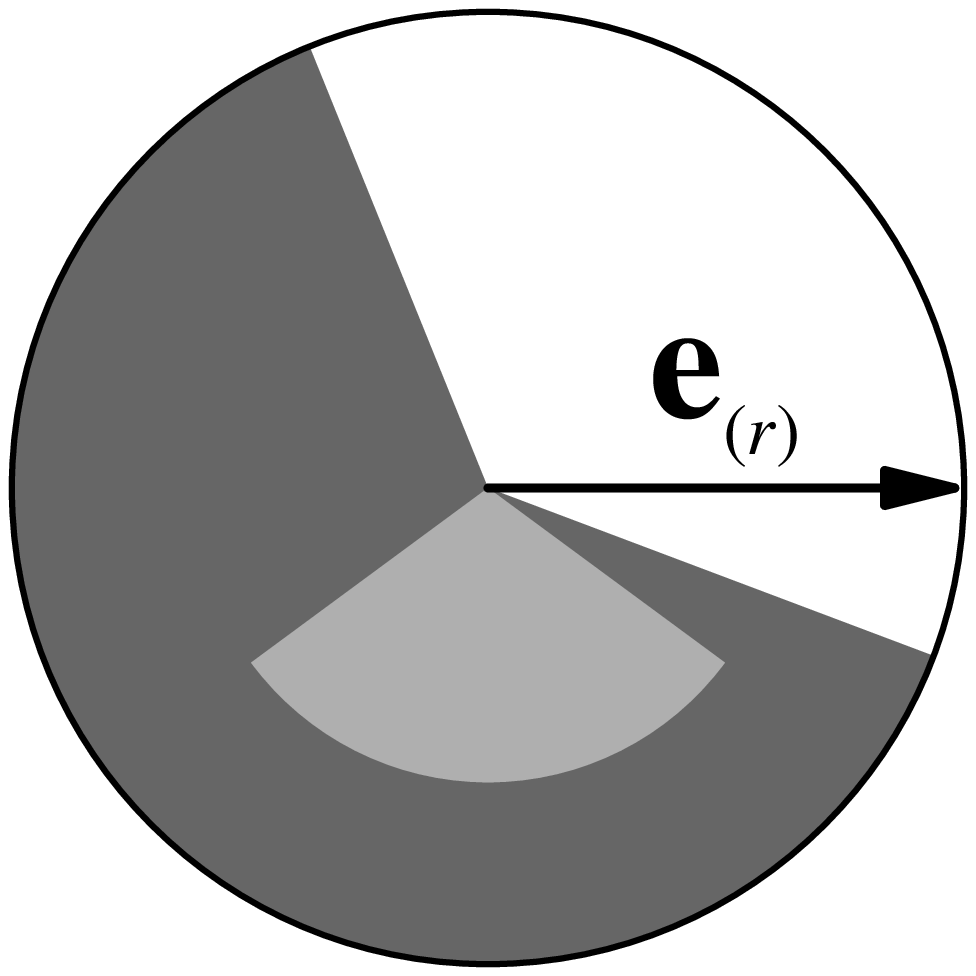}
\end{minipage}

\medskip

\begin{minipage}[b]{.19\hsize}
\mbox{}
\end{minipage}\hfill%
\begin{minipage}[b]{.19\hsize}
\mbox{}
\end{minipage}\hfill%
\begin{minipage}[b]{.19\hsize}
\mbox{}
\end{minipage}\hfill%
\begin{minipage}[b]{.19\hsize}
\centering\leavevmode\epsfxsize=\hsize \epsfbox{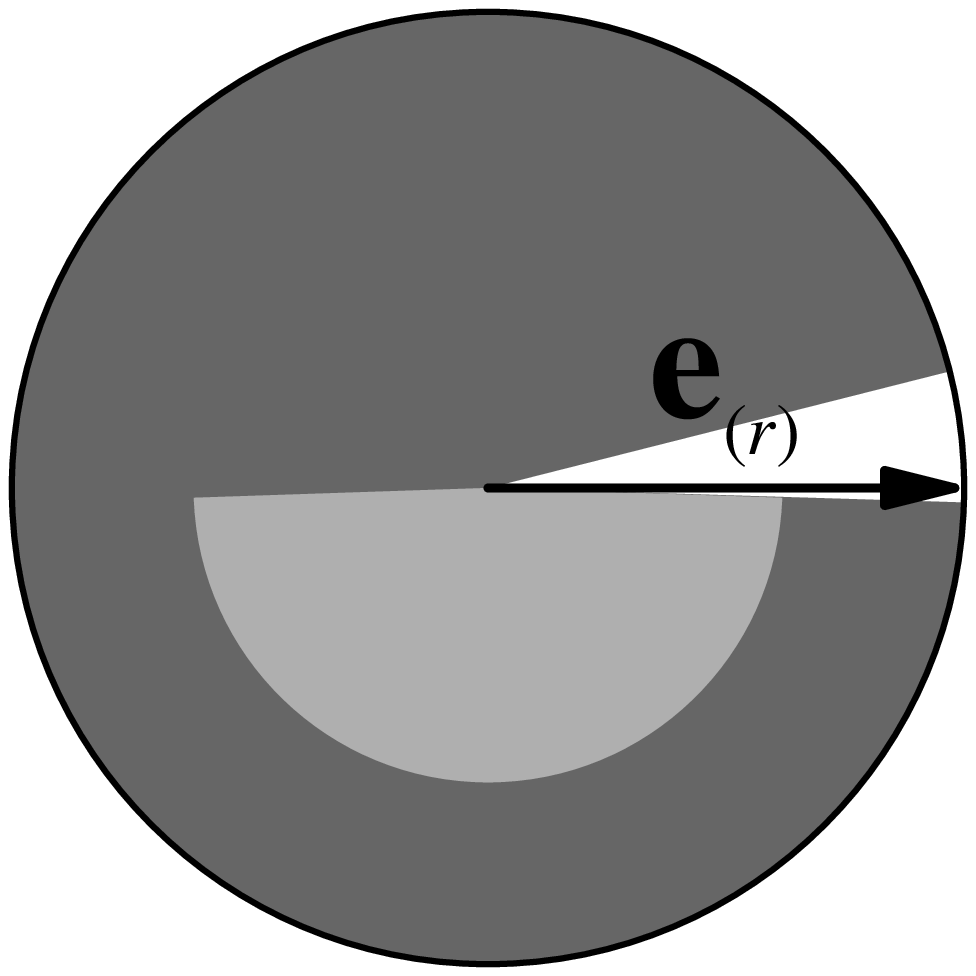}
\end{minipage}\hfill%
\begin{minipage}[b]{.19\hsize}
\centering\leavevmode\epsfxsize=\hsize \epsfbox{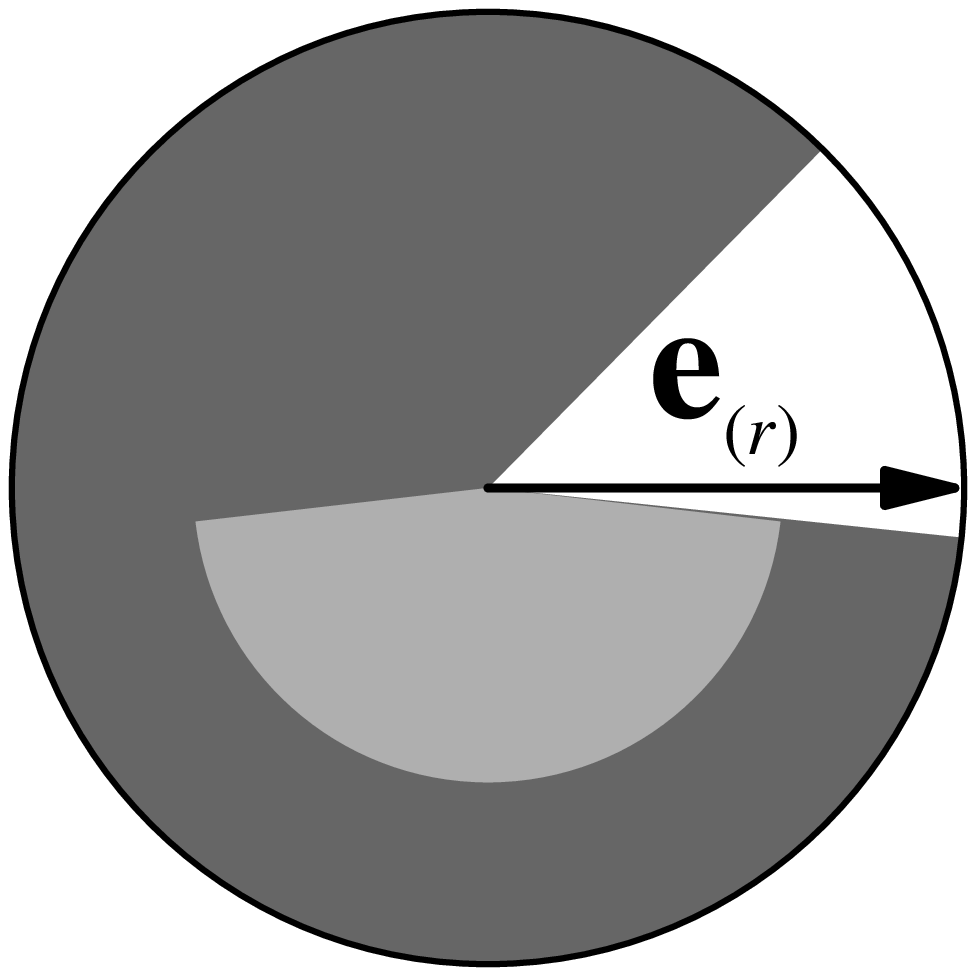}
\end{minipage}

\medskip

\rule{\textwidth}{.8pt}

\medskip 

\centering
\begin{minipage}[b]{.19\hsize}
\centering $r=r_{{\ri min}-,2}$\par\smallskip\par
\leavevmode\epsfxsize=\hsize \epsfbox{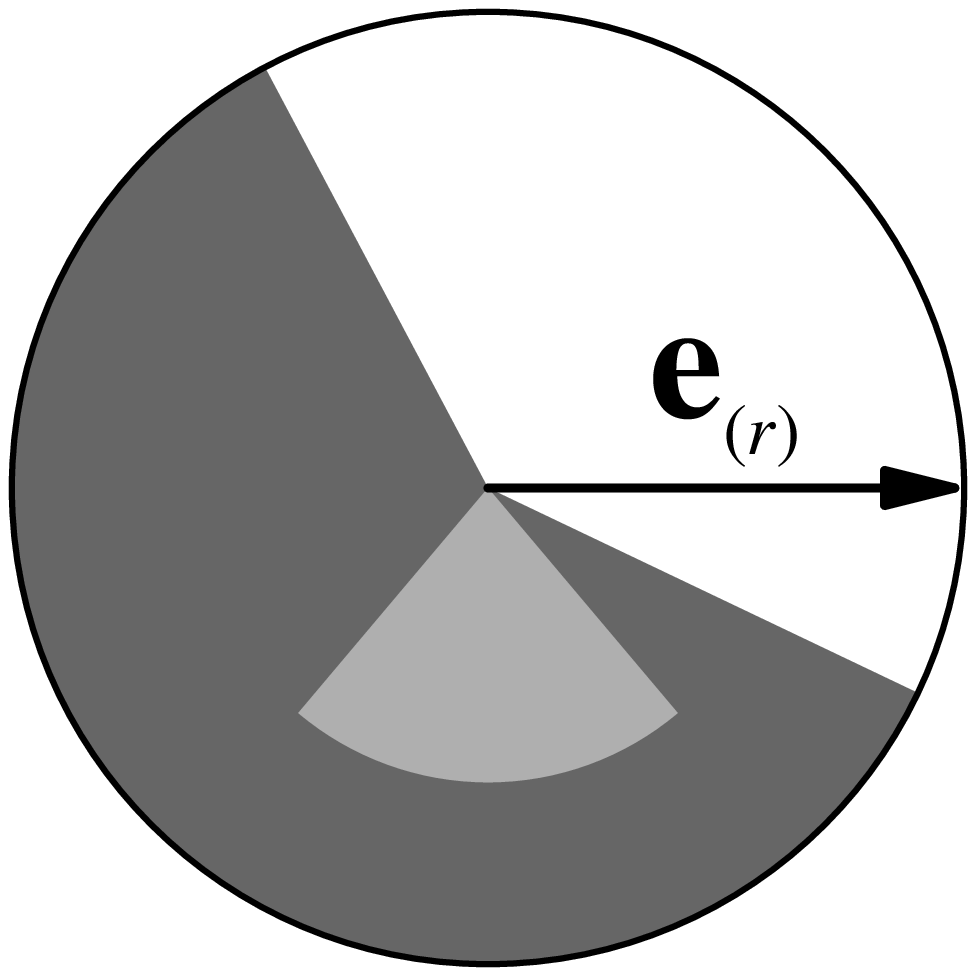}
\end{minipage}\hfill%
\begin{minipage}[b]{.19\hsize}
\centering $r=1.94=r_{{\ri h}+,3}+\delta$\par\smallskip\par
\leavevmode\epsfxsize=\hsize \epsfbox{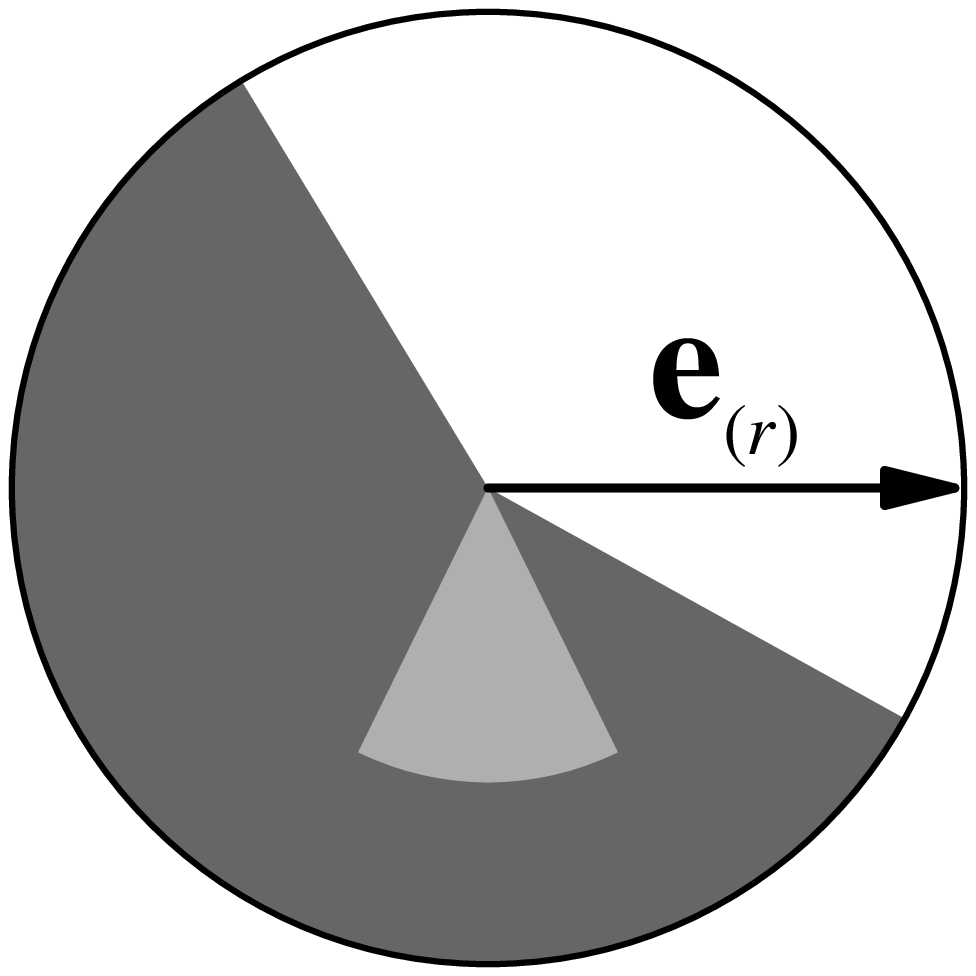}
\end{minipage}\hfill%
\begin{minipage}[b]{.19\hsize}
\centering $r=2$\par\smallskip\par
\leavevmode\epsfxsize=\hsize \epsfbox{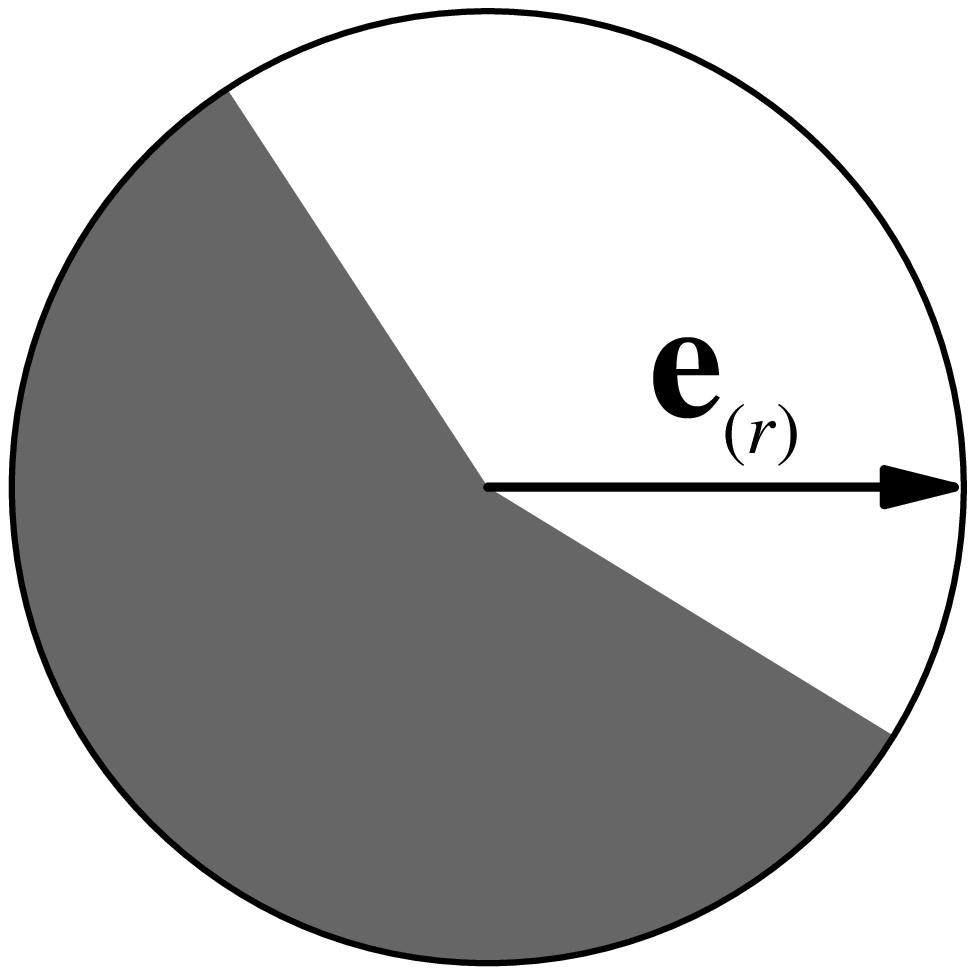}
\end{minipage}\hfill%
\begin{minipage}[b]{.19\hsize}
\centering $r=r_{{\ri min}-,3}$\par\smallskip\par
\leavevmode\epsfxsize=\hsize \epsfbox{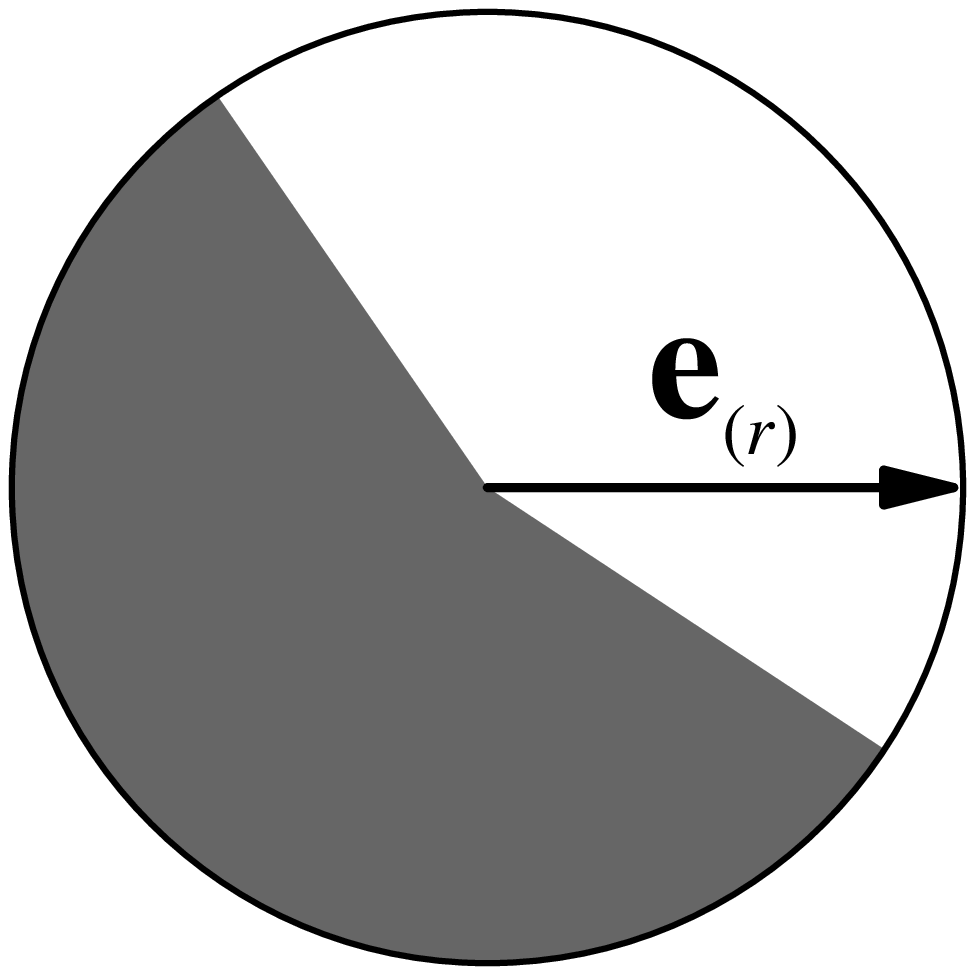}
\end{minipage}\hfill%
\begin{minipage}[b]{.19\hsize}
\centering $r=2.2$\par\smallskip\par
\leavevmode\epsfxsize=\hsize \epsfbox{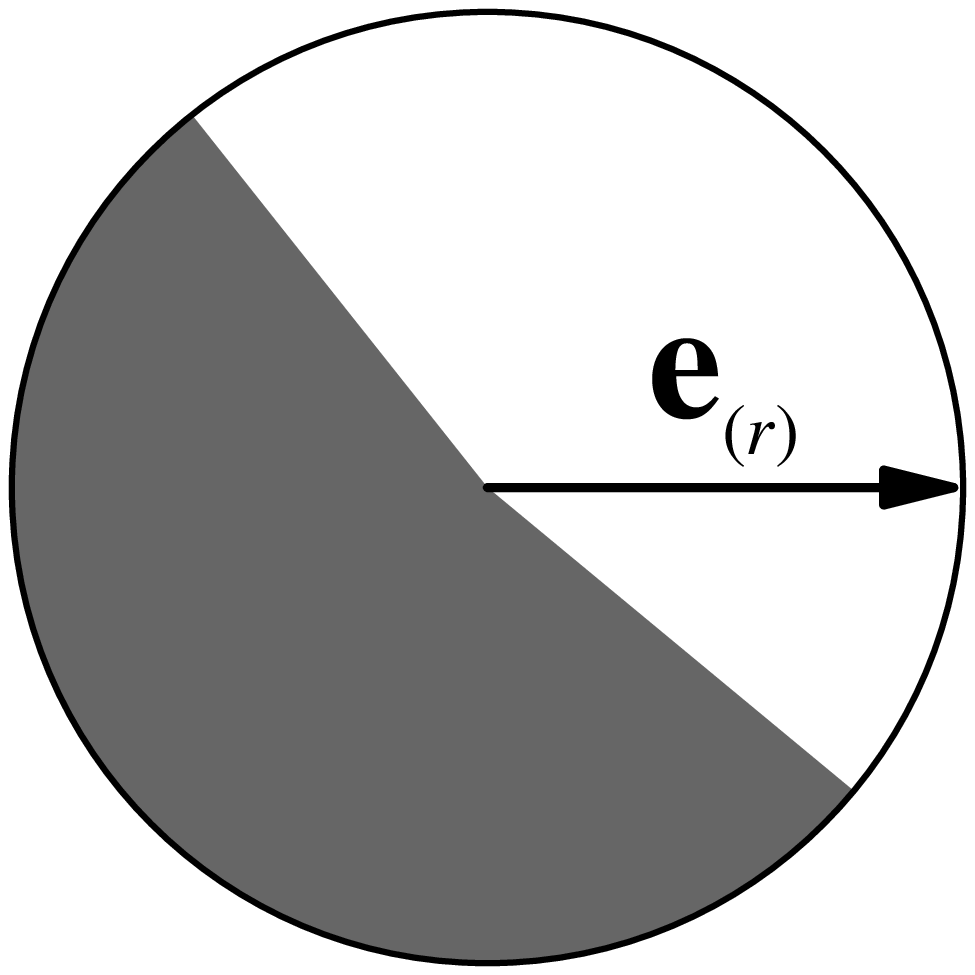}
\end{minipage}

\medskip

\begin{minipage}[b]{.19\hsize}
\centering\leavevmode\epsfxsize=\hsize \epsfbox{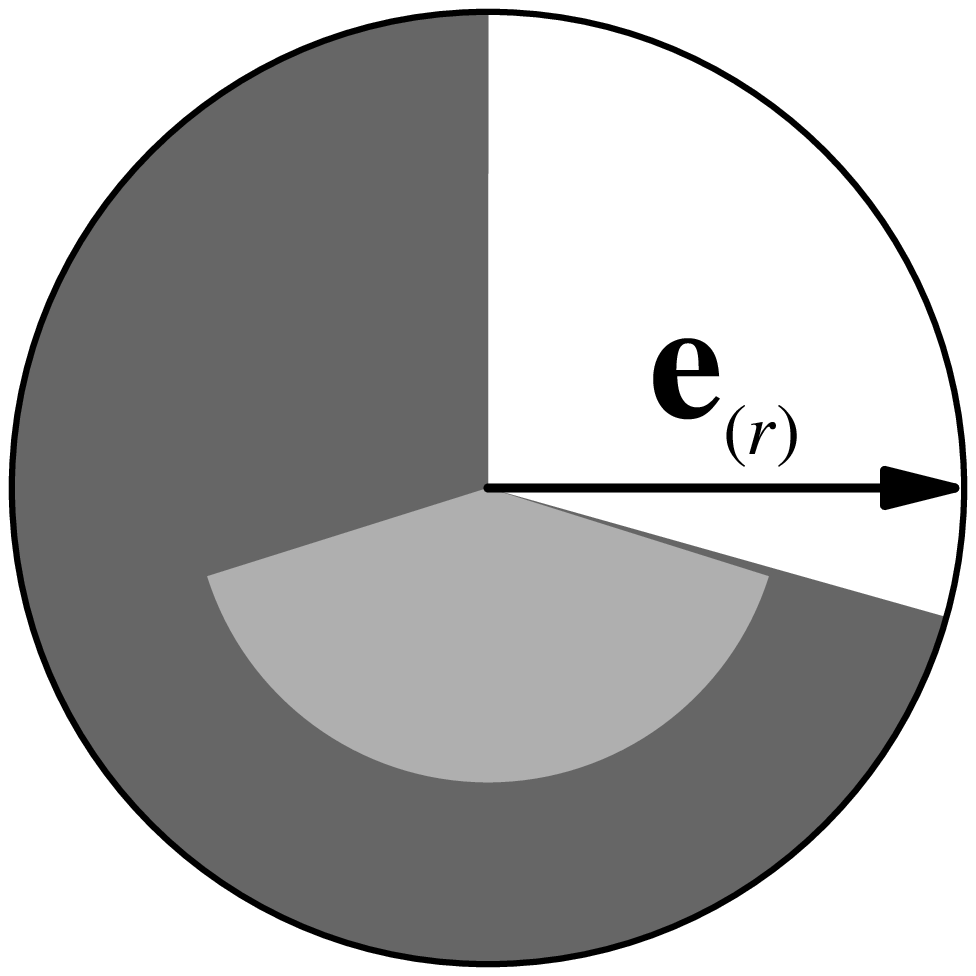}
\end{minipage}\hfill%
\begin{minipage}[b]{.19\hsize}
\centering\leavevmode\epsfxsize=\hsize \epsfbox{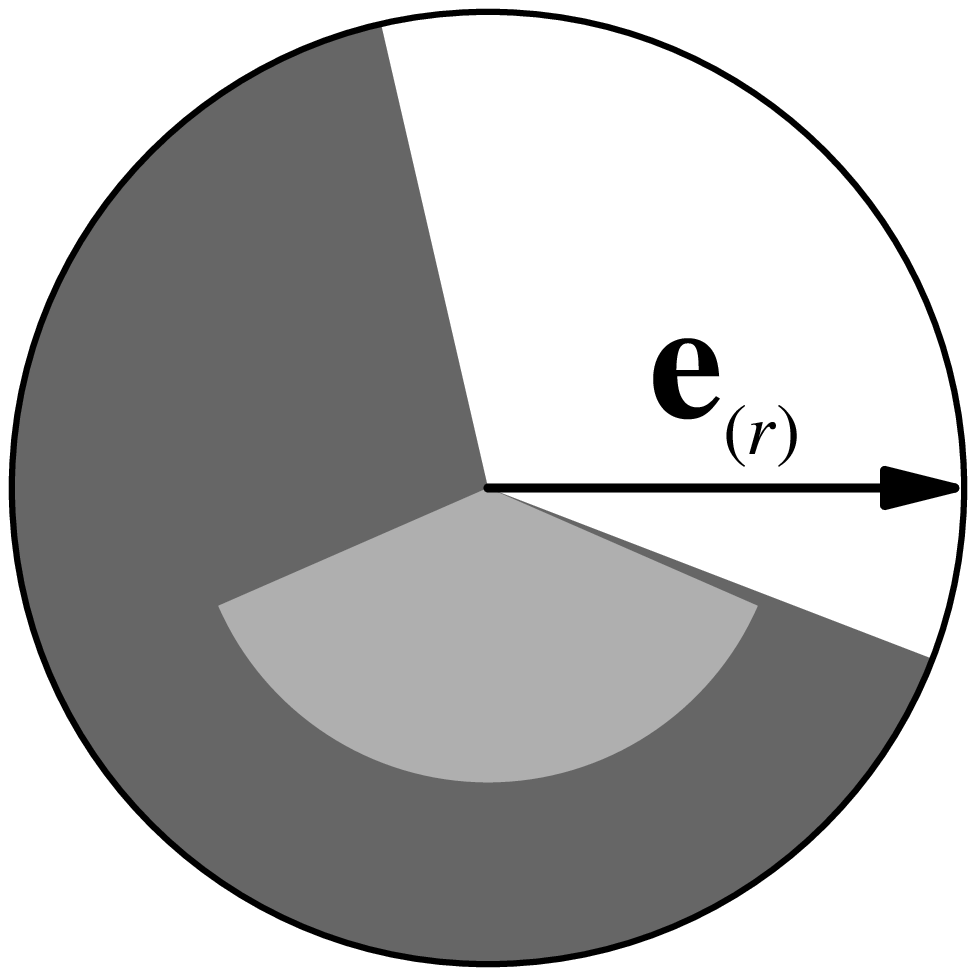}
\end{minipage}\hfill%
\begin{minipage}[b]{.19\hsize}
\centering\leavevmode\epsfxsize=\hsize \epsfbox{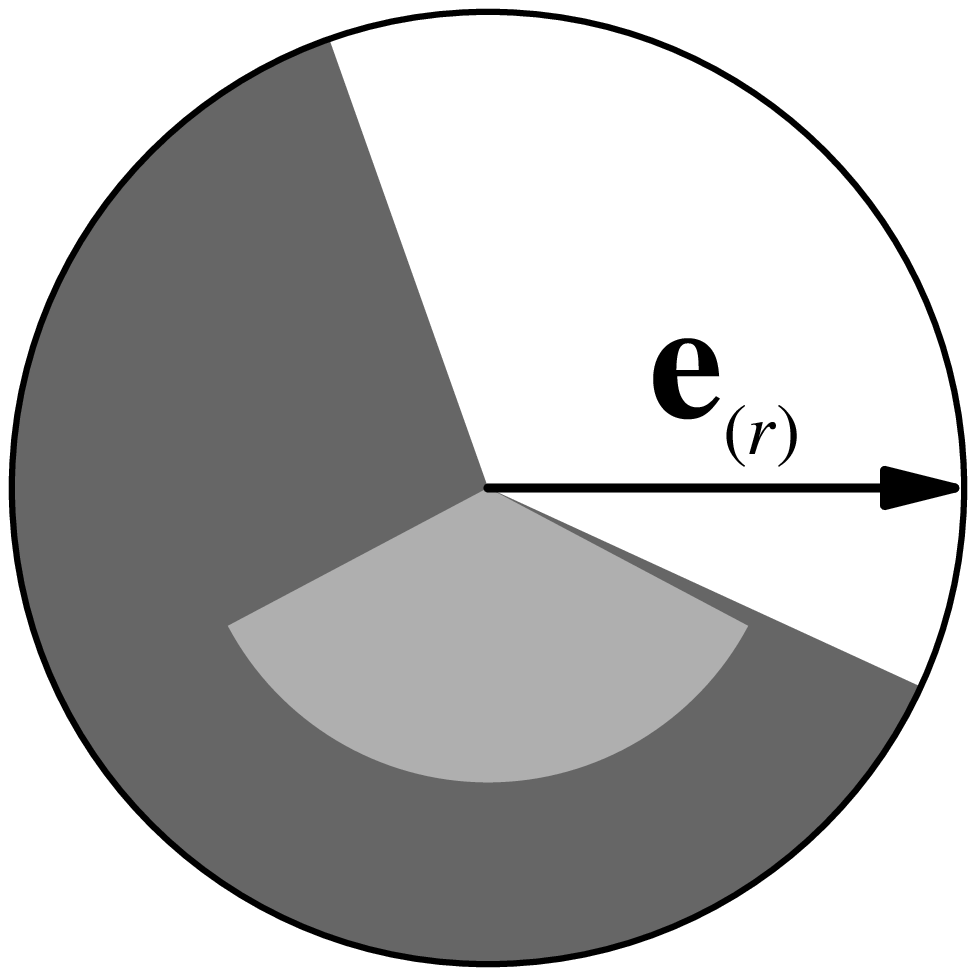}
\end{minipage}\hfill%
\begin{minipage}[b]{.19\hsize}
\centering\leavevmode\epsfxsize=\hsize \epsfbox{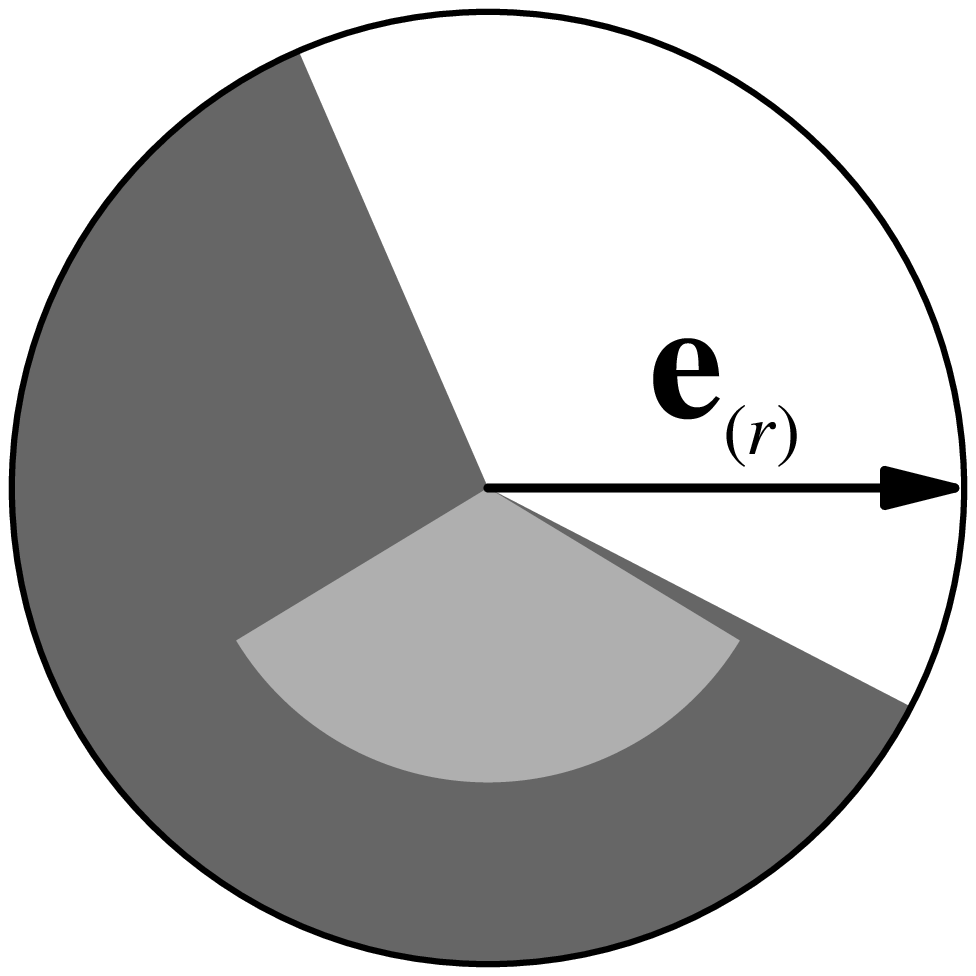}
\end{minipage}\hfill%
\begin{minipage}[b]{.19\hsize}
\centering\leavevmode\epsfxsize=\hsize \epsfbox{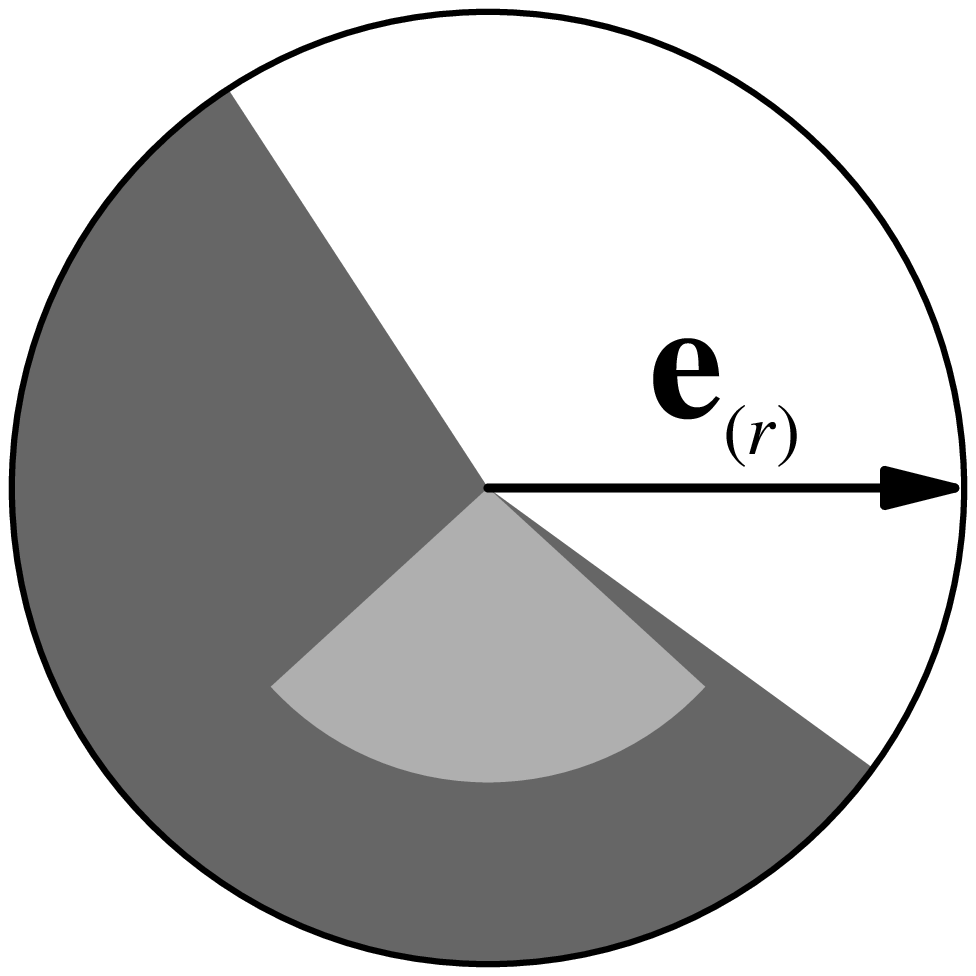}
\end{minipage}

\medskip

\begin{minipage}[b]{.19\hsize}
\mbox{}
\end{minipage}\hfill%
\begin{minipage}[b]{.19\hsize}
\centering\leavevmode\epsfxsize=\hsize \epsfbox{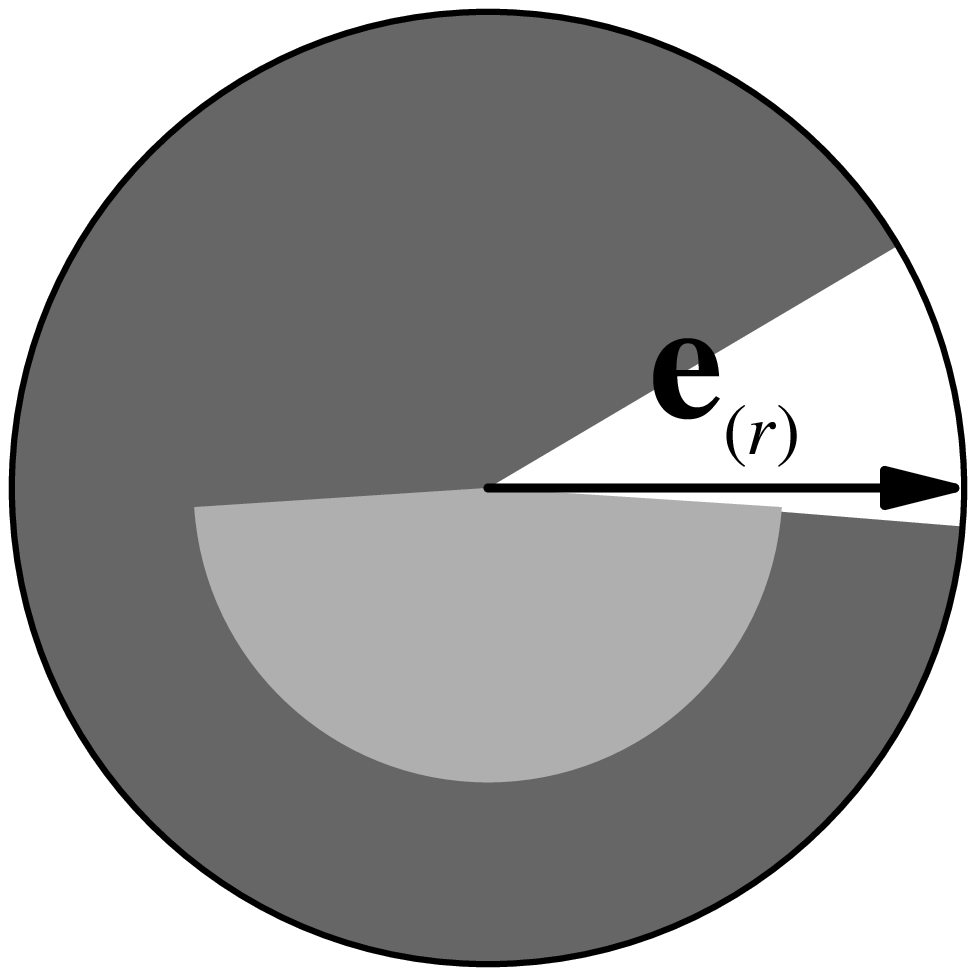}
\end{minipage}\hfill%
\begin{minipage}[b]{.19\hsize}
\centering\leavevmode\epsfxsize=\hsize \epsfbox{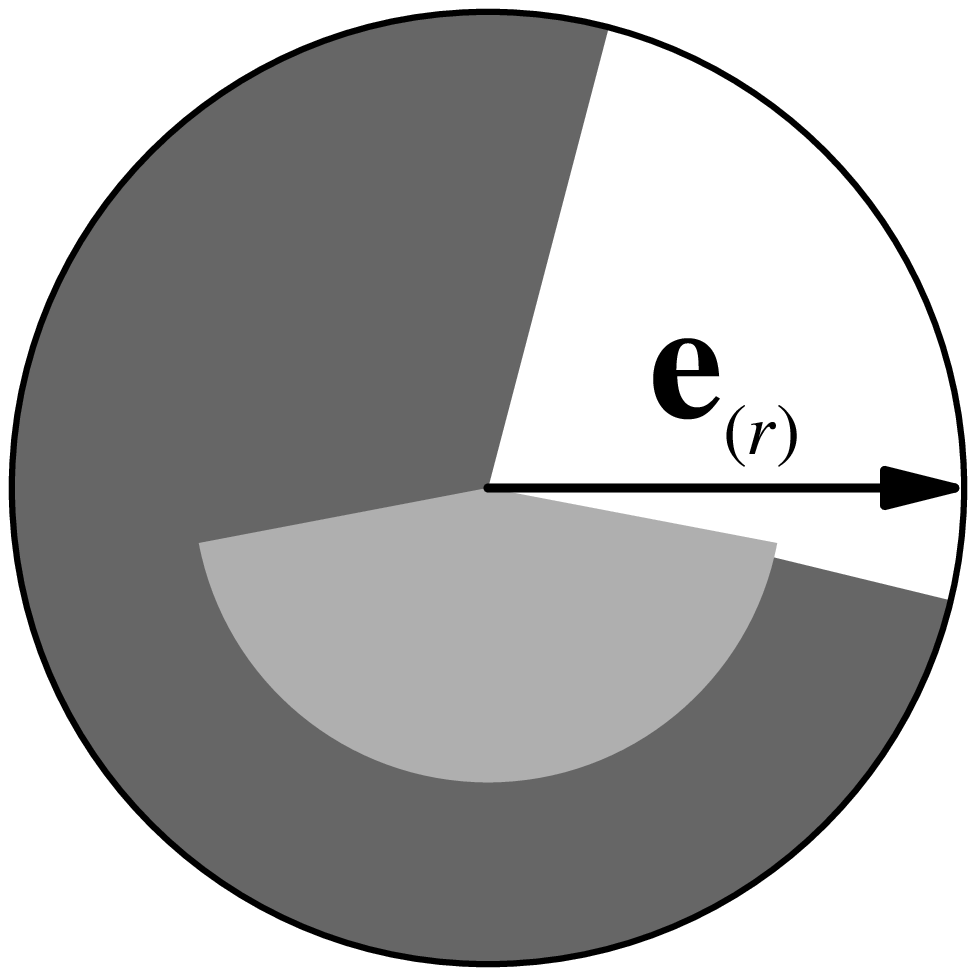}
\end{minipage}\hfill%
\begin{minipage}[b]{.19\hsize}
\centering\leavevmode\epsfxsize=\hsize \epsfbox{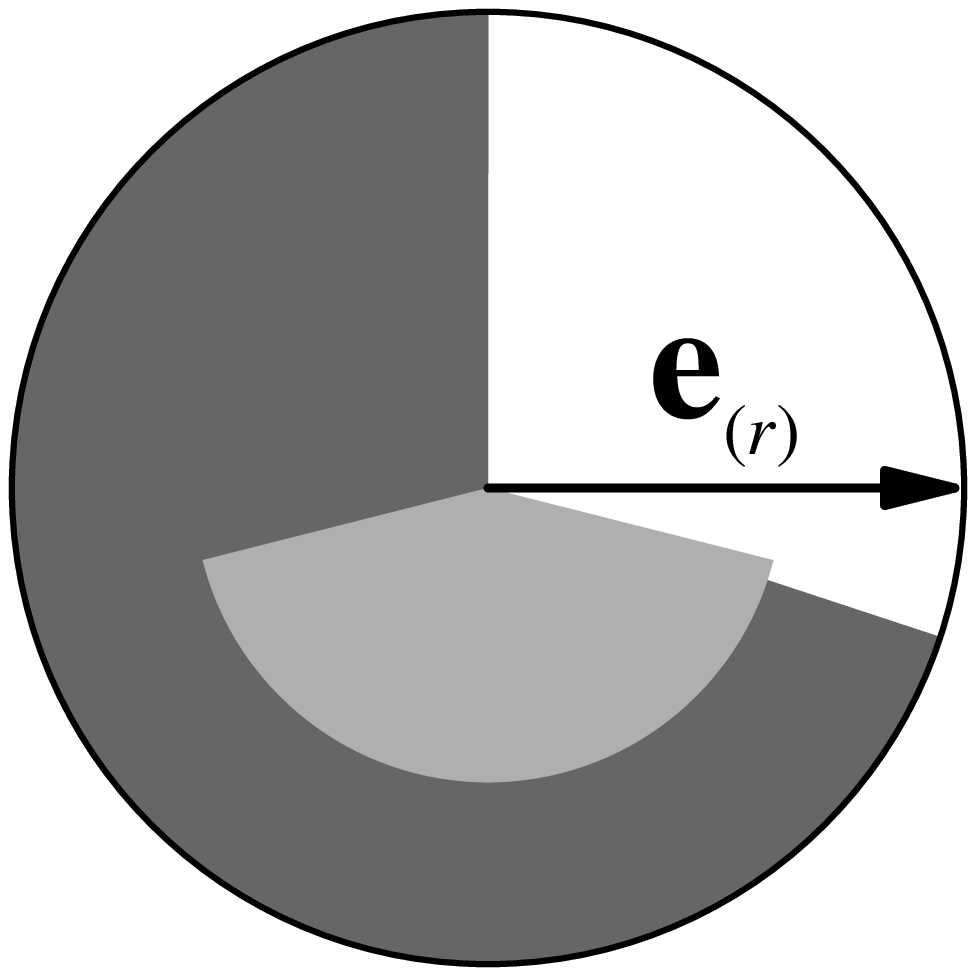}
\end{minipage}\hfill%
\begin{minipage}[b]{.19\hsize}
\centering\leavevmode\epsfxsize=\hsize \epsfbox{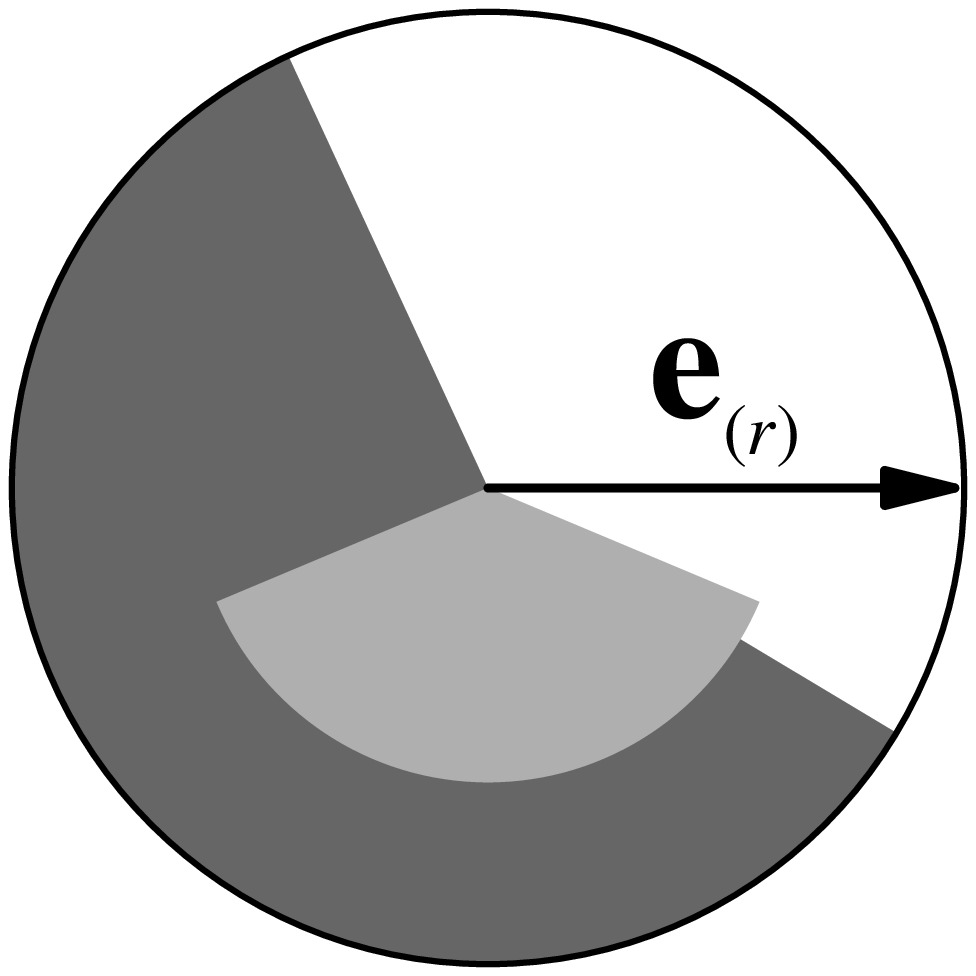}
\end{minipage}

\medskip

\rule{\textwidth}{.8pt}

\smallskip

{(\it Figure continued)}
\end{figure}

\begin{figure}[p]
\footnotesize
\rule{\textwidth}{.8pt}

\medskip 

\centering
\begin{minipage}[b]{.19\hsize}
\centering $r=2.4$\par\smallskip\par
\leavevmode\epsfxsize=\hsize \epsfbox{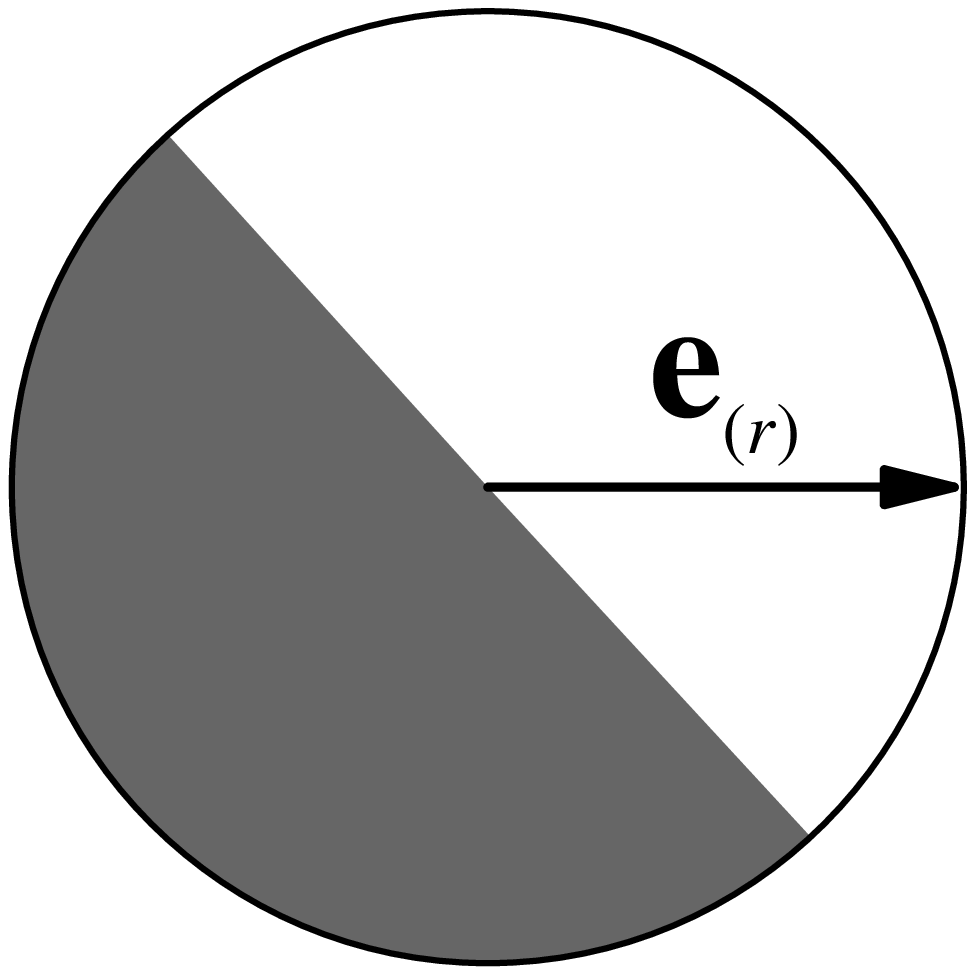}
\end{minipage}\hfill%
\begin{minipage}[b]{.19\hsize}
\centering $r=2.5$\par\smallskip\par
\leavevmode\epsfxsize=\hsize \epsfbox{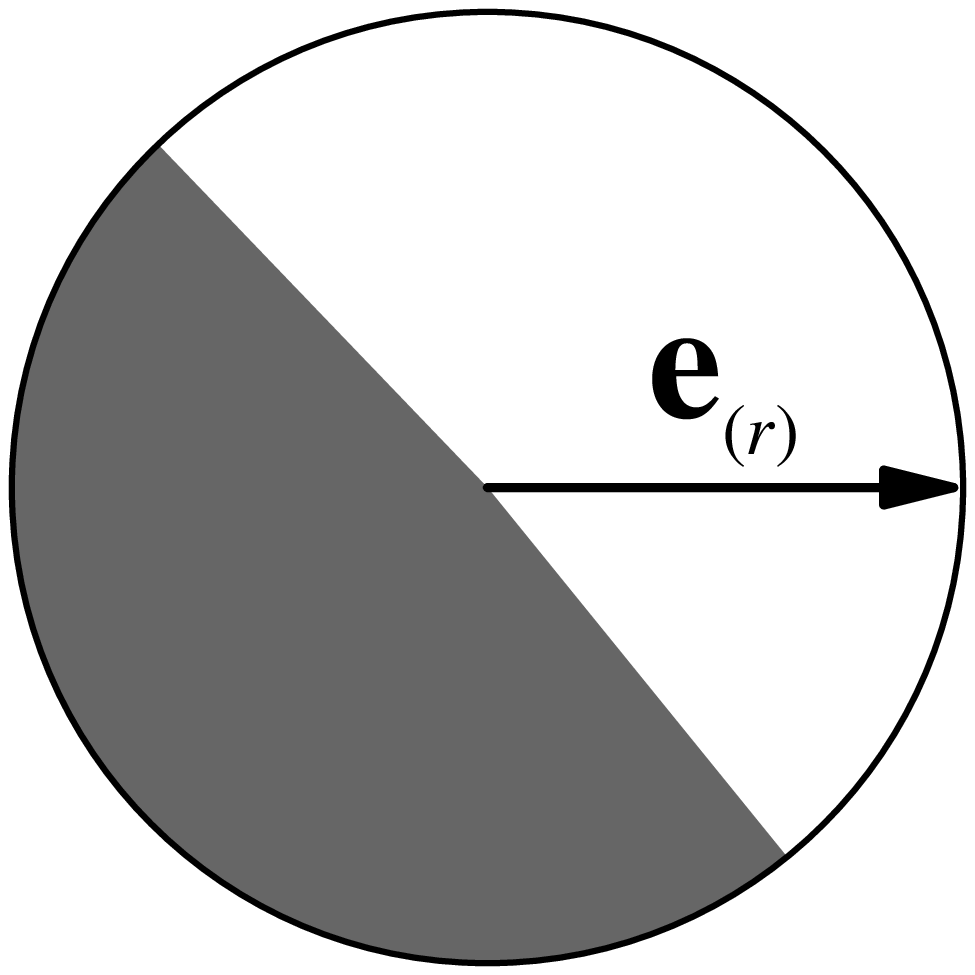}
\end{minipage}\hfill%
\begin{minipage}[b]{.19\hsize}
\centering $r=2.7$\par\smallskip\par
\leavevmode\epsfxsize=\hsize \epsfbox{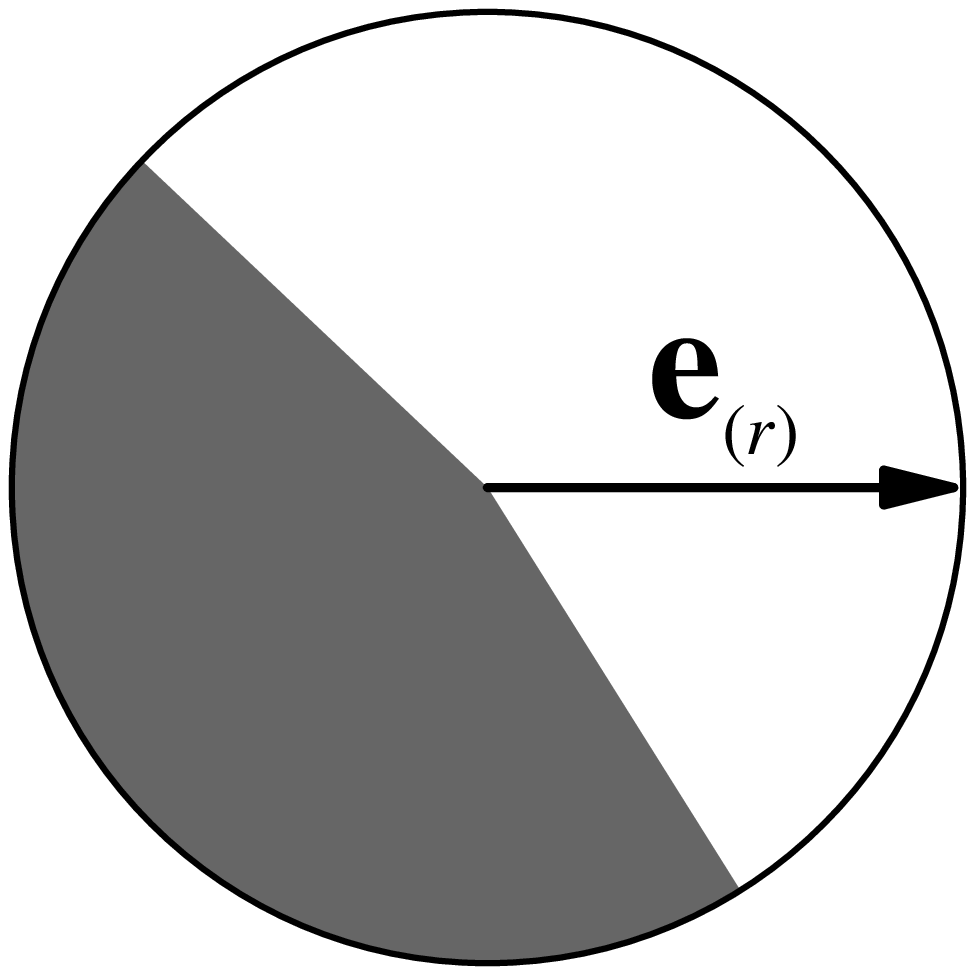}
\end{minipage}\hfill%
\begin{minipage}[b]{.19\hsize}
\centering $r=r_{{\ri max}+,3}$\par\smallskip\par
\leavevmode\epsfxsize=\hsize \epsfbox{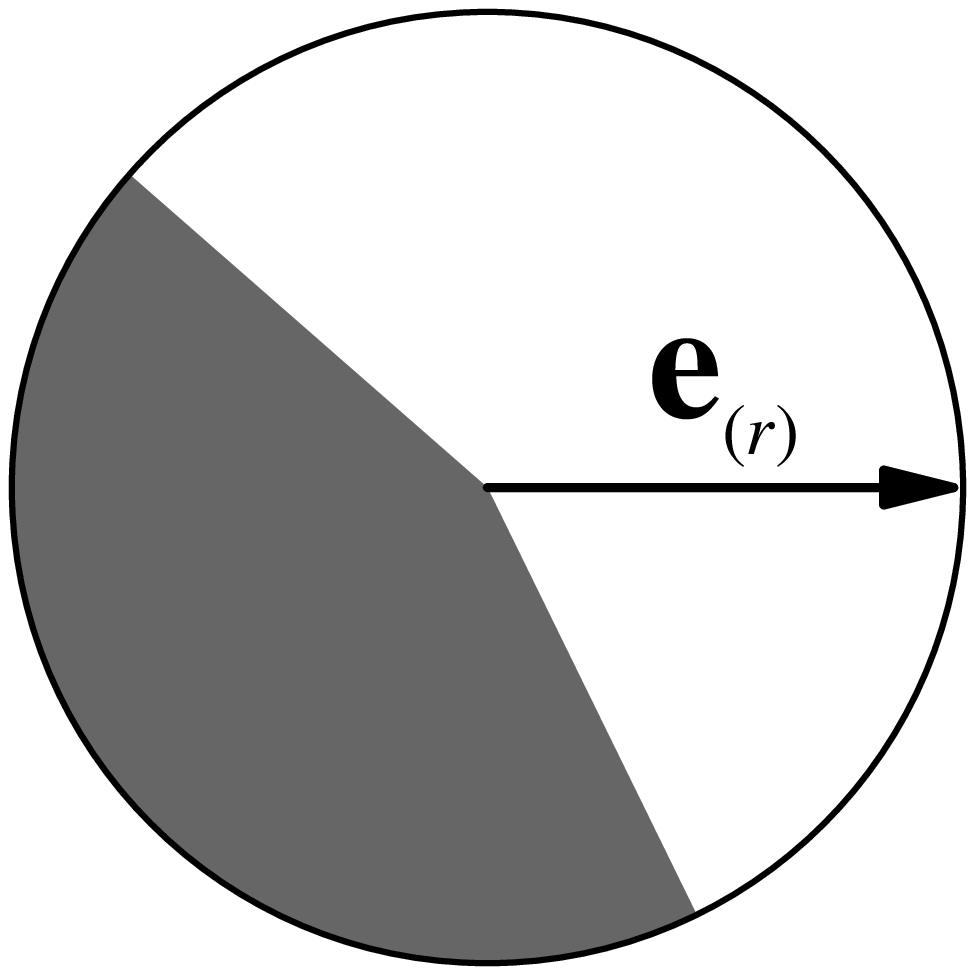}
\end{minipage}\hfill%
\begin{minipage}[b]{.19\hsize}
\centering $r=3$\par\smallskip\par
\leavevmode\epsfxsize=\hsize \epsfbox{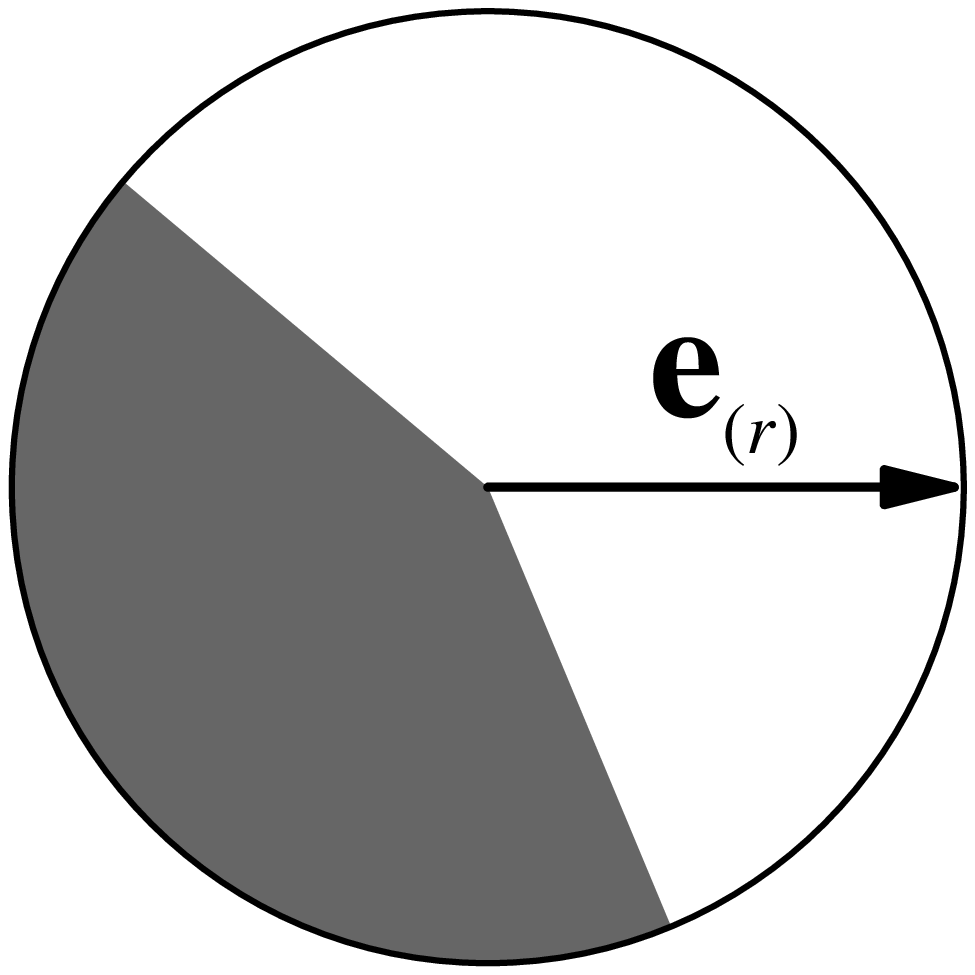}
\end{minipage}

\medskip

\begin{minipage}[b]{.19\hsize}
\centering\leavevmode\epsfxsize=\hsize \epsfbox{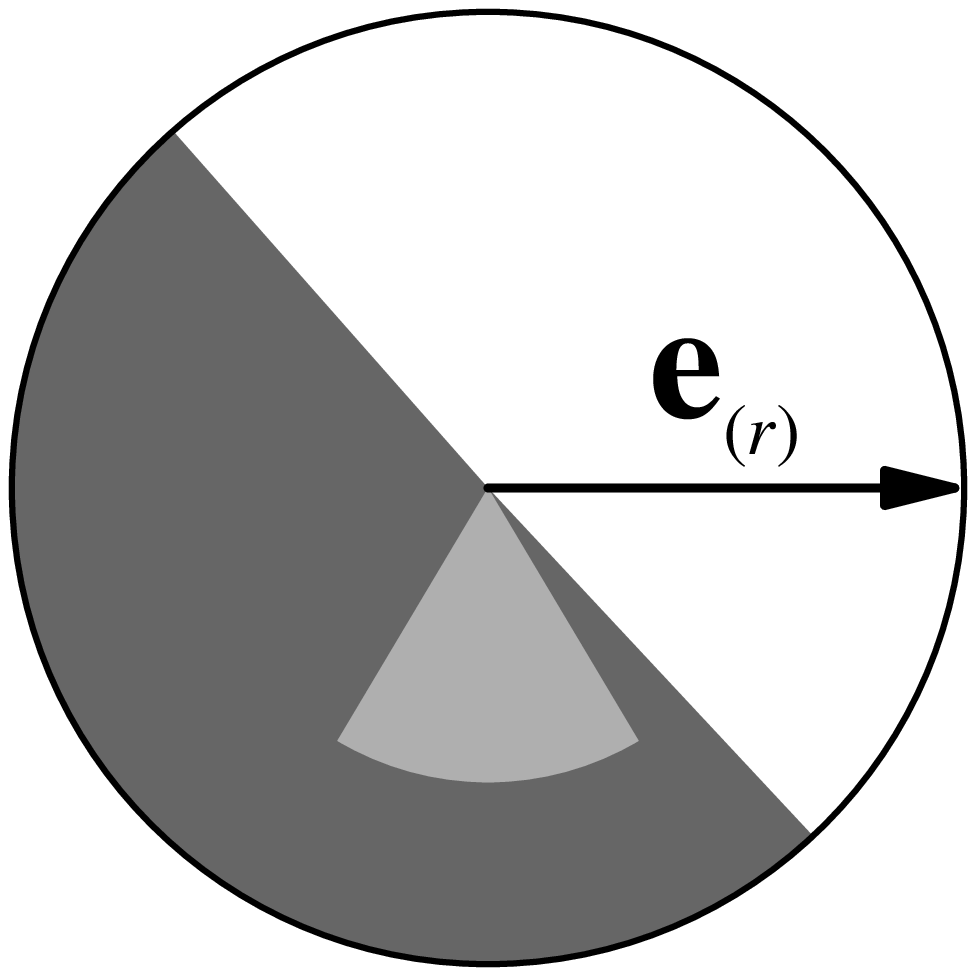}
\end{minipage}\hfill%
\begin{minipage}[b]{.19\hsize}
\centering\leavevmode\epsfxsize=\hsize \epsfbox{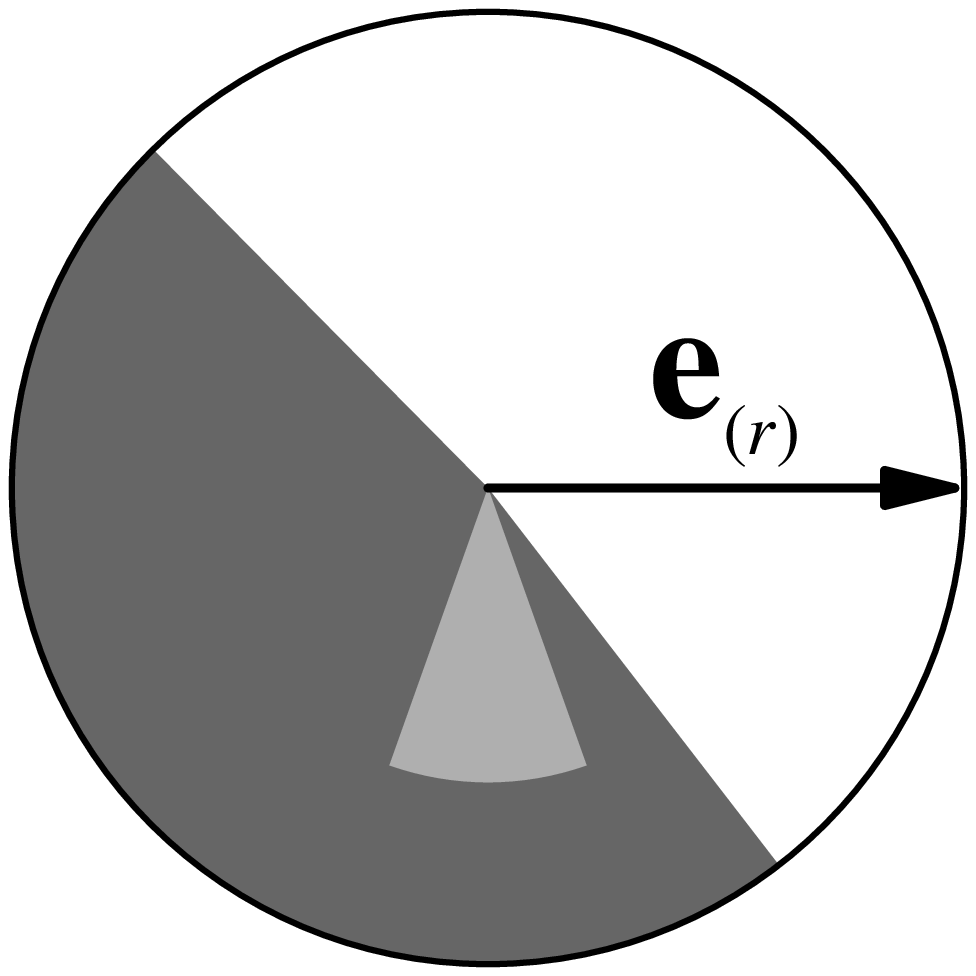}
\end{minipage}\hfill%
\begin{minipage}[b]{.19\hsize}
\centering\leavevmode\epsfxsize=\hsize \epsfbox{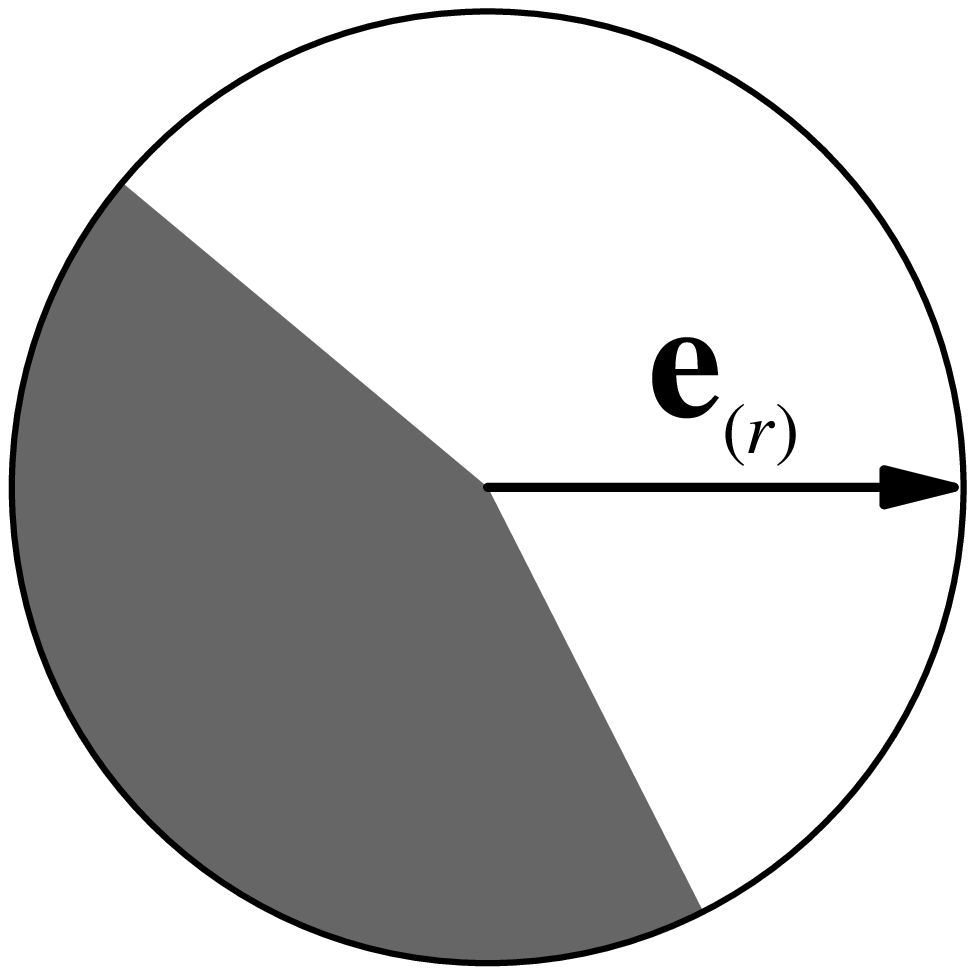}
\end{minipage}\hfill%
\begin{minipage}[b]{.19\hsize}
\centering\leavevmode\epsfxsize=\hsize \epsfbox{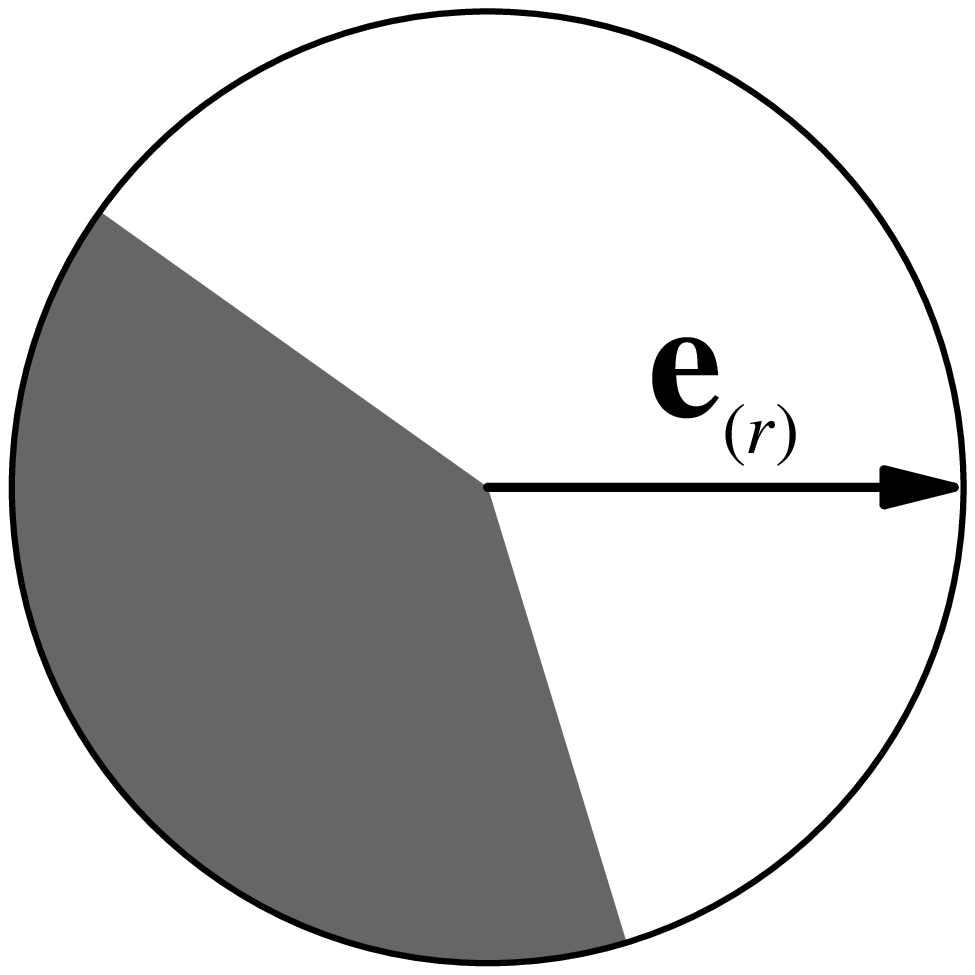}
\end{minipage}\hfill%
\begin{minipage}[b]{.19\hsize}
\centering\leavevmode\epsfxsize=\hsize \epsfbox{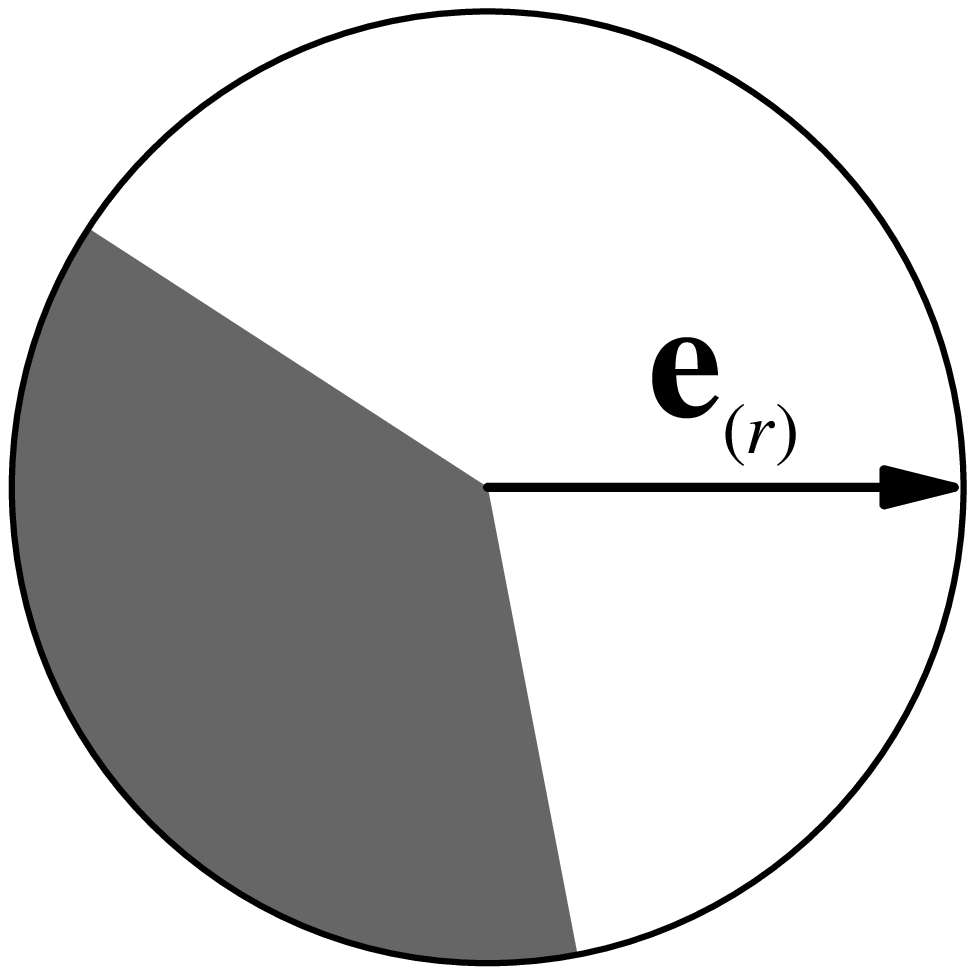}
\end{minipage}

\medskip

\begin{minipage}[b]{.19\hsize}
\centering\leavevmode\epsfxsize=\hsize \epsfbox{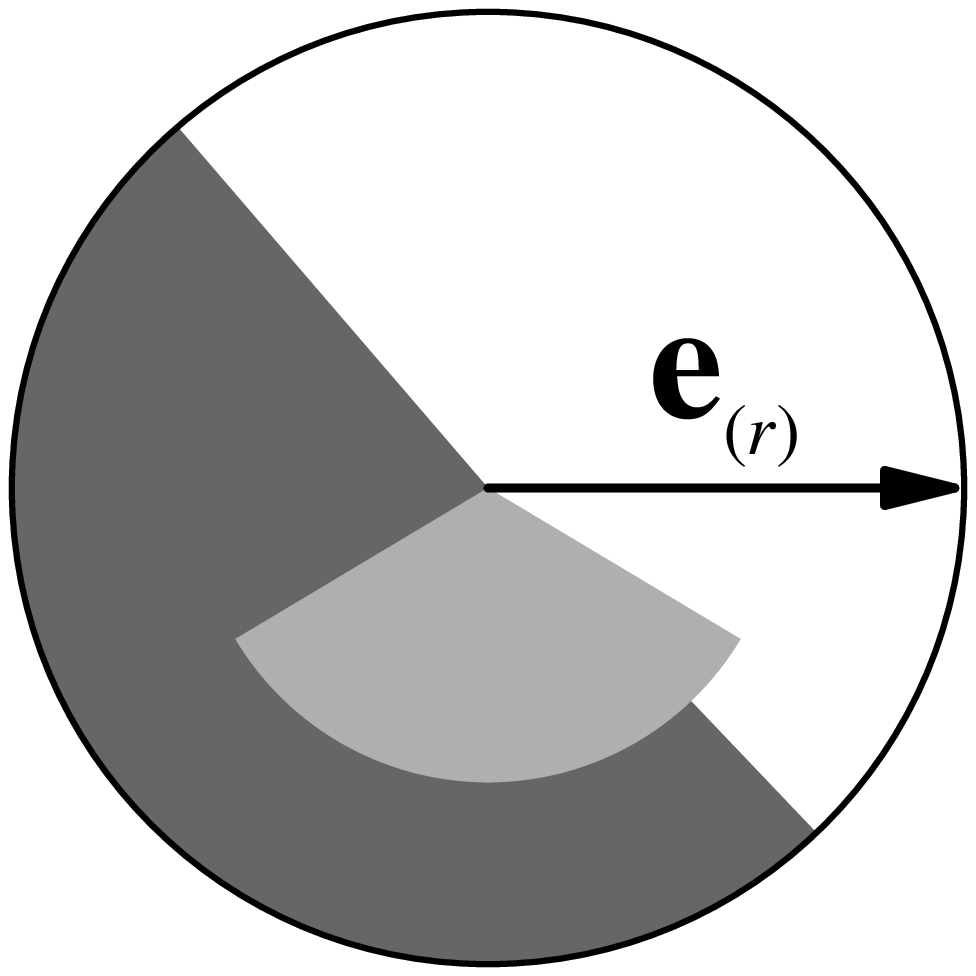}
\end{minipage}\hfill%
\begin{minipage}[b]{.19\hsize}
\centering\leavevmode\epsfxsize=\hsize \epsfbox{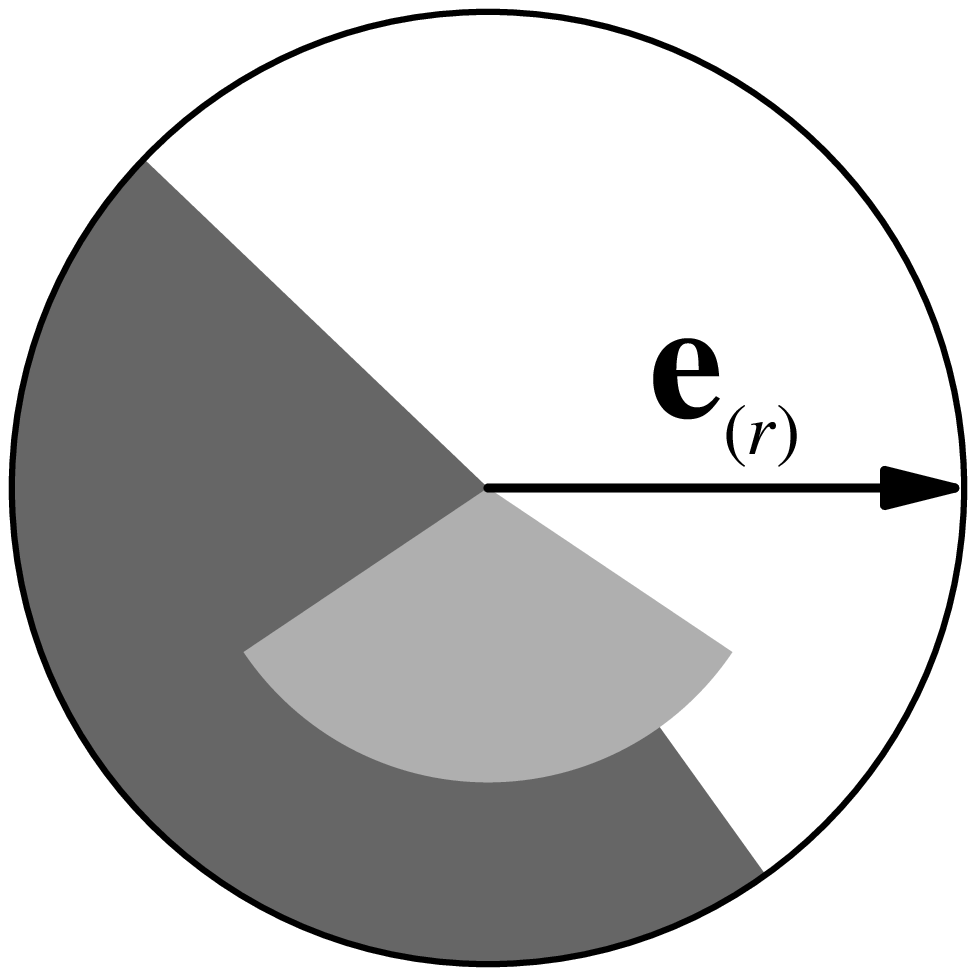}
\end{minipage}\hfill%
\begin{minipage}[b]{.19\hsize}
\centering\leavevmode\epsfxsize=\hsize \epsfbox{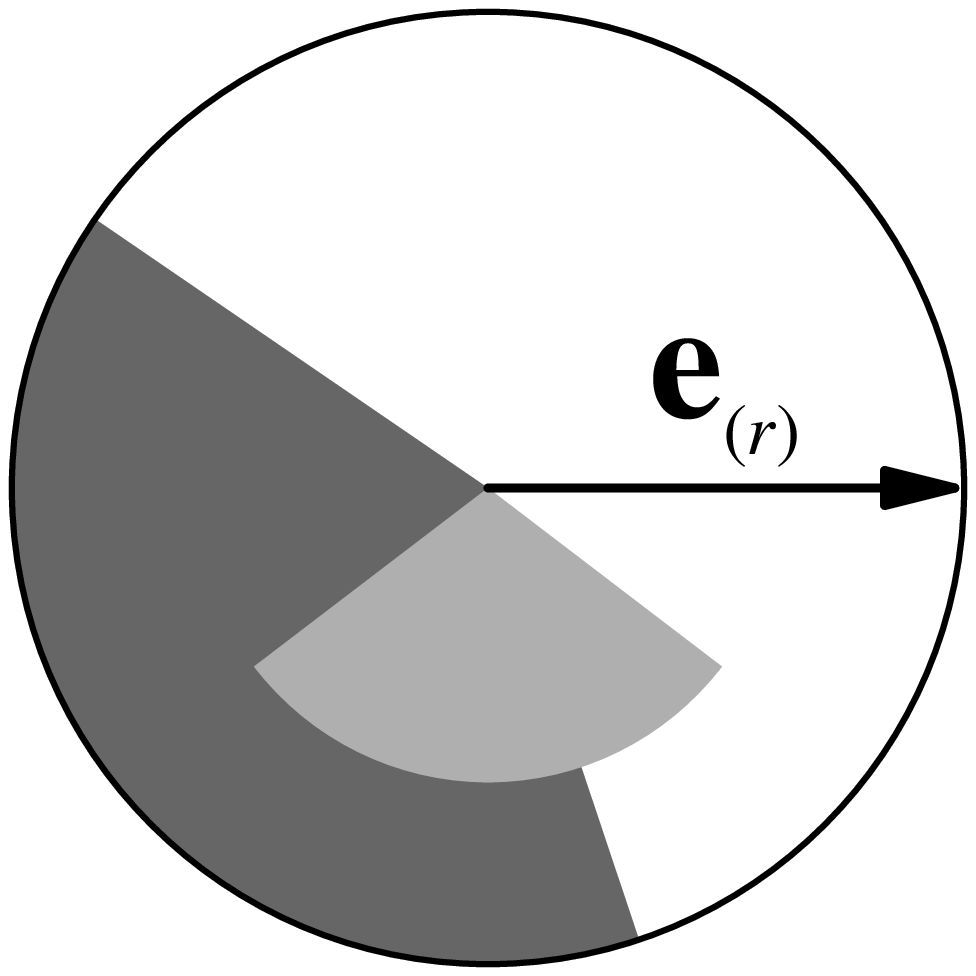}
\end{minipage}\hfill%
\begin{minipage}[b]{.19\hsize}
\centering\leavevmode\epsfxsize=\hsize \epsfbox{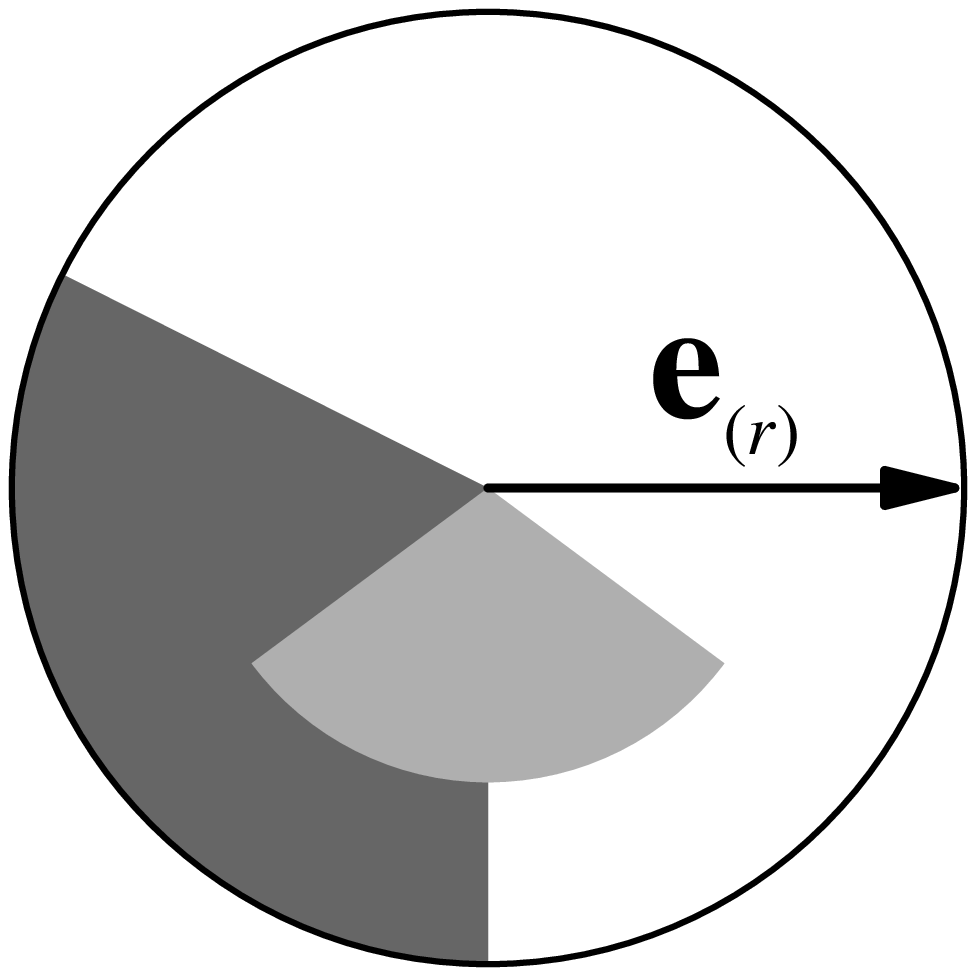}
\end{minipage}\hfill%
\begin{minipage}[b]{.19\hsize}
\centering\leavevmode\epsfxsize=\hsize \epsfbox{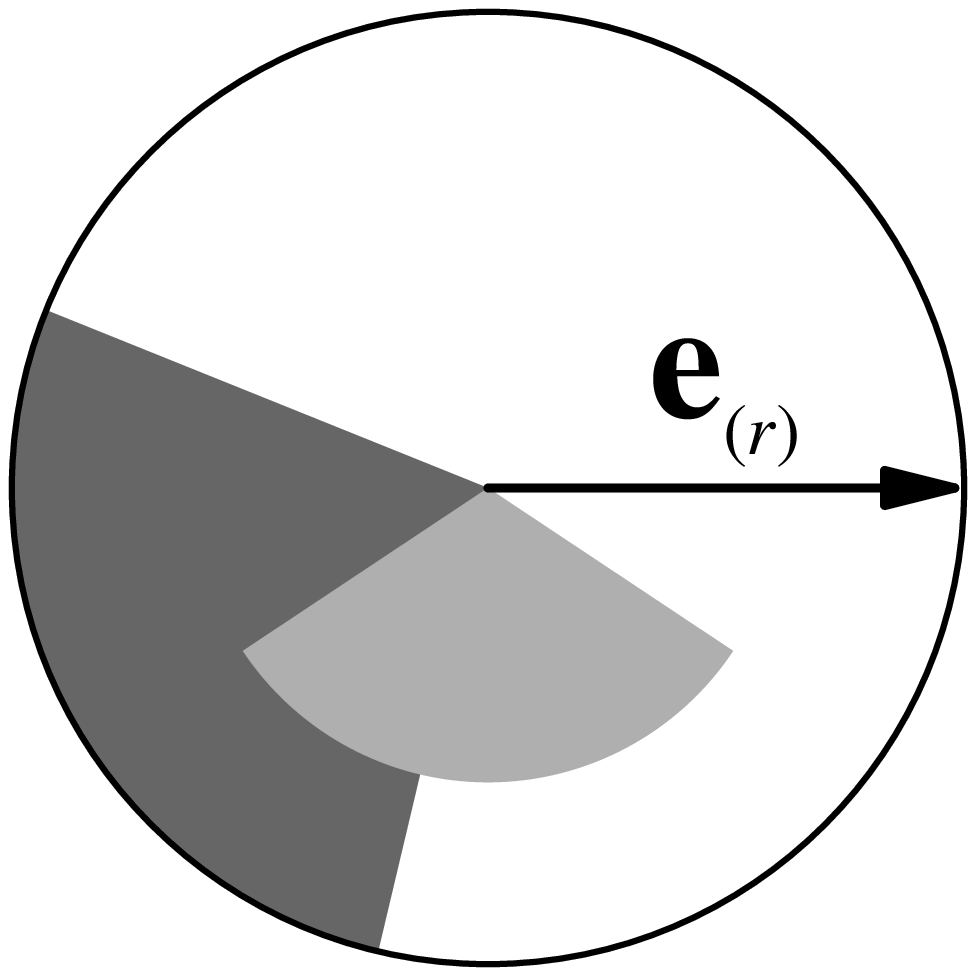}
\end{minipage}

\medskip

\rule{\textwidth}{.8pt}

\medskip 

\centering
\begin{minipage}[b]{.19\hsize}
\centering $r=r_{{\ri max}+,2}$\par\smallskip\par
\leavevmode\epsfxsize=\hsize \epsfbox{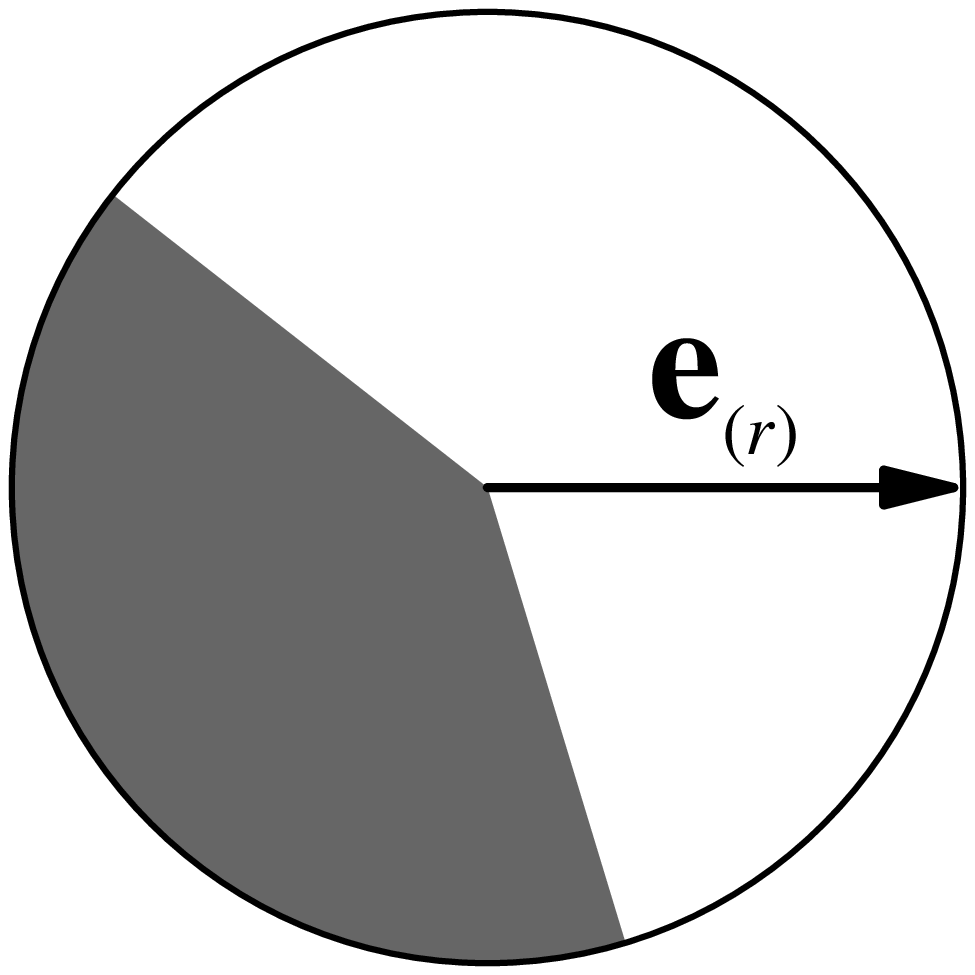}
\end{minipage}\hfill%
\begin{minipage}[b]{.19\hsize}
\centering $r=3.3=r_{{\ri c},3}-\delta$\par\smallskip\par
\leavevmode\epsfxsize=\hsize \epsfbox{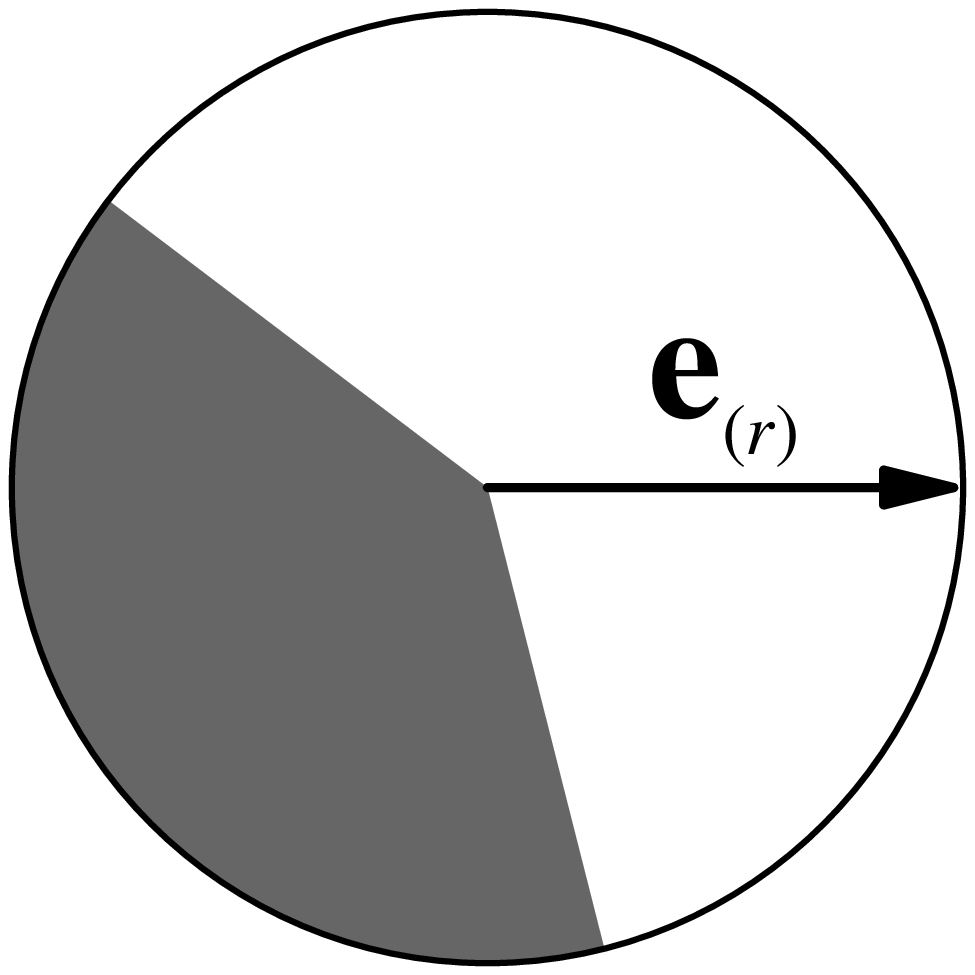}
\end{minipage}\hfill%
\begin{minipage}[b]{.19\hsize}
\centering $r=3.6$\par\smallskip\par
\leavevmode\epsfxsize=\hsize \epsfbox{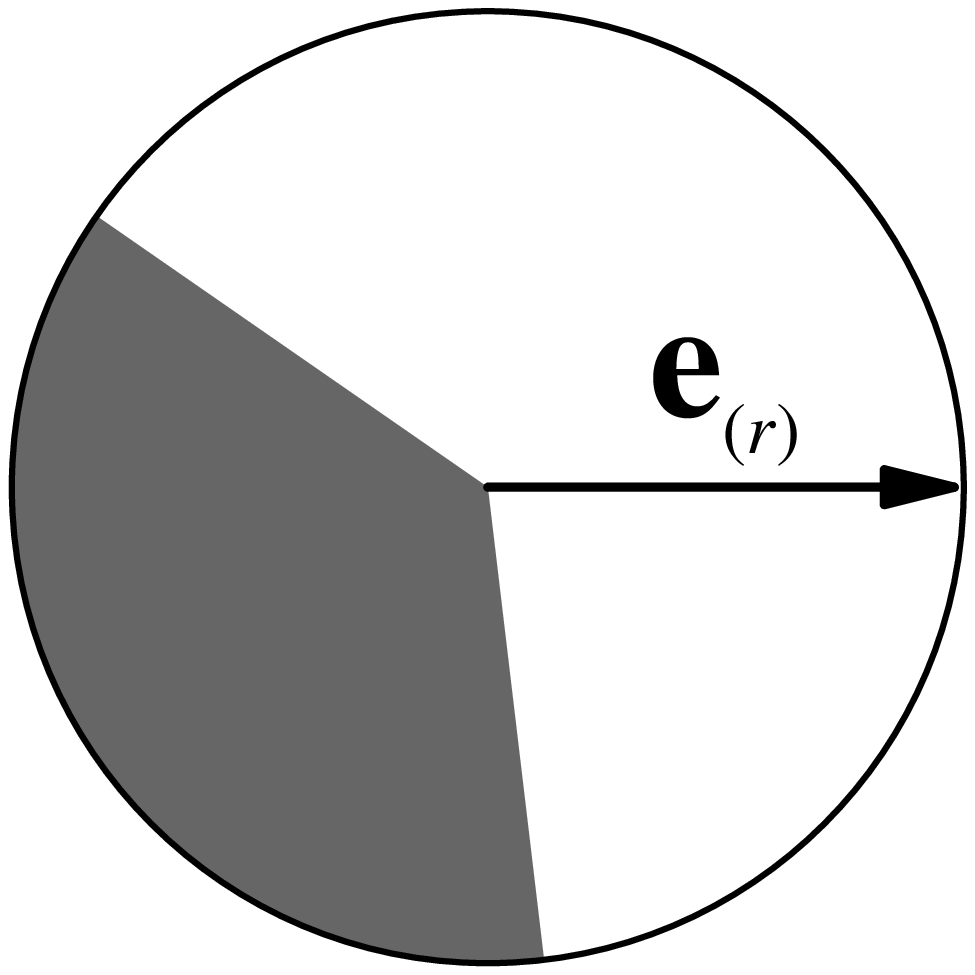}
\end{minipage}\hfill%
\begin{minipage}[b]{.19\hsize}
\centering $r=r_{{\ri max}+,1}$\par\smallskip\par
\leavevmode\epsfxsize=\hsize \epsfbox{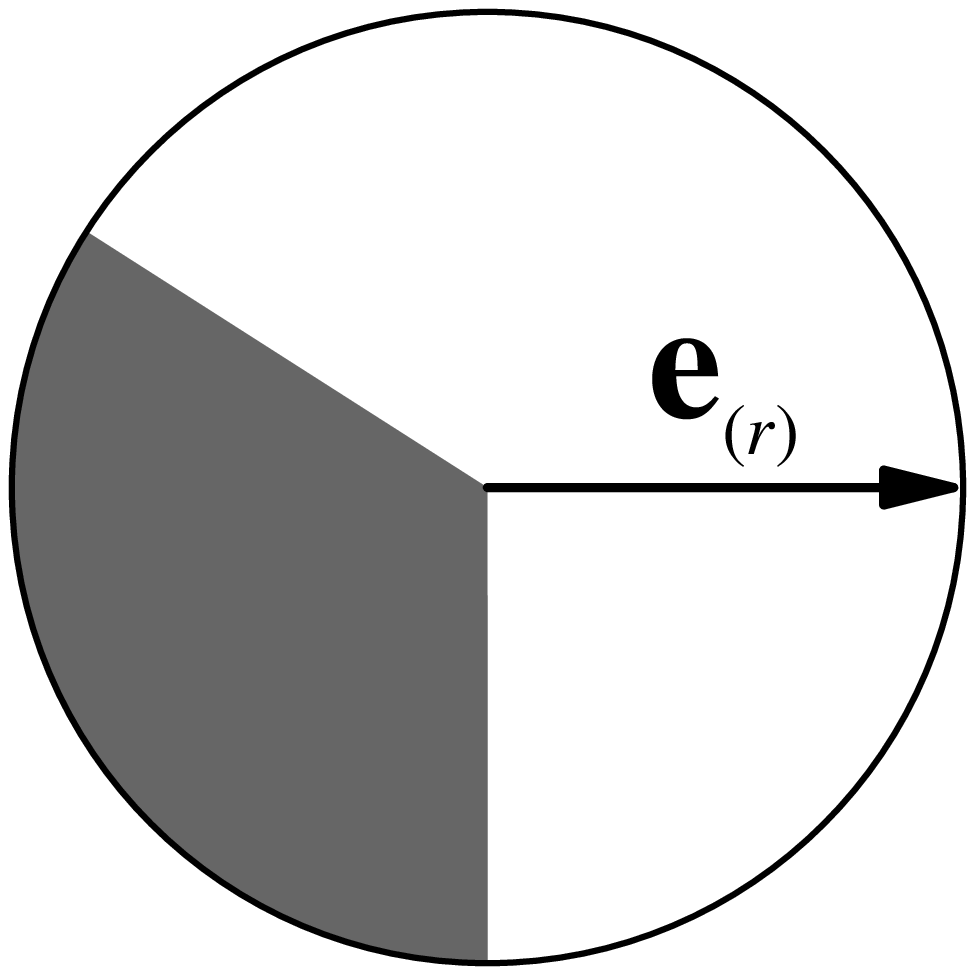}
\end{minipage}\hfill%
\begin{minipage}[b]{.19\hsize}
\centering $r=4$\par\smallskip\par
\leavevmode\epsfxsize=\hsize \epsfbox{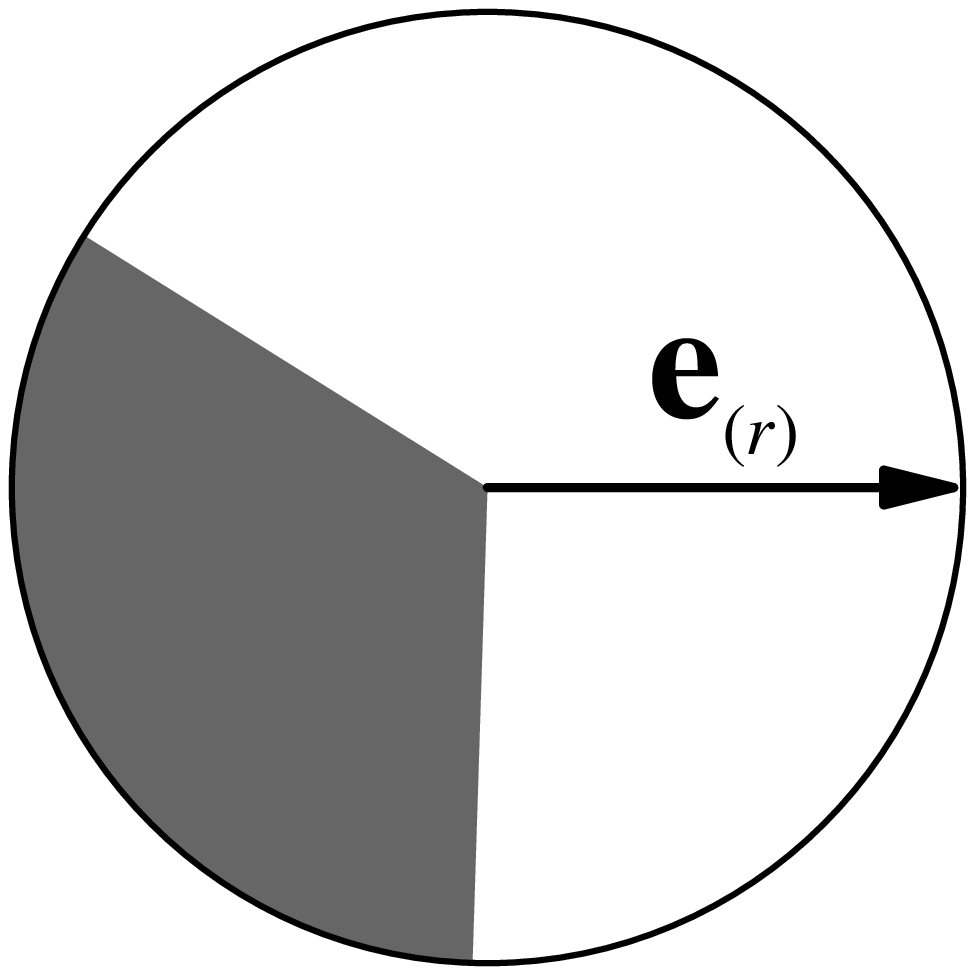}
\end{minipage}

\medskip

\begin{minipage}[b]{.19\hsize}
\centering\leavevmode\epsfxsize=\hsize \epsfbox{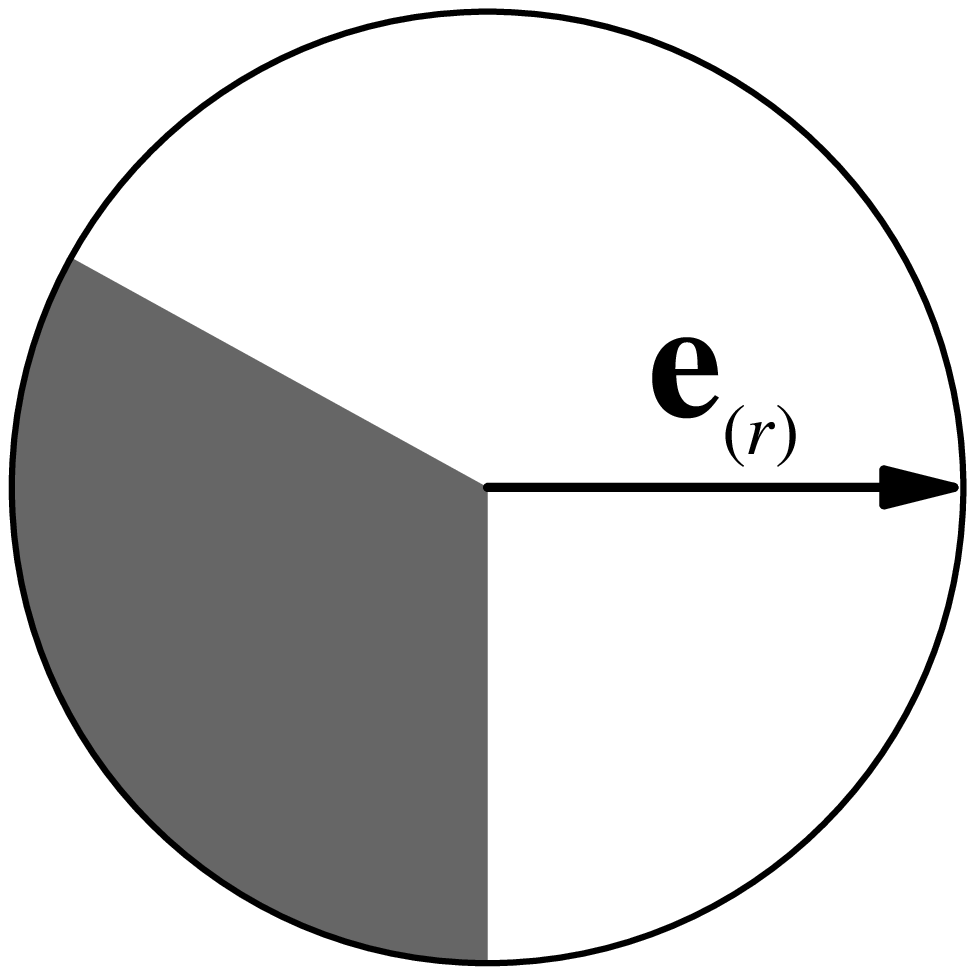}
\end{minipage}\hfill%
\begin{minipage}[b]{.19\hsize}
\centering\leavevmode\epsfxsize=\hsize \epsfbox{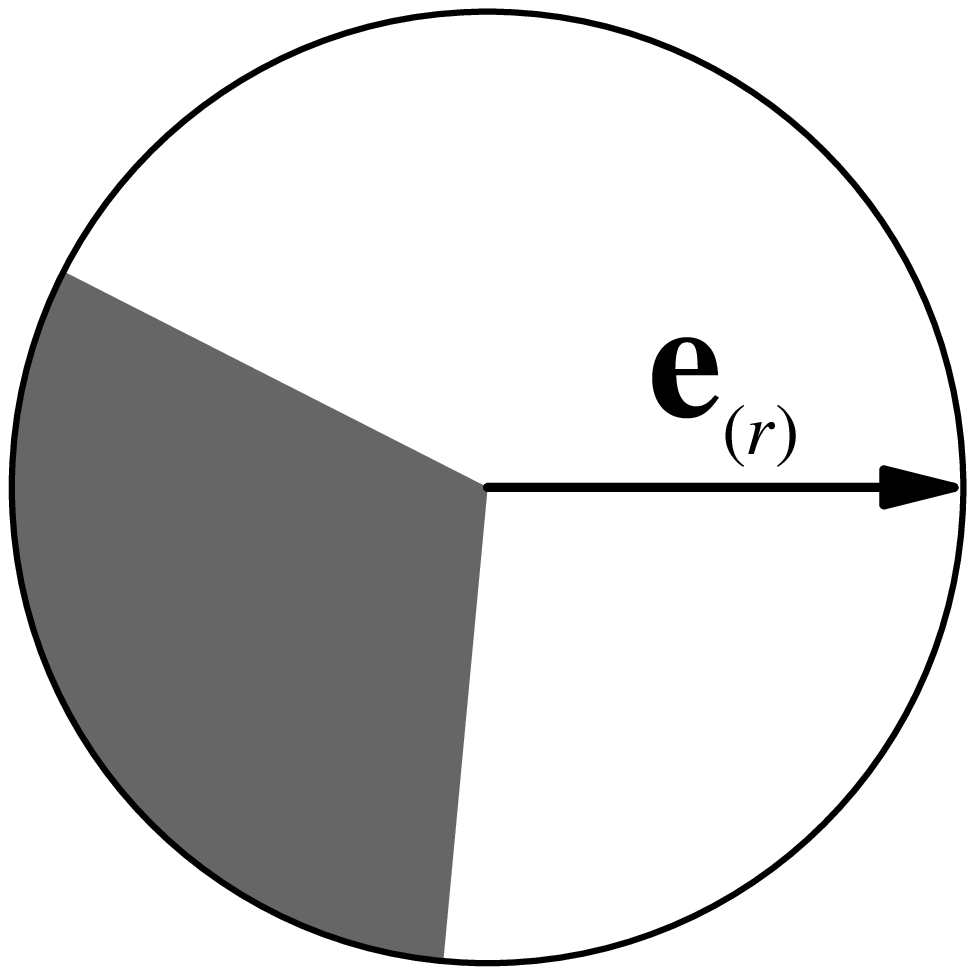}
\end{minipage}\hfill%
\begin{minipage}[b]{.19\hsize}
\centering\leavevmode\epsfxsize=\hsize \epsfbox{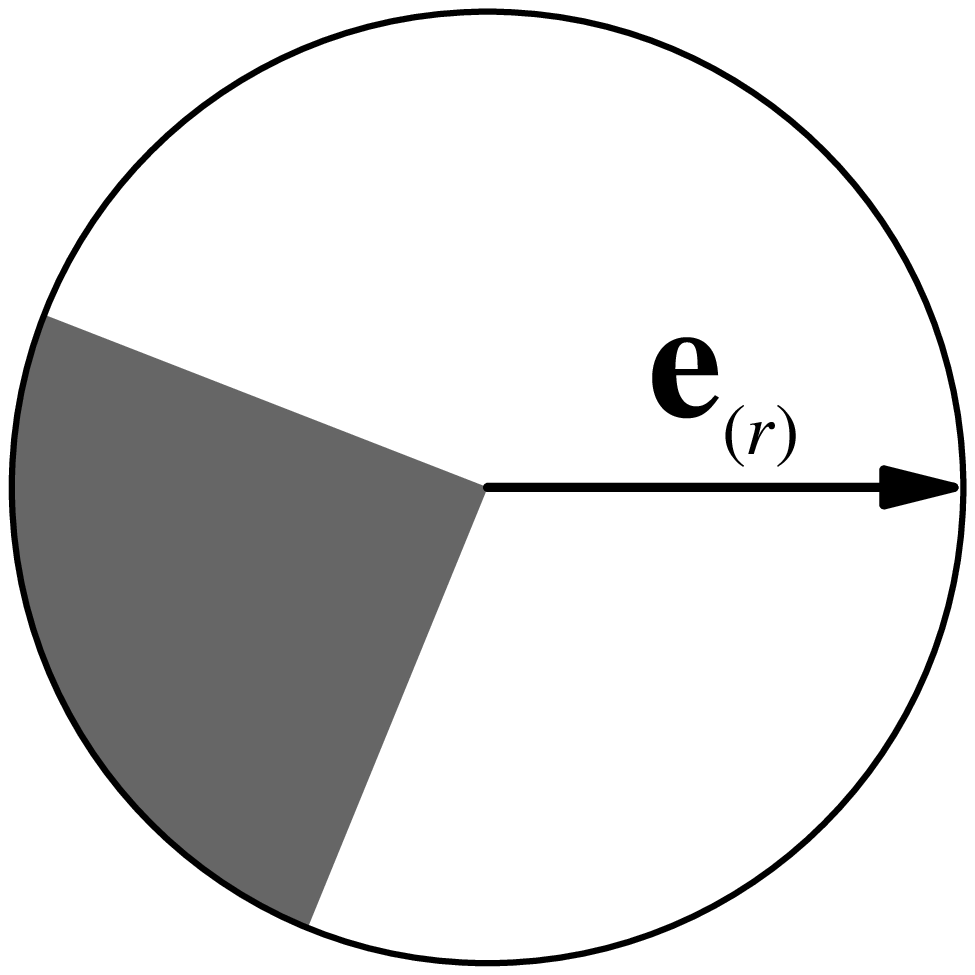}
\end{minipage}\hfill%
\begin{minipage}[b]{.19\hsize}
\centering\leavevmode\epsfxsize=\hsize \epsfbox{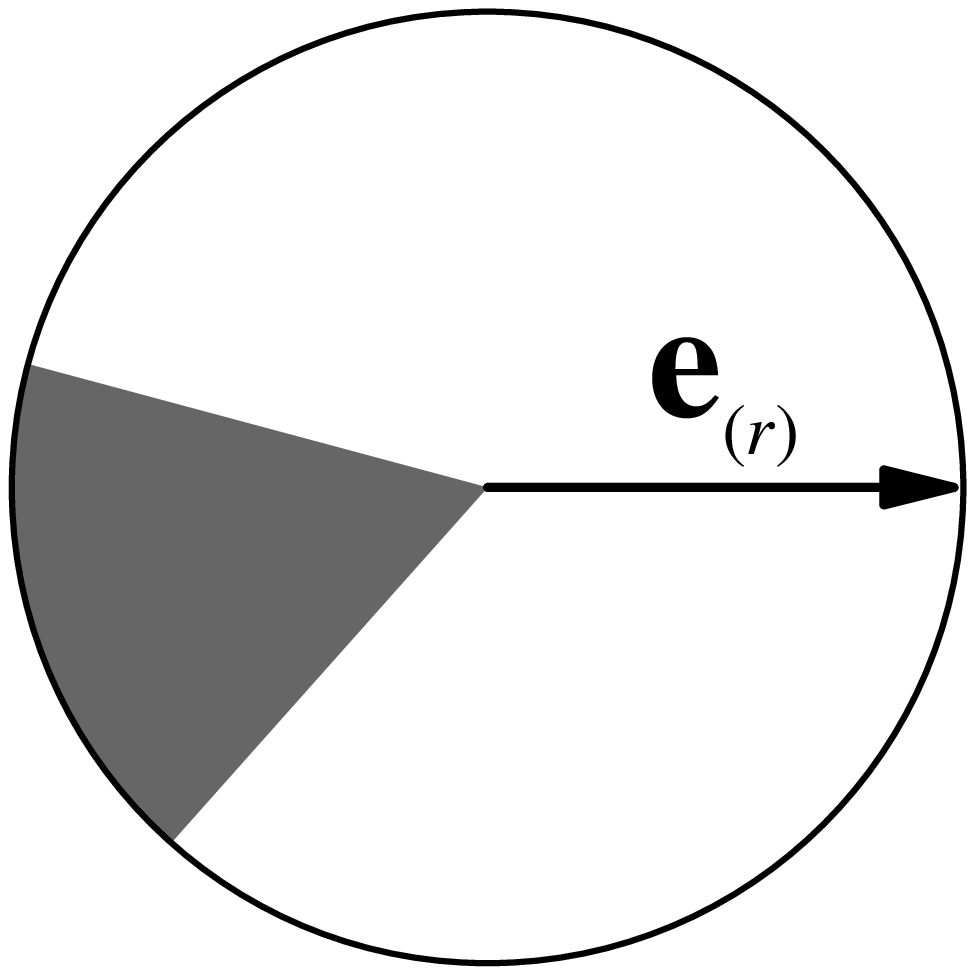}
\end{minipage}\hfill%
\begin{minipage}[b]{.19\hsize}
\centering\leavevmode\epsfxsize=\hsize \epsfbox{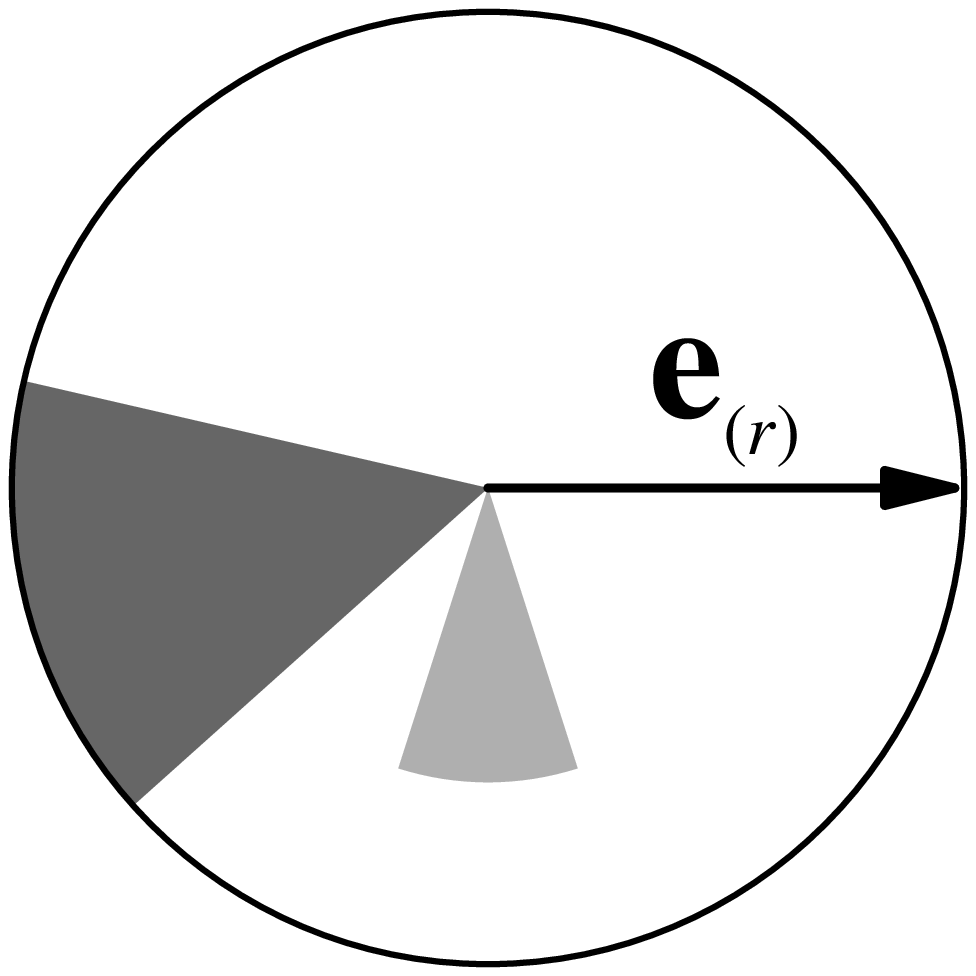}
\end{minipage}

\medskip

\begin{minipage}[b]{.19\hsize}
\centering\leavevmode\epsfxsize=\hsize \epsfbox{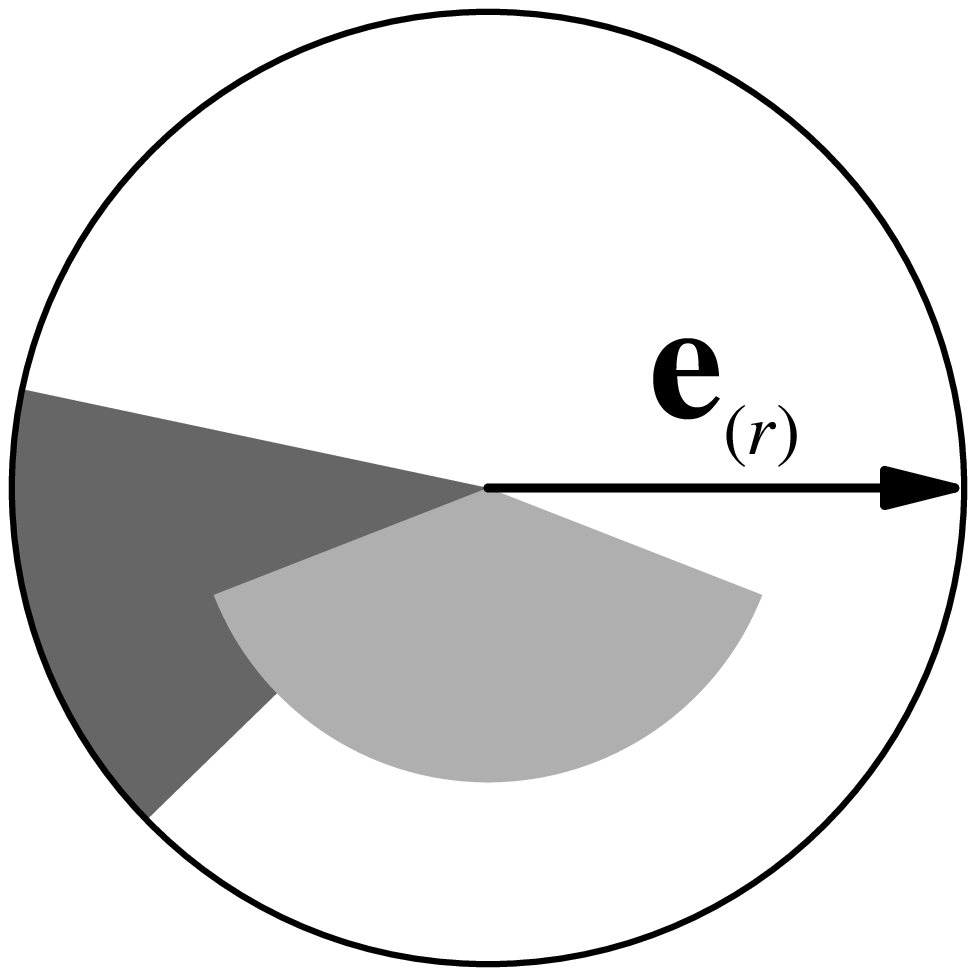}
\end{minipage}\hfill%
\begin{minipage}[b]{.19\hsize}
\centering\leavevmode\epsfxsize=\hsize \epsfbox{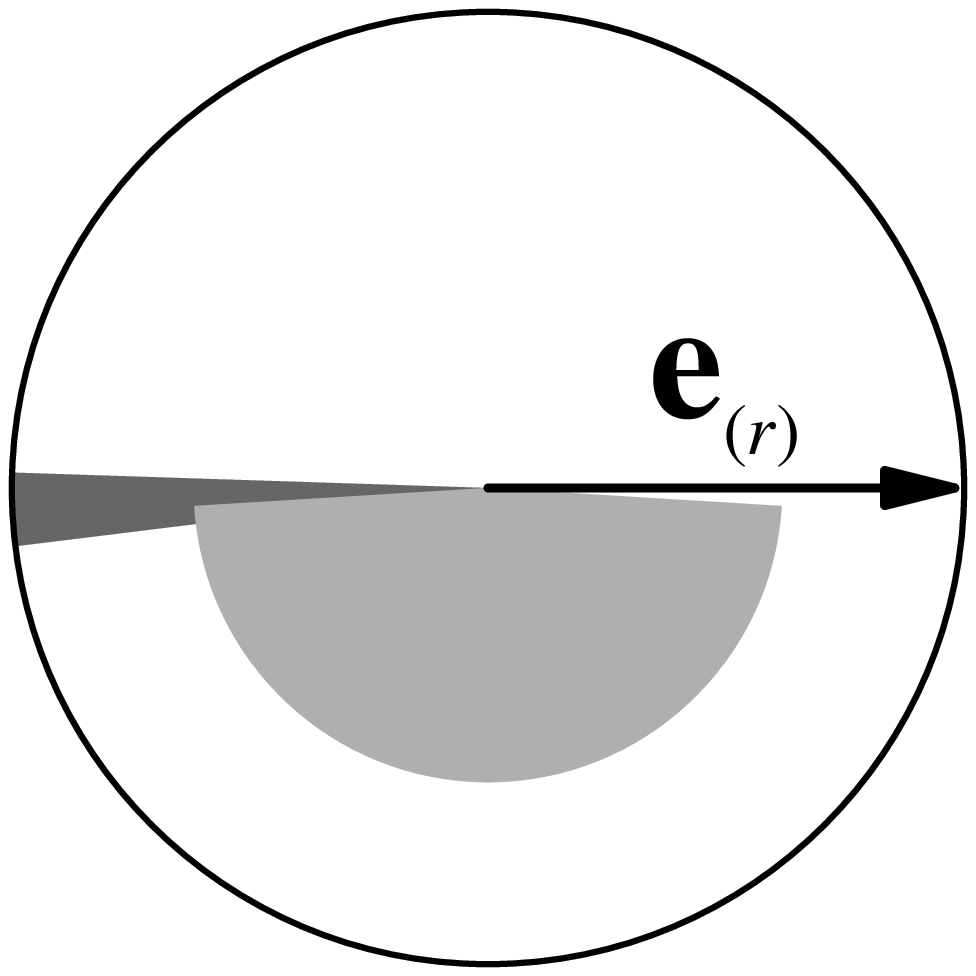}
\end{minipage}\hfill%
\begin{minipage}[b]{.19\hsize}
\mbox{}
\end{minipage}\hfill%
\begin{minipage}[b]{.19\hsize}
\mbox{}
\end{minipage}\hfill%
\begin{minipage}[b]{.19\hsize}
\mbox{}
\end{minipage}

\medskip

\rule{\textwidth}{.8pt}

\smallskip

{(\it Figure continued)}
\end{figure}

\begin{figure}[t]
\footnotesize

\rule{\textwidth}{.8pt}

\medskip 

\centering
\begin{minipage}[b]{.19\hsize}
\centering $r=4.2$\par\smallskip\par
\leavevmode\epsfxsize=\hsize \epsfbox{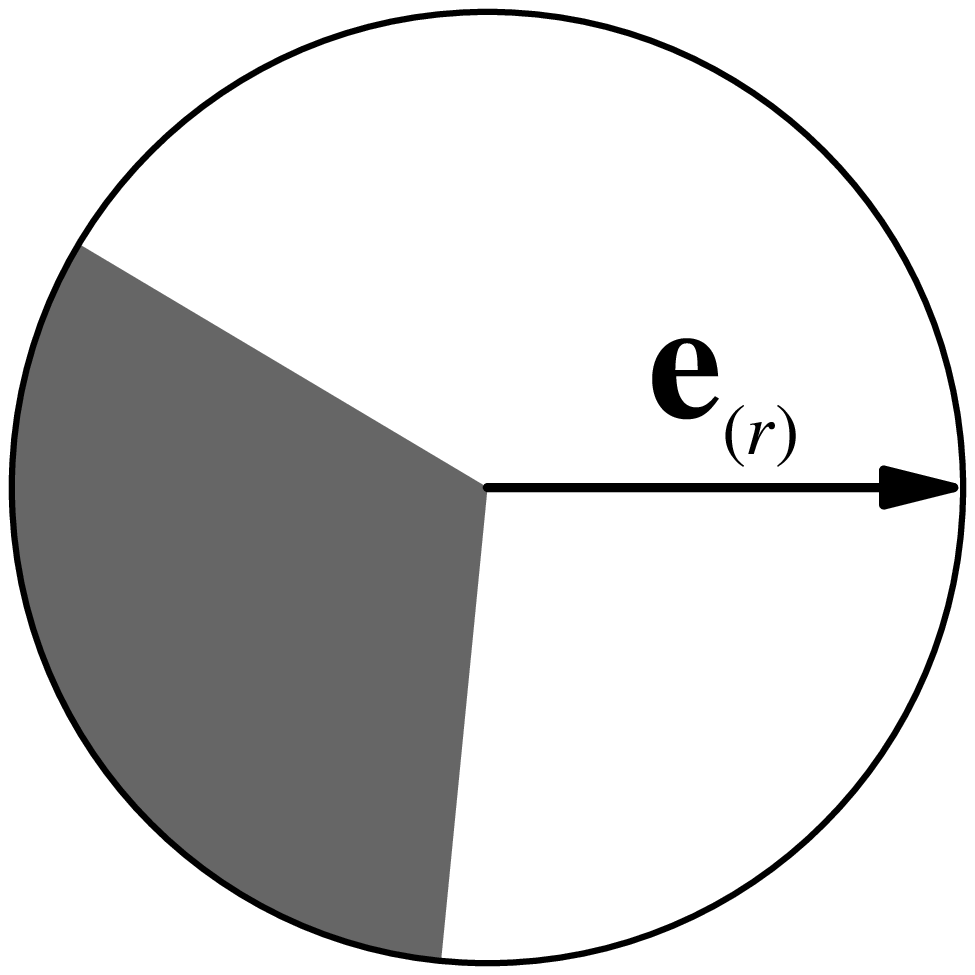}
\end{minipage}\hfill%
\begin{minipage}[b]{.19\hsize}
\centering $r=4.29=r_{{\ri c},2}-\delta$\par\smallskip\par
\leavevmode\epsfxsize=\hsize \epsfbox{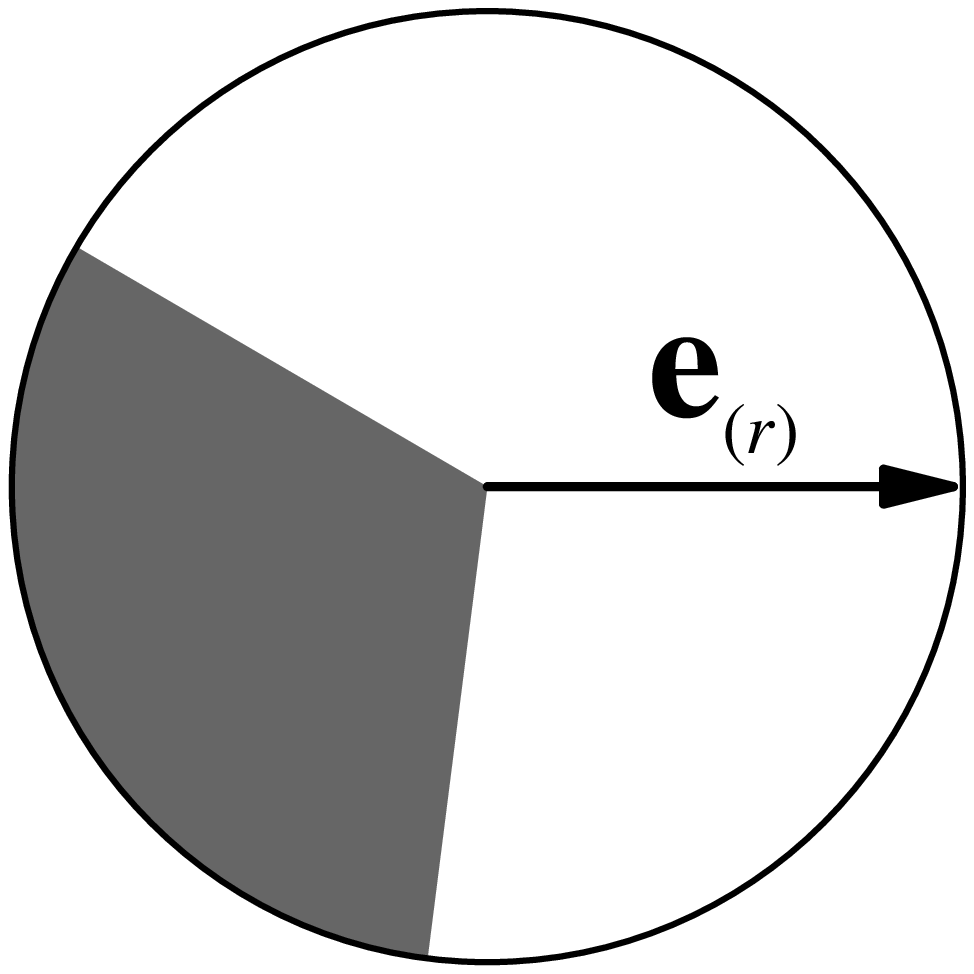}
\end{minipage}\hfill%
\begin{minipage}[b]{.19\hsize}
\centering $r=6$\par\smallskip\par
\leavevmode\epsfxsize=\hsize \epsfbox{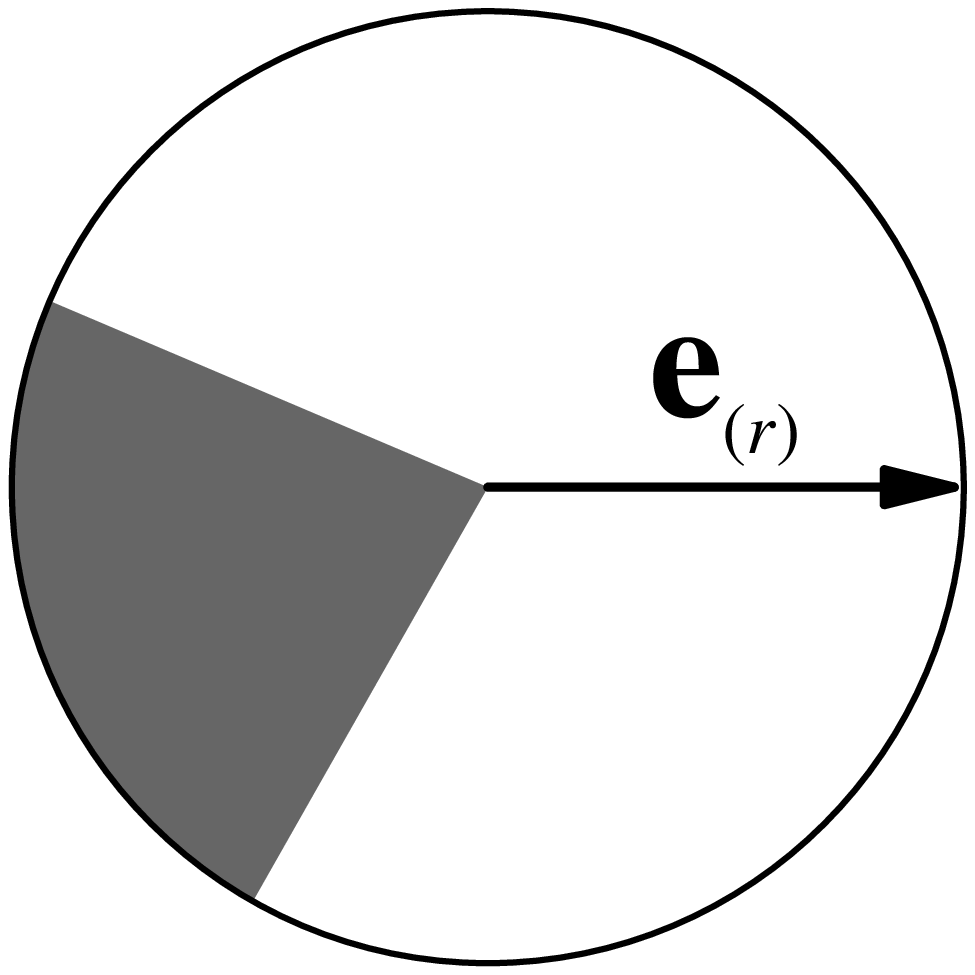}
\end{minipage}\hfill%
\begin{minipage}[b]{.19\hsize}
\centering $r=25$\par\smallskip\par
\leavevmode\epsfxsize=\hsize \epsfbox{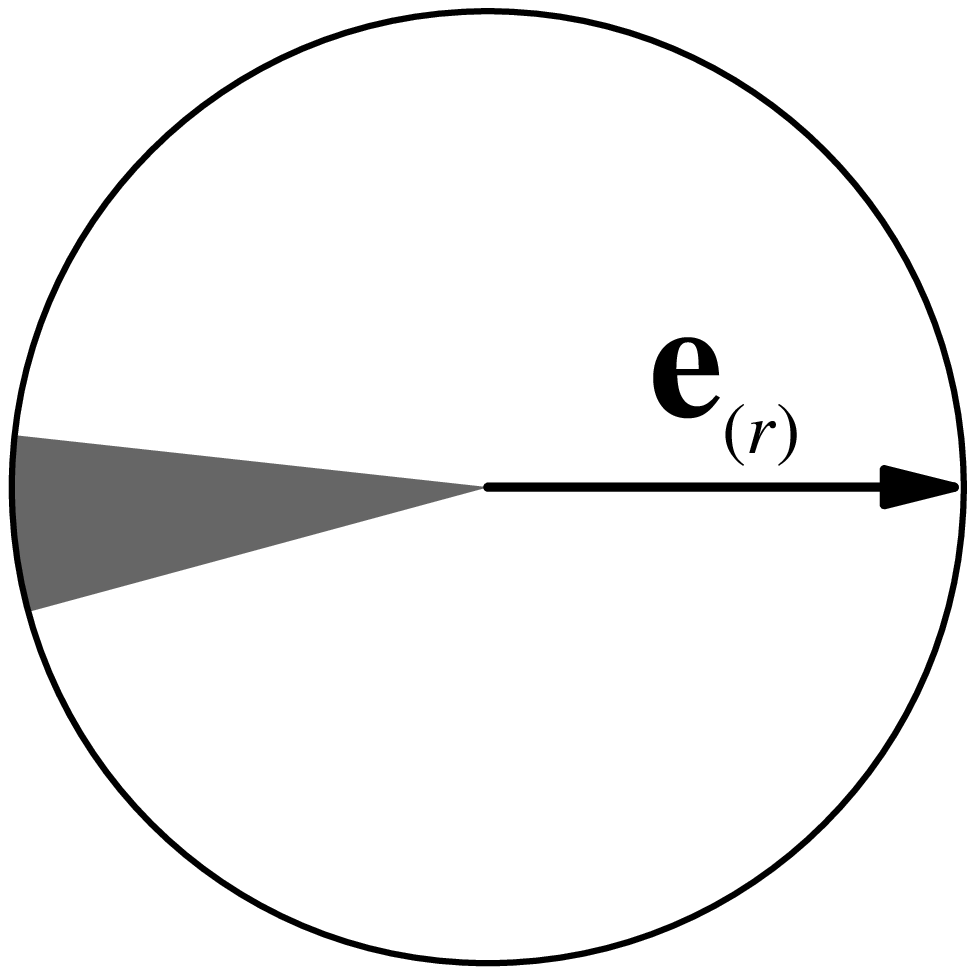}
\end{minipage}\hfill%
\begin{minipage}[b]{.19\hsize}
\centering $r=100={}$`$r_{{\ri c},1}-\delta$'\par\smallskip\par
\leavevmode\epsfxsize=\hsize \epsfbox{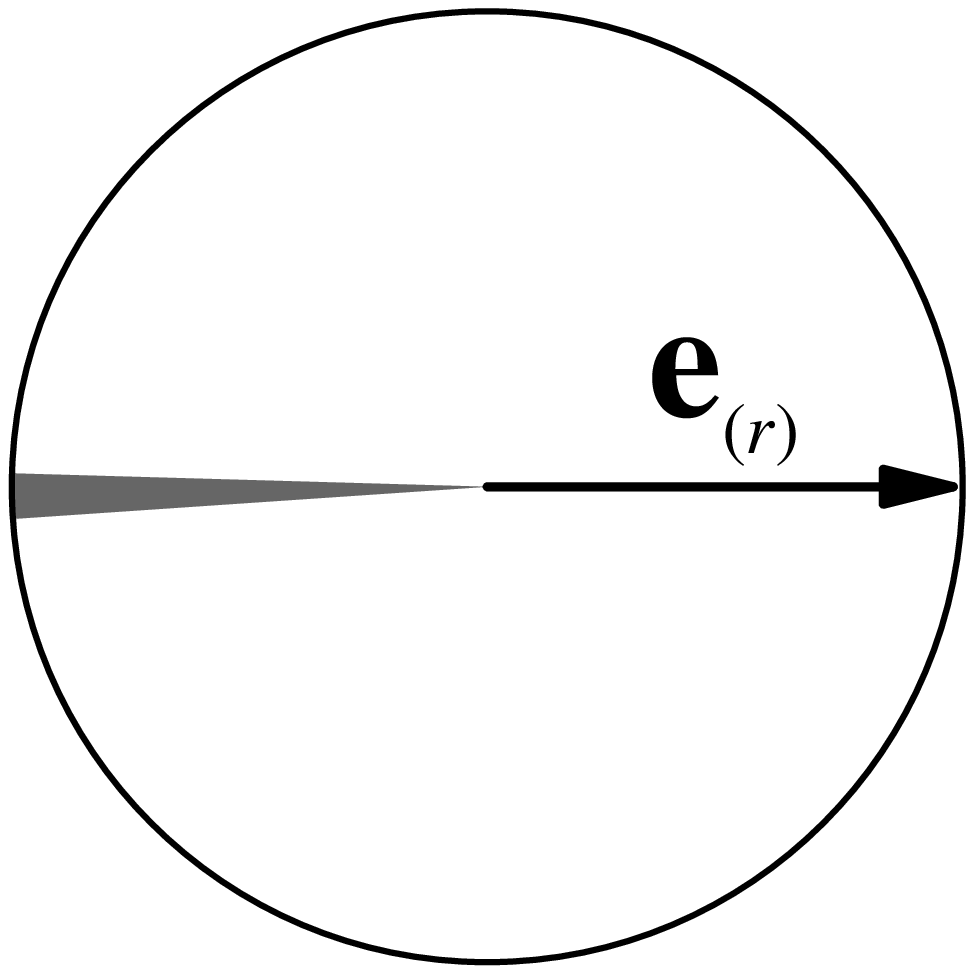}
\end{minipage}

\medskip

\begin{minipage}[b]{.19\hsize}
\centering\leavevmode\epsfxsize=\hsize \epsfbox{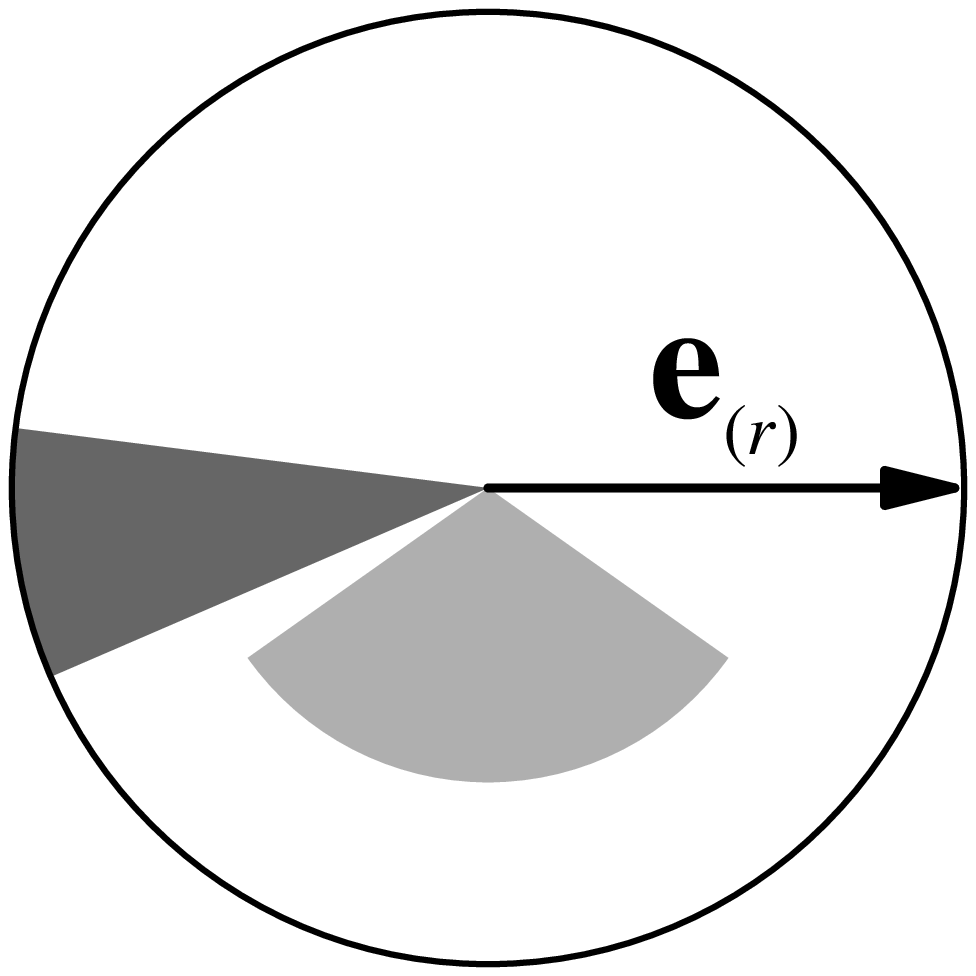}
\end{minipage}\hfill%
\begin{minipage}[b]{.19\hsize}
\centering\leavevmode\epsfxsize=\hsize \epsfbox{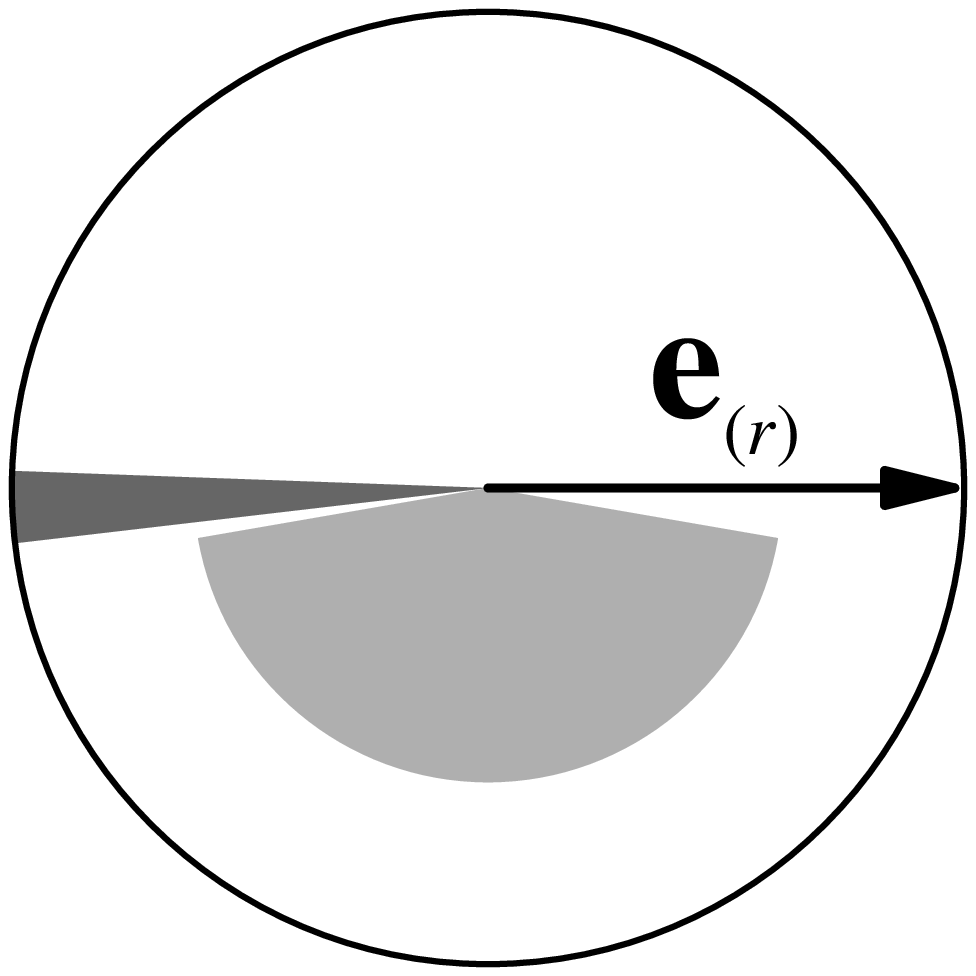}
\end{minipage}\hfill%
\begin{minipage}[b]{.19\hsize}
\mbox{}
\end{minipage}\hfill%
\begin{minipage}[b]{.19\hsize}
\mbox{}
\end{minipage}\hfill%
\begin{minipage}[b]{.19\hsize}
\mbox{}
\end{minipage}

\medskip

\rule{\textwidth}{.8pt}

\caption{Sectors of captured (dark gray) and escaping (white) equatorial
  photons for several values of the radius of a LNRF observer for fixed
  $e^2=0$, $a^2=0.81$, and (1)~$y=0$ (pure Kerr, top rows), (2)~$y=0.03$
  (\KdS\ with `ordinary' divergent barrier, middle row),
  (3)~$y=0.04$ (\KdS\ with \RRB, bottom row, if any).  The upper and bottom
  half-pies correspond to corotating and counterrotating equatorial
  photons, respectively.  Small light gray sectors (if present) mark the
  interval in which the impact parameter $X(\alpha)$ of the counterrotating
  equatorial photons is positive. In all the cases above, $\delta$ refers
  to an appropriately small number. The values of the radii $r_{\ri h+}$
  of the outer black-hole horizon, the radii $r_{\ri min-}$ of the local
  minima of $X_-$, the radii $r_{\ri max+}$ of the local maxima of $X_+$,
  and the radii $r_{\ri c}$ of the cosmological horizon, are summarized in
  Table~\protect\ref{t1}.}
\label{f15}
\end{figure}

In order to understand the nature of the black-hole spacetimes with a
\RRB, it is important to notice that there can
exist two angles $\psi_{\ri d\pm}$ symmetric with respect to $\psi =
3\pi/2$, for which the function $X(\psi;r,y,a)$, or the function
$\ell(\psi;r,y,a)$, diverges. The photons counterrotating relative to the
locally non-rotating observers at directional angles $\psi\in (\psi_{\ri
  d-}, \psi_{\ri d+})$ have positive impact parameter $X$ (or $\ell$),
contrary to the situations we are accustomed to from non-rotating
backgrounds. The angles of divergence are determined by the relation
\be
  \sin \psi_{\ri d\pm} =
    -\frac{r^2 \Delta^{1/2}_r}{A \Omega} =
    -\frac{r^2 \left[(1-yr^2) (r^2+a^2) - 2r\right]^{1/2}}
    {a \left[yr^2 (r^2+a^2) + 2r \right]}.
\ee
Of course, they exist only if the condition $|\sin \psi_{\ri d\pm}| \leq 1$
is satisfied. This condition implies inequality
\be
  \left[r-2-yr \left(r^2+a^2\right)\right] \left[r^3+a^2 (r+2)+ ya^2r
    \left(r^2+a^2\right)\right] \leq 0.
\ee
However, the condition $X_+(r;y,a) > 0$ implies the inequality
\be
  r\left[r-2-yr \left(r^2+a^2\right) \right] \leq 0.
\ee

\begin{table}[t]
\caption{The values of the radii $r_{\ri h+}$
  of the outer black-hole horizon, the radii $r_{\ri min-}$ of the local
  minima of $X_-$, the radii $r_{\ri max+}$ of the local maxima of $X_+$,
  and the radii $r_{\ri c}$ of the cosmological horizon
  (see Fig.\,\protect\ref{f15}).}
\label{t1}
\footnotesize
\begin{center}
\begin{tabular}{lccccc}
\hline
Case &
$y$  &
Outer BH&
Radius of the local &
Radius of the local &
Cosmological \\
{} &
   &
horizon &
minimum of $X_-$ &
maximum of $X_+$ &
horizon \\ \hline
1 &
0 &
$r_{{\ri h}+,1}=  1.435890$ &
$r_{{\ri min}-,1}=1.557855$ &
$r_{{\ri max}+,1}=3.910268$ &
$r_{{\ri c},1}=   \infty$ \\ \hline
2 &
0.03 &
$r_{{\ri h}+,2}=  1.728078$ &
$r_{{\ri min}-,2}=1.859046$ &
$r_{{\ri max}+,2}=3.201891$ &
$r_{{\ri c},2}=   4.298185$ \\ \hline
3 &
0.04 &
$r_{{\ri h}+,3}=  1.931337$ &
$r_{{\ri min}-,3}=2.045904$ &
$r_{{\ri max}+,3}=2.886223$ &
$r_{{\ri c},3}=   3.302544$ \\ \hline
\end{tabular}
\end{center}
\end{table}

Thus, we can conclude that the angles of divergence occur just at those
regions of rotating spacetimes, where the effective potential of the radial
photon motion $X_+(r;y,a) \geq 0$. Such situation appears in the Kerr
spacetimes between the outer horizon and the surface $r=2$. In the \KdS\ 
black-hole spacetimes with a divergent repulsive barrier it appears at the
vicinity of both the outer black-hole and cosmological horizons, while in
the black-hole spacetimes with a \RRB\ it appear
everywhere between the black-hole and cosmological horizons.

In the case of Kerr black-holes the divergent angles are located inside the
photon capture cone. For \KdS\ black holes with a divergent repulsive
barrier, the angles of divergence are located inside the photon capture cone
in vicinity of the black-hole horizon, while they are located inside the
photon escape cone in vicinity of the cosmological horizon. For \KdS\ 
black-hole spacetimes with a \RRB, a new phenomenon
arises: region between the angles of divergence enters both the escape and
capture cones at each radius between the horizons (see~Fig.\,\ref{f15}).

\section{The azimuthal motion} \label{azimot}

\begin{figure}[b]
\centering\leavevmode
\epsfxsize=.8\hsize \epsfbox{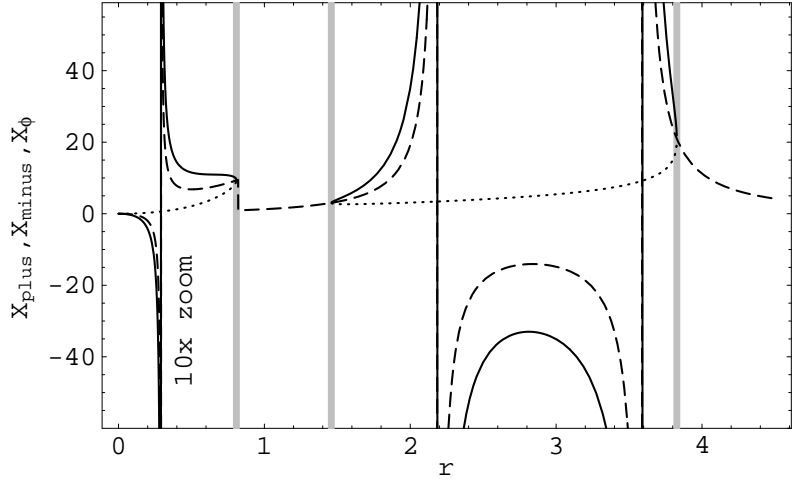}
\caption{The turning points of the azimuthal equatorial motion of
  photons. The function $X_{\phi}(r;y,a,e)$ (dashed curve) is drawn along
  with the effective potential of the radial motion $X_+(r;y,a,e)$ (solid
  curve) and $X_-(r;y,a,e)$ (dotted curve). The turning points of the
  azimuthal motion can appear only in regions where $X_+(r;y,a,e) < 0$.
  This example is drawn for $y = 0.036$, $a^2 = 0.49$, $e^2 = 0.5$ (class
  Ia); the horizons are depicted as thick gray bars. The portion below the
  inner black-hole horizon is vertically zoomed to enable distinguish the
  curves clearly.}
\label{f16}
\end{figure}

\begin{figure}[p]
\let\figlabsize=\small
\centering\leavevmode
\begin{minipage}[b]{0.32\hsize}
\centering\leavevmode
\epsfxsize=\hsize \epsfbox{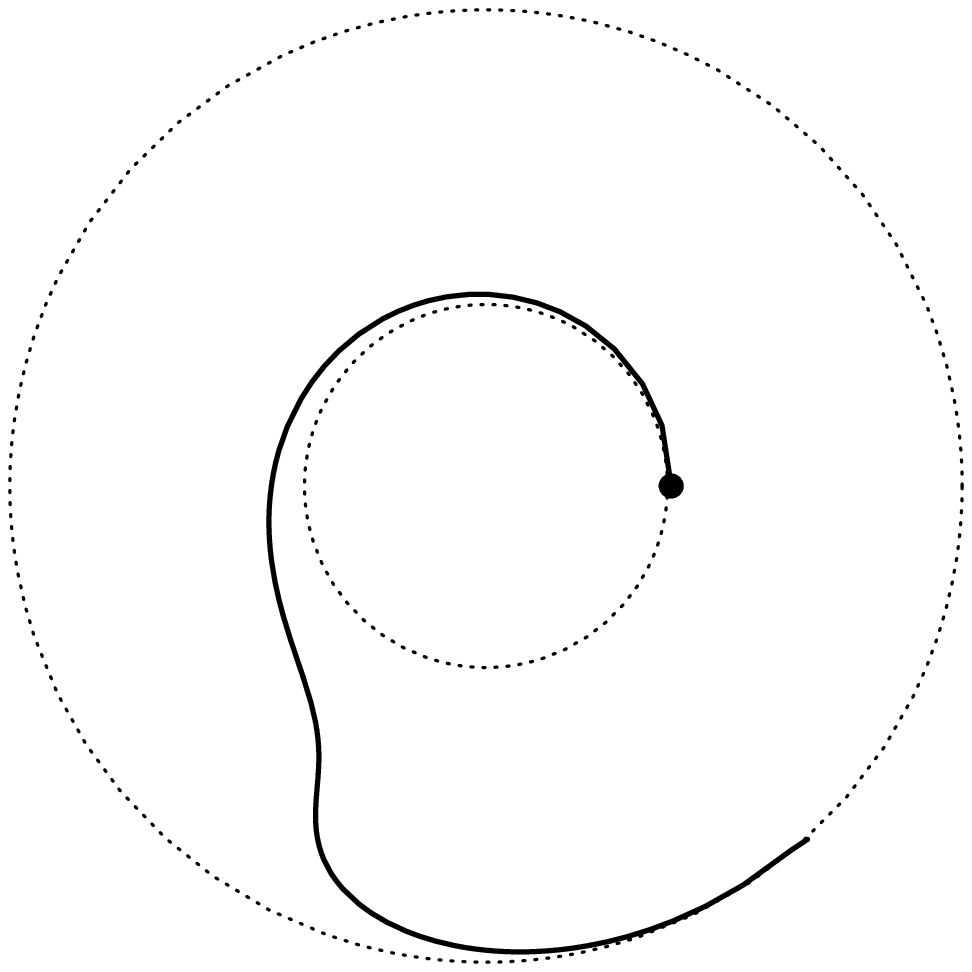}
\end{minipage}\hfill%
\begin{minipage}[b]{0.32\hsize}
\centering\leavevmode
\epsfxsize=\hsize \epsfbox{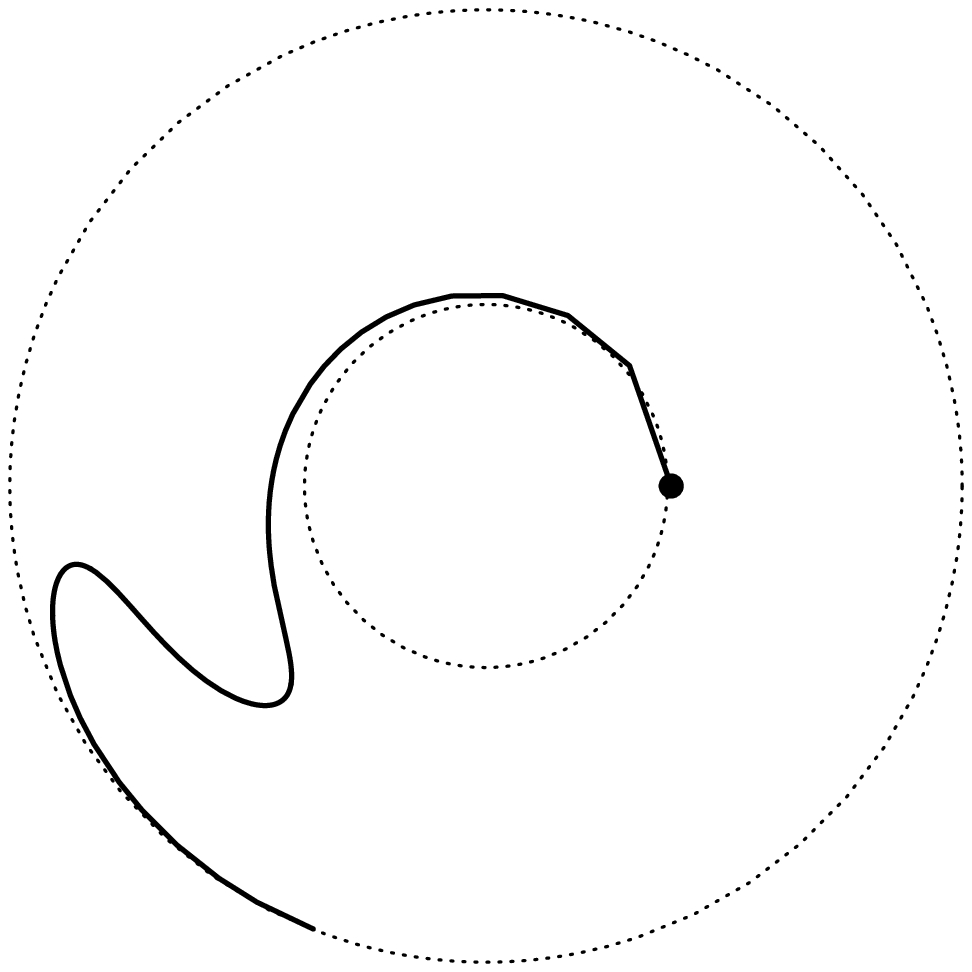}
\end{minipage}\hfill%
\begin{minipage}[b]{0.32\hsize}
\centering\leavevmode
\epsfxsize=\hsize \epsfbox{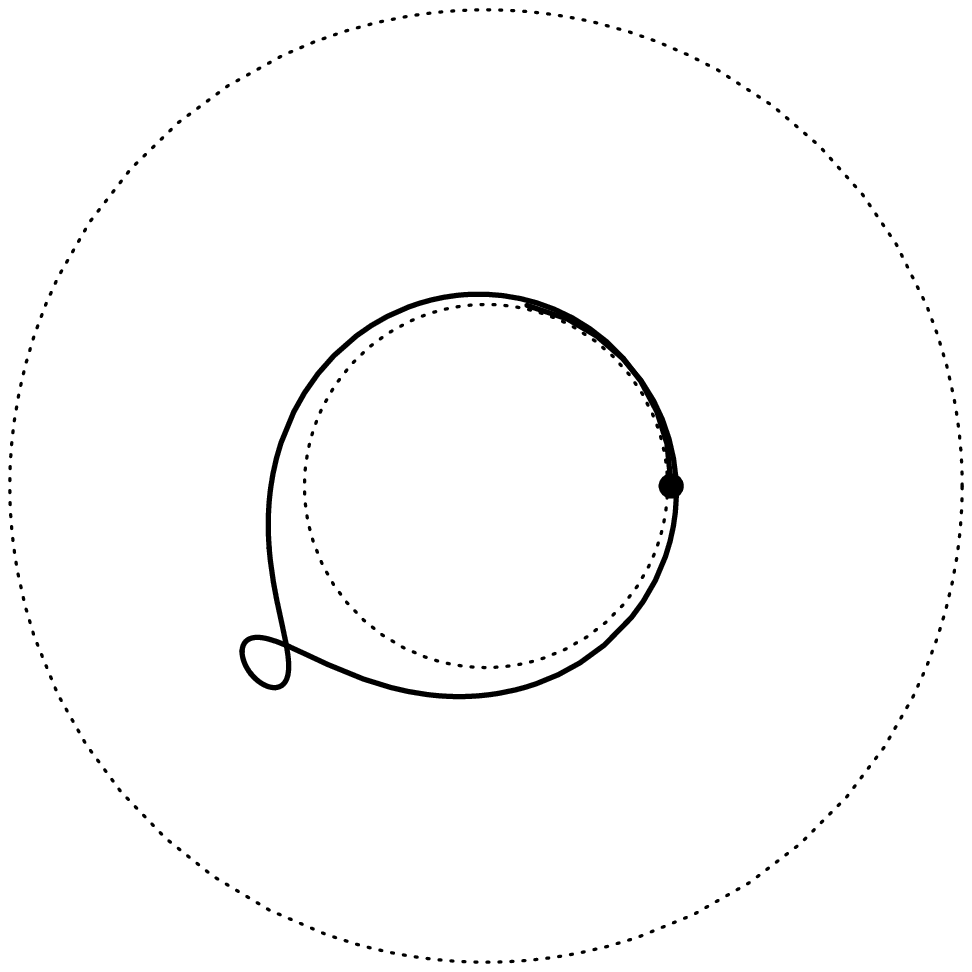}
\end{minipage}
\par\vskip 4mm\par
\begin{minipage}[b]{0.32\hsize}
\centering\leavevmode
\epsfxsize=\hsize \epsfbox{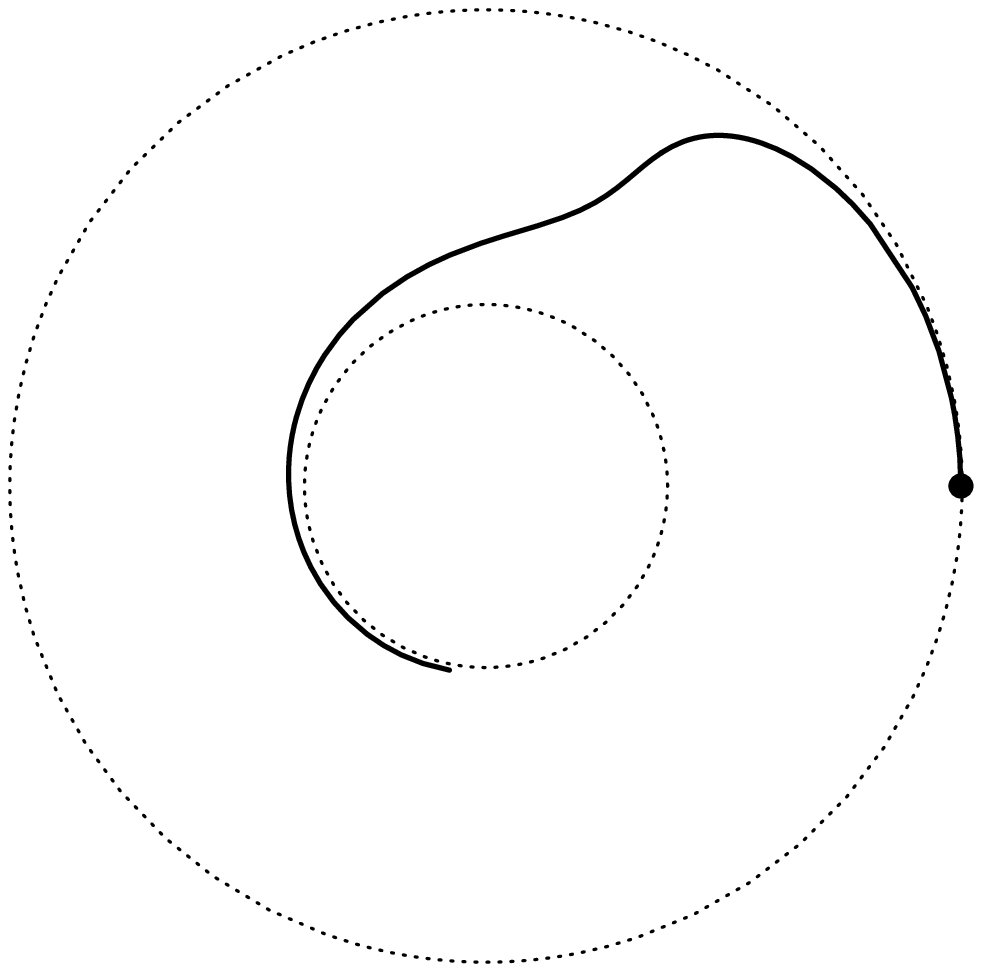}
\end{minipage}\hfill%
\begin{minipage}[b]{0.32\hsize}
\centering\leavevmode
\epsfxsize=\hsize \epsfbox{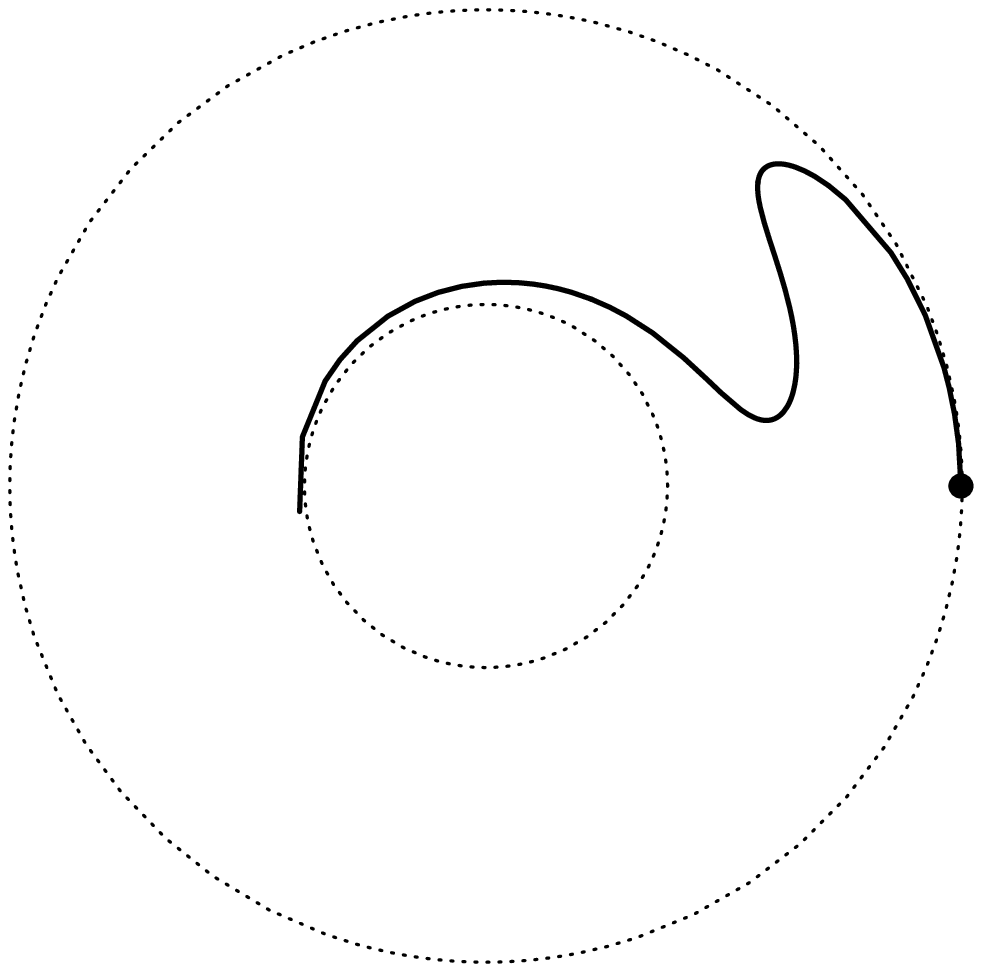}
\end{minipage}\hfill%
\begin{minipage}[b]{0.32\hsize}
\centering\leavevmode
\epsfxsize=\hsize \epsfbox{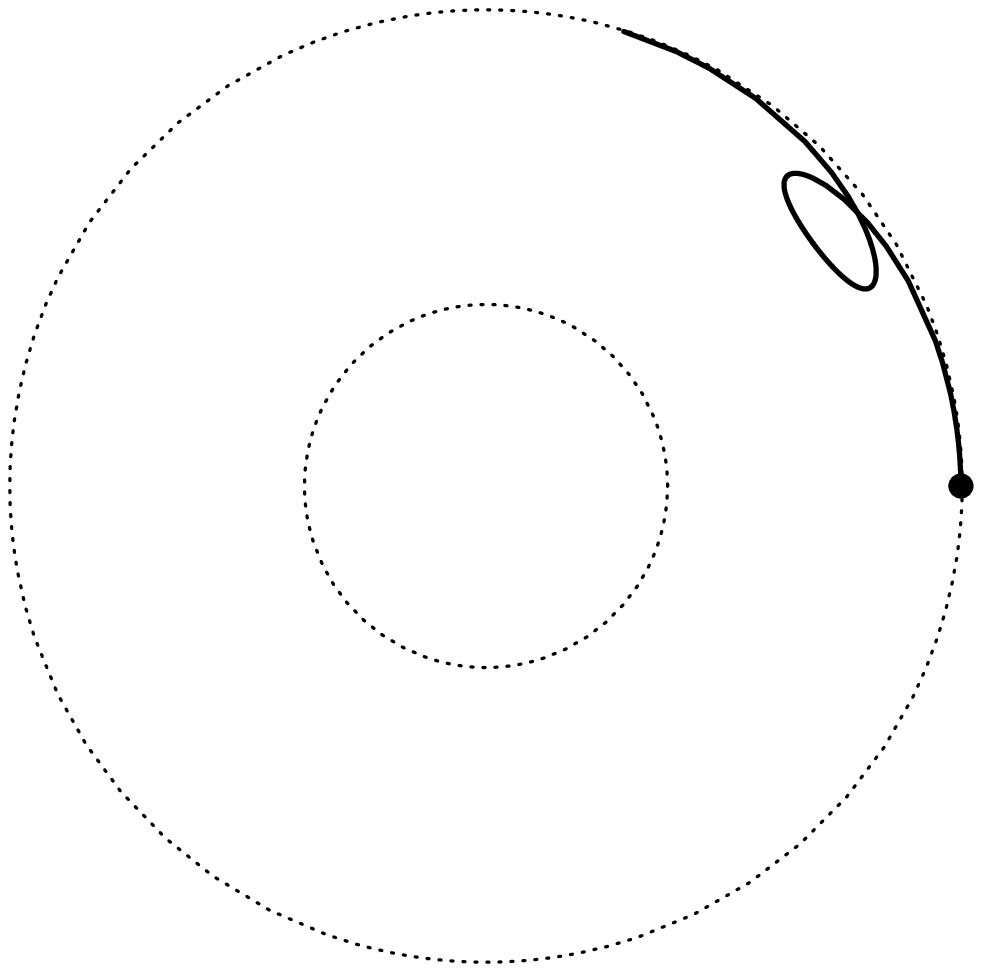}
\end{minipage}
\par\vskip 4mm\par
\begin{minipage}[b]{0.32\hsize}
\centering\leavevmode
\epsfxsize=.7\hsize \epsfysize=1.22\hsize \epsfbox{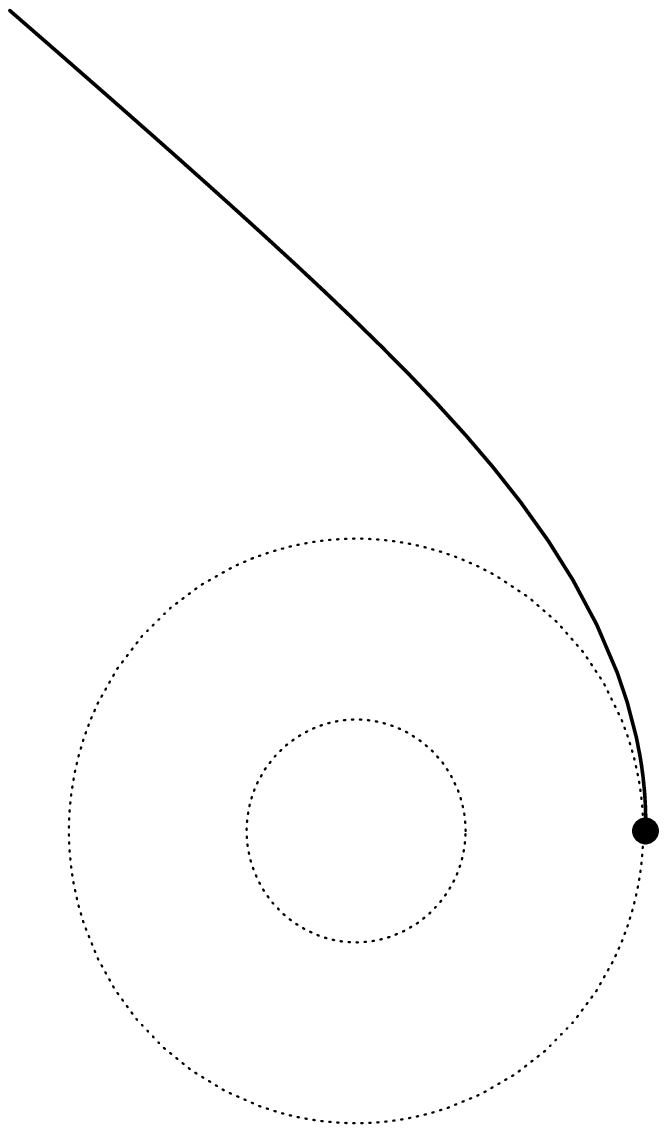}
\end{minipage}\hfill%
\begin{minipage}[b]{0.32\hsize}
\centering\leavevmode
\centering\leavevmode
\epsfxsize=.9\hsize \epsfysize=0.64\hsize \epsfbox{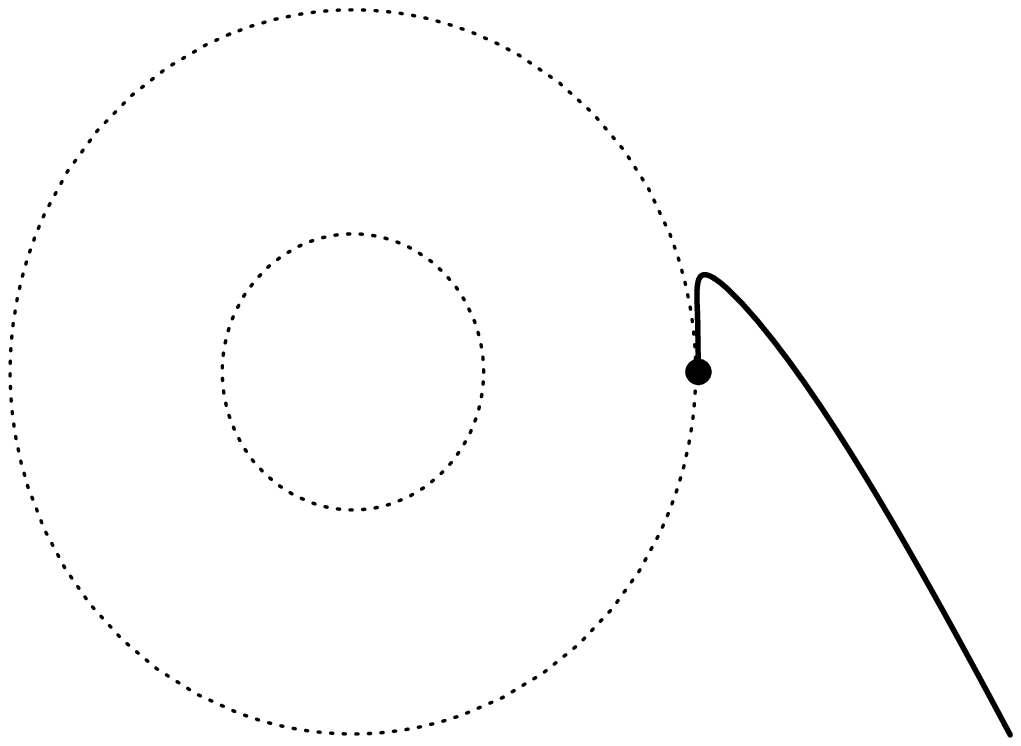}
\end{minipage}\hfill%
\begin{minipage}[b]{0.32\hsize}
\mbox{}
\end{minipage}
\caption{Trajectories of the equatorial photon motion
  obtained by numerical solving the equations of motion. The two upper rows
  depict the trajectories in the region between the outer black-hole and
  cosmological horizons, the bottom row depicts the trajectories in the
  region beyond the cosmological horizon. The trajectories with both $r$
  and $\phi$ monotonously increasing/decreasing are collected in the left
  column, the trajectories with turning point only in $\phi$ are collected
  in the middle column, and trajectories with turning points both in $r$
  and $\phi$ are collected in the right column (note that there is no such
  trajectory beyond the cosmological horizon). The values of spacetime
  parameters are $y = 0.036$, $a^2 = 0.49$, $e^2 = 0.5$ (class Ia), the
  value of impact parameter $X$ and the initial conditions are the
  following. The two top rows -- $X = -10$ (left), $X = -30$ (middle), $X =
  -50$ (right), the top row has $r_0 = r_{\ri h+} + \delta$, $\dot{r}_0 >
  0$, the middle row has $r_0 = r_{\ri c} - \delta$, $\dot{r}_0 < 0$. The
  bottom row -- $X = 30$ (left), $X = 10$ (middle), $r_0 = r_{\ri c} +
  \delta$, $\dot{r}_0 > 0$. (Here, $\delta$ refers to an appropriately
  small number.)}
\label{f17}
\end{figure}

The equation of the azimuthal motion in the equatorial plane can be written
in the form
\be
  \oder{\phi}{\lambda} =
    \frac{I^2}{r^2 \Delta^2}
    \left[(\Delta_r - a^2) X+ ar^2 \right].              \label{aziequamot}
\ee
Therefore, turning points of the azimuthal motion (where $\d \phi/\d
\lambda = 0$) are determined by the condition
\be
  X = X_{\phi}(r;y,a,e) \equiv
    \frac{ar^2}{a^2 - \Delta_r}.
\ee

There is only one zero point of $X_{\phi}(r;y,a,e)$, which is located at
$r=0$ for any values of parameters $y$, $a$, $e$. Divergences of
$X_{\phi}(r;y,a,e)$ are determined by the relation
\be
  y_{\ri d(\phi)}(r;a,e) \equiv
   \frac{r^2 - 2r + e^2}{r^2 (r^2+a^2)} \equiv y_{\ri d} (r;a,e).
\ee
Therefore, the divergent points of $X_{\phi}(r;y,a,e)$ coincide with the
divergent points of $X_+(r;y,a,e)$. Since
\be
  \pder{X_{\phi}}{r} = \frac{2ar}{(a^2 - \Delta_r)^2}
    (-yr^4 + r - e^2),
\ee
the extrema of $X_{\phi}(r;y,a,e)$ are given by the relation independent of
the parameter $a$
\be
  y = y_{\ri ex(\phi)}(r;e) \equiv \frac{r-e^2}{r^4}.
\ee
We can write
\be
  X_{\phi}(r;y,a,e) = \frac{a}{a+ \sqrt{\Delta r}} X_+(r;y,a,e).
\ee
Because $a/(a+ \sqrt{\Delta r}) < 1$, we can conclude that
\be
  X_{\phi}(r;y,a,e) \leq X_+ (r;y,a,e),
\ee
so that the turning points of the azimuthal motion must be located in the
regions forbidden by the conditions of the radial motion, if $X_+ > 0$.
However, they can exist in the regions where $X_+ < 0$.

We give an example of the behavior of the function $X_{\phi}$ (and function
$X_{\pm}$) in the case of the spacetimes of the class Ia in
Fig.\,\ref{f16}.

By combining the azimuthal equation of motion (\ref{aziequamot}) with the
radial one (\ref{radequamot}), we obtain the equation for trajectories of
the equatorial motion in the form
\be
  \oder{\phi}{r} = \pm \frac{I}{\Delta_r}
    \frac{(\Delta_r - a^2) X + ar^2}
    {\sqrt{(a^2 - \Delta_r)X^2 - 2ar^2 X + r^4}};          \label{EqPhiMot}
\ee
the $+(-)$ sign corresponds to the outward (inward) motion. The trajectory
equation was integrated for typical values of the impact parameter $X$ in
the case of spacetimes of the class Ia. (The integral (\ref{EqPhiMot}) can
be expressed in terms of elliptic integrals, but the expressions are too
complex.) We illustrate the typical trajectories in Fig.\,\ref{f17}; notice
the most interesting trajectories with the turning point of the azimuthal
motion, which can be both with or without the turning point of the radial
motion.

\section{Concluding remarks} \label{concl}

The analysis of the effective potential of the radial motion of photons in
the equatorial plane of the \KN\ spacetimes with a non-zero cosmological
constant enables us to separate these spacetimes into eighteen classes
according to qualitatively different character of the effective potential
reflecting appropriately the properties of the geometry.

From the behavior of the effective potential, one can easily achieve some
general conclusions about the equatorial photon motion.
\begin{enumerate}
\item In any class of the \KN\ spacetimes with $y \neq 0$ the ring
  singularity (at $r=0$, $\theta= \pi/2$) can be reached by photons with
  impact parameter $X=0$ (or $\ell = a$). No other photons can reach the
  ring singularity.
  
\item Outside the outer black-hole horizon, two unstable photon circular
  orbits always exist. Additional two circular photon orbits can exist
  under the inner horizon, the innermost being stable, the other being
  unstable.  This behavior holds for both asymptotically \dS, and \adS\ 
  black holes. Naturally, this property holds also for the \KN\ spacetimes
  with $y = 0$.
  
\item There can exist naked-singularity spacetimes (with both $y>0$, $y<0$)
  containing no circular photon orbit.
  
\item If the naked-singularity spacetimes contain four (or two) circular
  photon orbits, then two (one) of them are stable, while the others are
  unstable.
  
\item In some parts of the field of rotating black holes, there exists an
  unusual relation between directional angles of equatorial photons as
  measured by locally non-rotating observers, and their impact parameters.
  Namely, locally counterrotating photons have positive values of the
  impact parameter. In the field of Kerr black holes, this phenomenon is
  limited to vicinity of the black-hole horizon and all such photons must
  be captured by the black hole. For the \KNdS\ holes with a divergent
  repulsive barrier, this phenomenon is limited to vicinity of the
  black-hole horizon (with all such photons being captured by the hole) and
  to vicinity of the cosmological horizon (with all such photons escaping
  through the cosmological horizon). However, for the \KNdS\ holes with a
  \RRB, this phenomenon appears at all radii
  between the black-hole and cosmological horizon, and at all radii such
  photons are partly captured by the hole and partly escape through the
  cosmological horizon. Further, the existence of the \RRB\ is directly
  related to the fact that photons with positive impact parameter $X$ can
  be counterrotating relative to the locally non-rotating frames in the
  complete stationary region between the outer black-hole and cosmological
  horizons. All photons with positive impact parameter lying above the
  \RRB\ are counterrotating in locally non-rotating frames. We probably
  could expect special optical effects connected with the \RRB\ of the
  photon motion.
  
\item A \RRB\ exists also for the non-equatorial motion of photons in the
  \KNdS\ black hole spacetimes with a \RRB\ of the equatorial photon
  motion. One can see it directly, if along with the parameter $X$ a new
  impact parameter $q$ is introduced in such a way that it disappears for
  the equatorial motion.  The effective potential of the radial motion can
  then be given in the form \cite{Stuchlik-Hledik}
  \be
    X_{\pm}(r;q,y,a,e) \equiv
      \frac{ar^2 \pm \sqrt{\Delta_r \left[r^4+q(a^2 - \Delta_r) \right]}}
      {a^2 - \Delta_r}.
  \ee
  Clearly, the properties of divergent points of this function are just the
  same as for the effective potential of the equatorial photon motion.
  
  The classification of the \KN\ spacetimes with $y \neq 0$, introduced in
  the analysis of the equatorial photon motion, can be useful also for the
  analysis of the non-equatorial photon motion. Particularly, the phenomena
  of the restricted repulsive barrier is also relevant for the
  non-equatorial motion.
  
  A combined discussion of the radial and latitudinal motion enables us to
  determine photon escape cones of local observers, and, further, to make
  calculations of various optical phenomena.

\item Turning points of the azimuthal motion can occur only at the region,
  where the inequality $X_+(r;y,a,e) < 0$ holds. Trajectories with an
  azimuthal turning point can also have a radial turning
  point. Trajectories of photons beyond the cosmological horizon can have a
  turning point of the azimuthal motion, however, naturally, no turning
  point of the radial motion.
\end{enumerate}

\section*{Acknowlwdgements}

This work has been supported by the GA\v{C}R Grant No.202/99/0261, by the
Committee for Collaboration of Czech Republic with CERN and by the Bergen
Computational Physics Laboratory project, an EU Research Infrastructure at
the University of Bergen, Norway, supported by the European Community --
Access to Research Infrastructure Action of the Improving Human Potential
Programme. The authors would like to acknowledge the perfect hospitality
and excellent working conditions at the CERN's Theory Division and the
Institute of Physics of the University of Bergen.

\end{document}